\documentclass[]{aa} 

\usepackage{graphicx, textcomp}
\usepackage{multirow}
\usepackage{natbib}
\usepackage[switch]{lineno}
\usepackage[]{threeparttable}
\usepackage{txfonts}
 \usepackage{hyperref}
 \hypersetup{
      colorlinks   = true,
      citecolor    = blue
 }

\begin{document}

   \title{Jet collimation in NGC\,315 and other nearby AGN}

\author{B. Boccardi\inst{1,2} 
\and M. Perucho\inst{3,4}
\and C. Casadio\inst{5,1}
\and P. Grandi\inst{2}
\and D. Macconi\inst{2,6}
\and E. Torresi\inst{2}
\and S. Pellegrini\inst{2,6}
\and T.P. Krichbaum\inst{1}
\and M. Kadler\inst{7}
\and G. Giovannini\inst{6,8}
\and V. Karamanavis \inst{9}
\and L. Ricci \inst{1}
\and E. Madika\inst{1} 
\and U. Bach\inst{1}
\and E. Ros \inst{1}
\and M. Giroletti\inst{8}
\and J.A. Zensus \inst{1}
}
\institute{Max-Planck-Institut f\"{u}r Radioastronomie, Auf dem H\"{u}gel 69, D-53121 Bonn, Germany \and INAF -- Osservatorio di Astrofisica e Scienza dello Spazio di Bologna, Via Gobetti 101, I-40129 Bologna, Italy  \and Departament d'Astronomia i Astrof\'{\i}sica, Universitat de
Val\`encia, C/ Dr. Moliner, 50, 46100, Burjassot, Val\`encia, Spain. \and
Observatori Astron\`omic, Universitat de Val\`encia, C/
Catedr\`atic Jos\'e Beltr\'an 2, 46980, Paterna, Val\`encia, Spain.
 \and Foundation for Research and Technology - Hellas, IESL \& Institute of Astrophysics, Voutes, 7110 Heraklion, Greece  \and  Dipartimento di Fisica e Astronomia, Universit\`a degli Studi di Bologna, Via Gobetti 93/2, I-40129 Bologna, Italy  \and Institut f\"ur Theoretische Physik und Astrophysik, Universit\"at W\"urzburg, Emil-Fischer-Str. 31, 97074 W\"urzburg, Germany \and INAF – Istituto di Radioastronomia, Bologna, Via Gobetti 101, I-40129 Bologna, Italy   \and Fraunhofer-Institut für Hochfrequenzphysik und Radartechnik FHR, Fraunhoferstra{\ss}e 20, 53343 Wachtberg, Germany }
                 


   \date{Received October 7, 2020 /
 accepted December 29, 2020}

 
  \abstract
   {}
{The collimation of relativistic jets in galaxies is a poorly understood process. Detailed radio studies of the jet collimation region have been performed so far in few individual objects, providing important constraints for jet formation models. However, the extent of the collimation zone as well as the nature of the external medium possibly confining the jet are still debated.}
{In this article we present a multi-frequency  and multi-scale analysis of the radio galaxy NGC\,315, including the use of mm-VLBI data up to 86 GHz, aimed at revealing the evolution of the jet collimation profile. We then consider results from the literature to compare the jet expansion profile in a sample of 27 low-redshift sources, mainly comprising radio galaxies and BL\,Lacs, classified based on the accretion properties as low-excitation (LEG) and high-excitation (HEG) galaxies.}
{The jet collimation in NGC\,315 is completed on sub-parsec scales. A transition from a parabolic to conical jet shape is detected at $z_{\rm t}=0.58\pm0.28$ parsecs or ${\sim}5\times10^3$ Schwarzschild radii ($R_{\rm S}$) from the central engine, a distance which is much smaller than the Bondi radius, $r_{\rm B}{\sim}92$\,$\rm pc$, estimated based on X-ray data. The jet in this and in few other LEGs in our sample may be initially confined by a thick disk extending out to ${\sim}10^3$-$10^4$\,$R_{\rm S}$.  A comparison between the mass-scaled jet expansion profiles of all sources indicates that jets in HEGs are surrounded by thicker disk-launched sheaths and collimate on larger scales with respect to jets in LEGs. These results suggest that disk winds play an important role in the jet collimation mechanism, particularly in high-luminosity sources. The impact of winds for the origin of the FRI/FRII dichotomy in radio galaxies is also discussed.}
   {}


   \maketitle

\section{Introduction}
Extragalactic jets are collimated outflows of relativistic plasma emanating from the center of  active galaxies \citep[see][for a recent review]{2019ARA&A..57..467B}. On parsec scales, they are observed to propagate with opening angles as small as a fraction of a degree \citep[e.g.,][]{2009A&A...507L..33P}, a striking feature which can be preserved up to kilo-parsec distances from the central engine. This high degree of collimation is thought to be achieved gradually, through physical mechanisms which are, at present, still unclear.  

Magneto-hydrodynamic simulations of jet launching  \citep{2004ApJ...611..977M, 2006ApJ...641..103H,2006MNRAS.368.1561M, 2011MNRAS.418L..79T} predict the formation of a light, relativistic outflow powered by the rotational energy of the black hole, as described in the work of \cite{1977MNRAS.179..433B}, as well as of a heavier and mildly relativistic wind powered by the accretion disk, as originally proposed by \cite{1982MNRAS.199..883B}. 
At their base, these outflows are expected to be slow and broad, with opening angles of tens of degrees. Along the collimation region, the bulk flow is accelerated by magnetic pressure gradients, thus the magnetic energy which dominates at the jet base is converted into kinetic energy. This description applies to magnetically-dominated cold jets, in which the acceleration and collimation mechanisms are shown to be necessarily co-spatial \citep[][and references therein]{2012rjag.book...81K}. In hot jets (internal energy-dominated) thermal acceleration can also occur in a conical flow which expands adiabatically in an external medium with a steep density gradient \citep[e.g.,][]{2007MNRAS.382..526P}. Thermal acceleration could, at least initially, play a role \citep{2004ApJ...605..656V}, particularly for electron-positron jets and for winds powered by hot accretion flows. The magnetic field, helically wrapped around the jet axis, is thought to contribute not only to the acceleration but also to the collimation, through the ``hoop stress'' exerted by its toroidal component and/or through the pressure of the poloidal component \citep[see][for a discussion] {1997MNRAS.288..333S}. However, it has been shown that magnetic self-collimation may not be sufficiently effective in relativistic flows to account for the observed collimation degree \citep[e.g.,][]{1994PASJ...46..123T, 1998MNRAS.299..341B}, and that some confinement from the ambient medium is required  \citep{2007MNRAS.380...51K,2009ApJ...698.1570L}. The nuclear environment in active galaxies is complex, and the nature of this confining medium is far from understood. At short distances from the black hole it may be the accretion disk itself or dense gas clouds in its surroundings, while at larger distances there could be contributions from the shocked cocoon surrounding the jet or from the interstellar medium. In addition, disk winds are likely to play a crucial role, as they may effectively confine the inner relativistic jet \citep[][]{2005MNRAS.357..918B, 2016MNRAS.461.2605G}.

Several observational studies have been performed with the aim of determining the extension of the acceleration and collimation region, thus providing constraints for theoretical models. 
Such studies were mostly focused on the analysis of low-luminosity radio galaxies, such as 
NGC\,6251 \citep{2016ApJ...833..288T}, 3C\,84 \citep[][]{2018NatAs...2..472G}, 3C\,270 \citep{2018ApJ...854..148N}, 3C\,264 \citep{2019A&A...627A..89B}, NGC\,1052 \citep{2020AJ....159...14N} and M\,87 \citep{2012ApJ...745L..28A,2016A&A...595A..54M, 2016ApJ...817..131H, 2018A&A...616A.188K, 2019MNRAS.489.1197N}, but some constraints for high-luminosity radio galaxies, like Cygnus\,A \citep{2016A&A...585A..33B,2016A&A...588L...9B, 2019ApJ...878...61N}, and blazars  \citep{2008A&A...488..905G,galaxies6010015,2018ApJ...860..141H, 2019ApJ...886...85A, 2020A&A...634A.112T} were also provided.  
Moreover, part of the MOJAVE sample was investigated by \cite{2017MNRAS.468.4992P,2020MNRAS.tmp.1273K}. \cite{2012ApJ...745L..28A} have first shown that the inner jet in M\,87 expands following a characteristic parabolic profile, and that this shape is preserved up to ${\sim}10^5$ Schwarzschild radii ($R_{\rm S}$) from the black hole. At larger distances, the collimation stops and the flow assumes a conical shape. The transition between these two regimes occurs in the proximity of the stationary feature HST-1 and of the Bondi radius $r_{\rm B}$. Assuming that the jet is confined by a spherical, hot accretion flow of the Bondi type \citep{1952MNRAS.112..195B}, it was proposed that the transition, as well as the recollimation at HST-1 \citep[e.g.,][]{2017MNRAS.465.1608L}, may be induced by a change in the ambient pressure profile beyond $r_{\rm B}$. A transition from parabolic to conical expansion at similar distances (${\sim}10^4-10^6$\,$R_{\rm S}$) was later observed in other sources, most recently in ten nearby objects in the MOJAVE\footnote{http://www.physics.purdue.edu/astro/MOJAVE/} sample \citep{2020MNRAS.tmp.1273K}, including both low-power (Fanaroff-Riley I, FRI) and high-power (Fanaroff-Riley II, FRII) radio galaxies, as well as BL Lacs. 

If the ambient medium does shape jets, however, one could expect to observe differences in the collimation profile of sources characterized, for instance, by different accretion modes. Most jets in FRII galaxies and flat spectrum radio quasars (FSRQs) are thought to be powered by radiatively efficient, geometrically thin accretion systems fed by cold gas \citep[][]{1995ApJ...451...88B, 1999ApJ...526...60S, 2004Sci...306..998G, 2006ApJ...642..113G, 2005ApJ...622L..97B, 2007MNRAS.376.1849H} and surrounded by a torus \citep{2006ApJ...647..161O}. On the contrary, FRIs and BLLacs are fed by hot, radiatively inefficient accretion flows  \citep[][]{1995ApJ...451...88B, 2000A&A...355..873C, 2006A&A...451...35B, 2006MNRAS.372...21A, 2007MNRAS.376.1849H, 2014ARA&A..52..529Y}, and generally lack a torus \citep[][]{1999A&A...349...77C, 2004ApJ...602..116W}. 
The transition from radiatively efficient to inefficient accretion mode occurs around Eddington ratios of 
${\sim}0.01$-$0.1$ \citep{1995ApJ...452..710N}, and is reflected by a change in the optical spectra, which, below this threshold, tend to lack strong high-excitation lines from the narrow-line region \citep{1994ASPC...54..201L,1997MNRAS.286..241J, 2012MNRAS.421.1569B, 2014ARA&A..52..589H}. The different energy output of low-excitation (LEG) and high-excitation (HEG) galaxies likely result in a different feedback on the environment \citep[jet mode vs. radiative mode, see e.g.,][]{2014ARA&A..52..589H}. While cold and hot disks are both expected to launch winds and a mildly relativistic jet sheath  \citep[e.g.,][]{2006ApJ...641..103H, 2006MNRAS.368.1561M, 2009PASJ...61L...7O, 2019MNRAS.487..550L}, the properties of these disk outflows and their impact on jet collimation in the two cases are not well constrained. 

Ultimately, understanding the role of the environment in the collimation process requires a detailed investigation of the connection between the properties of the jet base and those of the AGN in the vicinity of the black hole in different luminosity classes. Imaging of the innermost jet base has now become possible in several nearby sources thanks to very-long-baseline interferometry (VLBI) at millimeter wavelengths \citep[mm-VLBI, e.g.,][and references therein]{2017A&ARv..25....4B}; however, studies of the nuclear environment, for instance of the hot X-ray emitting gas within the black hole sphere of influence, are resolution-limited. The best chance is offered by nearby radio galaxies powered by very massive black holes, like M\,87, where both the black hole sphere of influence and the event horizon have a large apparent size. Based on this criterion, our team has identified a first sample of suitable sources for performing high-resolution studies of jet formation in different AGN classes (Boccardi et al. in prep.). In this article we present a detailed analysis of jet collimation in one of the best targets identified so far, the low-luminosity source NGC\,315.  Being close ($z = 0.0165$), quite bright (0.15-0.4 Jy in the mm-band), and hosting a billion solar masses black hole (Sect. 3.1), this giant radio galaxy could be imaged through mm-VLBI with a resolution as high as ${\sim}160$\,$R_{\rm S}$. In the second part of the paper, the results obtained for NGC\,315 are compared with those obtained for other nearby AGN. The article is structured as follows: in Sect. 2 we describe the NGC\,315 data set and its analysis; in Sect. 3 we present the jet collimation profile and investigate possible sources of confinement; in Sect. 4 we present a comparison between the properties of the collimation region in NGC\,315 and in other nearby jets, based on results obtained for the MOJAVE sample and for other objects in the literature; we summarize our conclusions in Sect. 5. Throughout the article we assume a $\Lambda$CDM cosmology with H$_\mathrm{0}$= 71 \rm{km} \rm{s}$^{-1}$ Mpc $^{-1}$, $\Omega_\mathrm{M}=0.27$, $\Omega_{\mathrm{\Lambda}}=0.73$ \citep{2009ApJS..180..330K}.

 \begin{figure}
    \centering
    \includegraphics[width=0.49\textwidth]{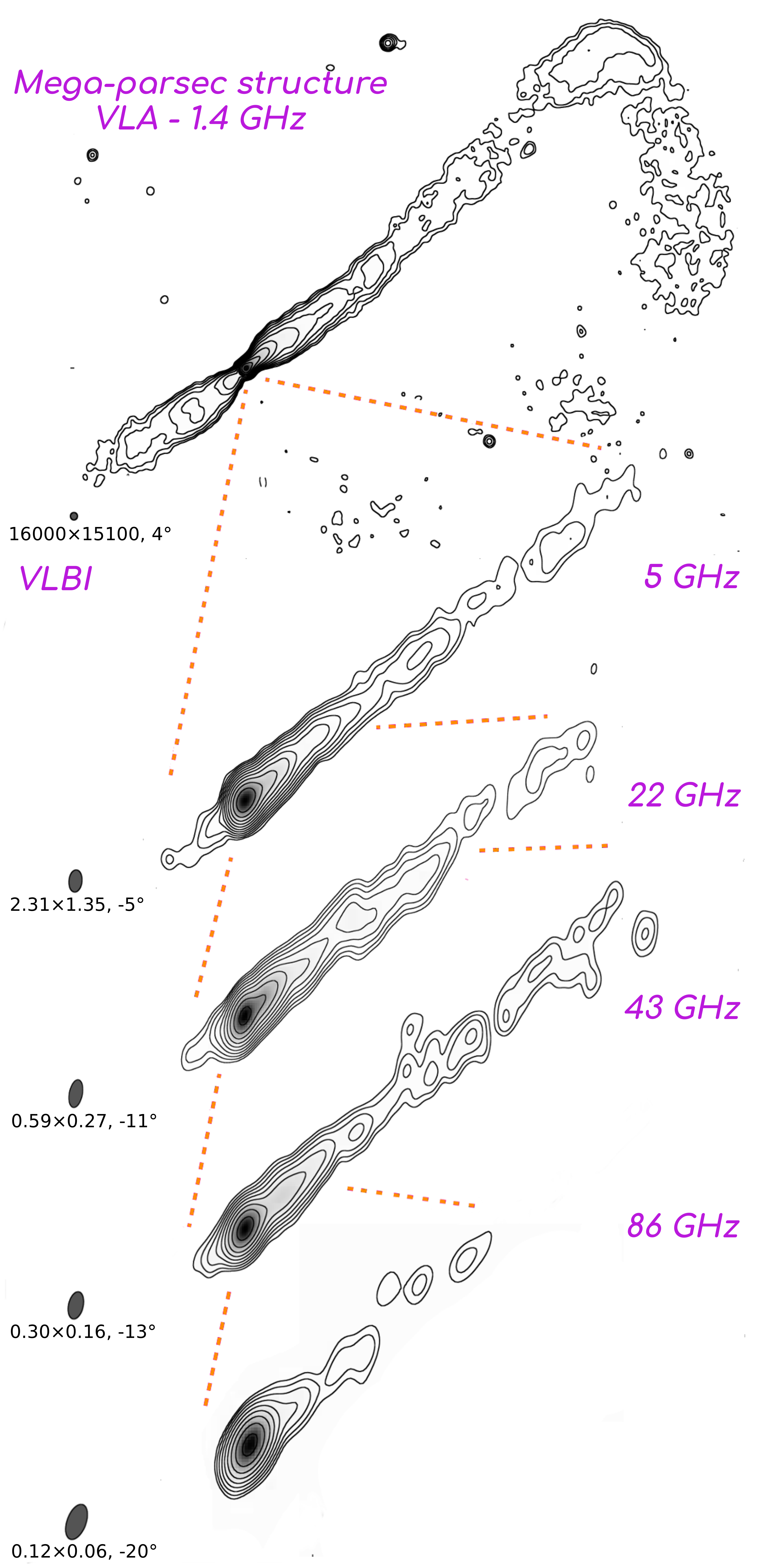}
    \caption{Jet structure of NGC\,315 from Mega-parsec to sub-parsec scales. The top image was obtained based on VLA data at 1.4 GHz, while the rest of the maps show the VLBI structure at different frequencies (5 GHz, 22 GHz, 43 GHz, 86 GHz) based on some of analyzed the data presented in Sect. 2, Table 1. The complete series of images considered in the article is presented in the Appendix.}
    \label{fig:my_label}
\end{figure}

 \begin{table*}
\centering
\footnotesize
\caption{Log of observations and characteristics of the VLBI clean maps forming the NGC\,315 multi-frequency data-set. Col. 1: Frequency. Col. 2: Project code. Col. 3: Date of observation. Col. 4: Array. VLBA - Very Long Baseline Array. * no data from Mauna Kea, ** no data from Saint Croix; GMVA - Global Millimeter-VLBI Array; EVN - European VLBI network; EF - Effelsberg; Y - Very Large Array; AR - Arecibo. Col. 5: Polarization.  Col. 6: Beam FWHM and position angle. Col.7: Equivalent circular beam. Col. 8: Total flux density. Col. 9: Peak flux density. Col. 10: Image noise. All values are for untapered data with uniform weighting.}
\label{my-label}
\begin{tabular}{cccccccccc}
\hline
\hline
 Freq.  & P.C.           & Date       & Array    & Pol & Beam & Eq. beam &$S_{\mathrm {tot}}$ & S$_{\mathrm {peak}}$& rms \\ 
 $\mathrm{[GHz]}$&&  & &&$\mathrm {[mas\times mas, deg]}$ & $\mathrm {[mas]}$& $\mathrm {[mJy]}$& $\mathrm {[mJy/beam]}$& $\mathrm {[mJy/beam]}$  \\ 
\hline
 1.4      & BB020    & 1994-11-26        &  VLBA,Y &        RR   &$8.30\times 4.73, 8$&6.27&373&192&0.20    \\ 
 1.6      & ES049B   & 2005-03-09        &  EVN      &        Dual &$16.90\times 12.40, -87$&14.48&691&506& 0.11   \\ \hline
 5.0      & BG0012   & 1994-11-15        &  VLBA   &        RR   &$2.48\times 1.29, -5$& 1.79 &598&328& 0.11  \\  
 5.0      & BC0051A  & 1995-10-28        &  VLBA   &        Dual &$2.21\times 1.34, -3$&1.72&604&314&0.08   \\      
 5.0      & BC0051B  & 1996-05-10        &  VLBA*,Y &        Dual &$3.65\times 2.56, -18$& 3.06&603&403&0.07  \\      
 5.0      & BC0051C  & 1996-10-07        &  VLBA,Y &  Dual &$2.31\times 1.35, -5$&1.77&596&300& 0.04  \\      
 5.0      & BG0170A  & 2007-12-02        &  VLBA**,Y,GB,EF &  Dual &$3.92\times 0.93, -11$&1.91&569&292&0.75   \\  \hline 
 8.4      & BG0012   & 1994-11-15        &  VLBA     &        RR   &$1.44\times 0.76, -7$&1.05&719&361&0.16   \\  
 8.4      & BL137A   & 2006-02-12        &  VLBA            & Dual & $1.57\times 1.02, -5$ &1.27 & 657 & 365 & 0.18 \\
 8.4      & BG0170B  & 2008-02-03        &  VLBA*,Y,GB,EF,AR&Dual &$1.15\times 0.63, -10$& 0.85 &675&304& 0.05  \\ \hline
 12.1     & BL137A   & 2006-02-12        &  VLBA      &       Dual &$1.11\times 0.70, -4$& 0.88 & 758 & 392  & 0.35 \\ \hline
 15.4     & BL137A   & 2006-02-12        &  VLBA      &       Dual &$0.84\times 0.57, -6$& 0.69 & 647 & 354  & 0.20 \\ \hline
 21.9     & BN0005A  & 1997-02-15        &  VLBA,Y &        RR   &$0.52\times 0.32, -1$&0.41&675&338&0.28 \\  
 22.2     & BG0170A  & 2007-12-02        &  VLBA**,Y,GB,EF &  Dual &$0.94\times 0.25, -22$&0.49&632&255&0.25   \\  
 22.3     & BB393C   & 2018-11-24        &  VLBA,Y,EF &     Dual &$0.59\times 0.27, -11$&0.40&447&184& 0.04   \\  \hline 
 43.2     & BG0170B  & 2008-02-03        &  VLBA*,Y,GB,EF &  Dual &$0.27\times 0.14, -20$&0.19&417&208&0.10   \\ 
 43.2     & BB393C   & 2018-11-24        &  VLBA,Y,EF &     Dual &$0.30\times 0.16, -13$&0.22&268&116& 0.05  \\  \hline  
 86.2     & MB010    & 2018-09-27        &  GMVA        &     Dual &$0.12\times 0.06, -20$&0.08&136&115& 0.14  \\  \hline 
\end{tabular}
\label{vlbi_obs}
\end{table*}

 \begin{table}
\centering
\footnotesize
\caption{Log of observations and basic characteristics of the archival VLA data sets of NGC\,315. Col. 1: Frequency. Col. 2: Array configuration. Col. 3: Date of observation. Col. 4: Beam FWHM and position angle.}
\label{my-label}
\begin{tabular}{cccc}
\hline
\hline
 Freq.  & Configuration & Date          &Beam  \\ 
 $\mathrm{[GHz]}$&  & &$\mathrm {[''\times'', deg]}$   \\ 
\hline
 1.4        & C&2001-07-17       &$16.0\times 15.1, 4$   \\ 
 4.9        & A&1999-07-26       &$0.4\times 0.3, -66$  \\ \hline
\end{tabular}
\label{vla_obs}
\end{table}

 \section {NGC\,315: data set and analysis}

The VLBI data set considered for the analysis of the parsec and sub-parsec scale jet structure in NGC\,315 comprises eighteen  observations at eight frequencies, spanning the range between 1.4 GHz and 86 GHz (Table 1). Several arrays were used in these observations: the Very Long Baseline Array (VLBA), the European VLBI Network (EVN), the High Sensitivity Array (HSA), and the Global Mm-VLBI Array (GMVA). Of these eighteen data sets, fourteen were calibrated in AIPS following the standard procedures, while the remaining four, specifically the data from February 2006 (at 8.4\,GHz, 12.1\,GHz, and 15\,GHz) and from March 2005 (at 1.6 GHz), are calibrated data obtained from the MOJAVE archive and from the EVN archive respectively. To examine the large scale expansion profile of the source, two Very Large Array (VLA) calibrated data set at 1.4 GHz and 5 GHz provided by the NRAO VLA Archive \footnote{http://archive.nrao.edu/nvas/} were also considered. All the data were imaged using DIFMAP (version 2.5e). The main information on the VLBI  observations and clean maps are reported in Table 1.  The basic properties of the VLA data sets are presented in Table 2. VLBI images of the source at 22\,GHz, 43\,GHz, and 86\,GHz are published for the first time in this article, while previous analyses at 1.4\,GHz, 5\,GHz, and 8\,GHz considering some of these data were presented by \cite{1999ApJ...519..108C, 2001ApJ...552..508G}.
All the images are shown in Figures 1-8 of the Appendix, while a multi-scale view of the source, including the Mega-parsec scale structure probed at 1.4 GHz by the VLA, is depicted in Figure 1. On parsec scales, the source presents a bright core and a straight one-sided jet. A faint counter-jet is often, but not always, detected, in agreement with previous results \citep{1999ApJ...519..108C, 2001ApJ...552..508G}. The jet base is bright and compact up to high frequencies, and it was imaged at 86 GHz with a resolution down to ${\sim}60$\,$\rm \mu as$. For a black hole mass $M_{\rm BH}{\sim}1.3\times10^9$\,$M_{\odot}$ \citep[][see Sect. 3.1]{2005ApJ...633...86S} this corresponds to ${\sim}160$\,$R_{\rm S}$. For a detailed description of the large scales radio properties of the source, we refer to \cite{2005MNRAS.363.1223C, 2006MNRAS.368...48L}

\subsection{Alignment of maps at different frequencies}

In order to combine the data obtained at each frequency and correctly reconstruct the jet expansion profile, it is necessary to refer all the measured distances to a common origin, ideally the central supermassive black hole. In each map, the origin coincides approximately with the position of the emission peak, which is however frequency-dependent due to synchrotron opacity effects at the jet base. Several methods have been developed for determining the opacity shift and aligning VLBI images \citep[see e.g.,][and references therein]{2009MNRAS.400...26O}. In this study we performed a 2D cross-correlation analysis taking into account optically thin regions of the jet at given pairs of frequencies (Table 3). We selected pairs of images from close-in-time observations, in order to minimize the uncertainties arising from flux variability and proper motion of the plasma. Same-day observations were available in all but two cases. The closest in time available observations at 1.4 GHz and 5 GHz are separated by 11 days (November 1994), a time short enough for estimating the core-shift, since structural changes in radio galaxies are observed over relatively long timescales (${\sim}$months). The core-shift analysis at 43\,GHz and 86\,GHz was not performed, since the closest-in-time observations are separated by two months, and the core-shift is expected to be small. 

Images in each pair were restored with a common circular beam, corresponding to the average equivalent beam $b_{\rm eq}=\sqrt{b_{\rm min}\times b_{\rm maj}}$ of the two images, where $b_{\rm min}$ and $b_{\rm maj}$ are the minor and major axis of the natural beam, respectively (see Column 7 in Table 1). The pixel size was set to one tenth of the beam FWHM, which we estimate to be comparable to the resolution limits in the brightest regions of our images. Before performing the cross-correlation, each image was slightly shifted so that the pixel with peak flux density was exactly centered at the origin. Therefore the error on the shift determination in $x$ and $y$ corresponds to the in quadrature sum of the error on the core alignment, equal to one pixel, and the error on the images alignment, also equal to one pixel, since the 2D cross-correlation algorithm cannot determine shifts smaller than this size.
The results obtained from the cross-correlation, including the correlation coefficients, are reported in Tables 3 and 4 for each frequency. In Figure 2 we show the dependence of the derived core positions $z_{\rm core}$ relative to the mm-VLBI core (43 GHz and 86 GHz), as a function of frequency. By fitting a power law of the form $z_{\rm core}\propto\nu^{-1/a}$, we obtain $a=0.84\pm0.06$. According to the Blandford \& Koenigl jet model, this result is in agreement with the expectations for a synchrotron self-absorbed conical jet in equipartition, as observed in other extra-galactic jets \citep[e.g.,][]{2011Natur.477..185H, 2011A&A...532A..38S}. Note that the result of the fit is strongly influenced by the 1.4 GHz data point and, as it will be shown in the following, the core at this frequency is indeed located in the conical jet region.  
\begin{table*}
\caption{Results of the 2-D cross correlation performed for the determination of the core-shift. Col. 1: Frequency pair. Col. 2: Date of observation. Col. 3: Circular diameter of the common beam used to restore the maps in each pair. Col. 4: Core-shift in right ascension. Col. 5: Core-shift in  declination. Col. 6: Radial shift. Col. 7: Correlation coefficient.}
 \centering
 \begin{tabular}{c c c c c c c}
 \hline
 \hline
 Freq. pair  & Date & Beam & $\Delta x$ & $\Delta y$ & $\Delta z$  & Correlation coefficient\\
 $\mathrm{[GHz]}$& &$\mathrm {[mas]}$& $\mathrm {[mas]}$ &$\mathrm {[mas]}$&$\mathrm {[mas]}$ & \\
 \hline
 1.4/5.0 & 1994-11&$4.0$&$3.20\pm0.57$&$-2.40\pm0.57$& $4.00\pm0.57$& 0.988\\  
 5.0/8.4 & 1994-11-15&$1.4$&$0.42\pm0.20$&$-0.42\pm0.20$& $0.59\pm0.20$& 0.989\\  
 8.4/12.1& 2006-06-15&$1.0$&$0.20\pm0.14$&$-0.20\pm0.14$&$0.28\pm0.14$&0.958\\
 8.4/15.4& 2006-06-15&$1.0$&$0.30\pm0.14$&$-0.30\pm0.14$&$0.42\pm0.14$&0.984\\ 
 8.4/43.2& 2008-02-03&$0.5$&$0.60\pm0.07$&$-0.50\pm0.07$& $0.78\pm0.07$&0.964\\
 22.3/43.2& 2018-11-24&$0.3$&$0.09\pm0.04$&$-0.09\pm0.04$& $0.13\pm0.04$&0.989\\ \hline
 \end{tabular}
 \label{Shift}
 \end{table*}
 \begin{table}
\caption{Total core-shift with respect to the mm-cores. Column 1: Frequency. Column 2: Core-shift in right ascension. Col. 3: Core-shift in  declination. Col. 4: Radial shift.}
 \centering
 \begin{tabular}{c c c c}
 \hline
 \hline
Freq.  & $x$ & $y$ & $z$\\
$\mathrm {[GHz]}$  & $\mathrm {[mas]}$  & $\mathrm {[mas]}$  & $\mathrm {[mas]}$ \\
\hline
1.4 & $4.22\pm0.60$ & $-3.32\pm0.60$ & $5.37\pm0.60$\\
5.0 & $1.02\pm0.21$ & $-0.92\pm0.21$ & $1.37\pm0.21$\\
8.4 & $0.60\pm0.07$ & $-0.50\pm0.07$ & $0.78\pm0.07$\\
12.1 & $0.40\pm0.16$ & $-0.30\pm0.16$ & $0.50\pm0.16$\\
15.4 & $0.30\pm0.16$ & $-0.20\pm0.16$ & $0.36\pm0.16$\\
22.3 & $0.09\pm0.04$ & $-0.09\pm0.04$ & $0.13\pm0.04$\\
43.2 & 0 & 0 & 0\\
86.2& 0 & 0 & 0\\
 \hline
 \end{tabular}
 \label{TotShift}
 \end{table}

\begin{figure}
\centering
\includegraphics[width=0.45\textwidth]{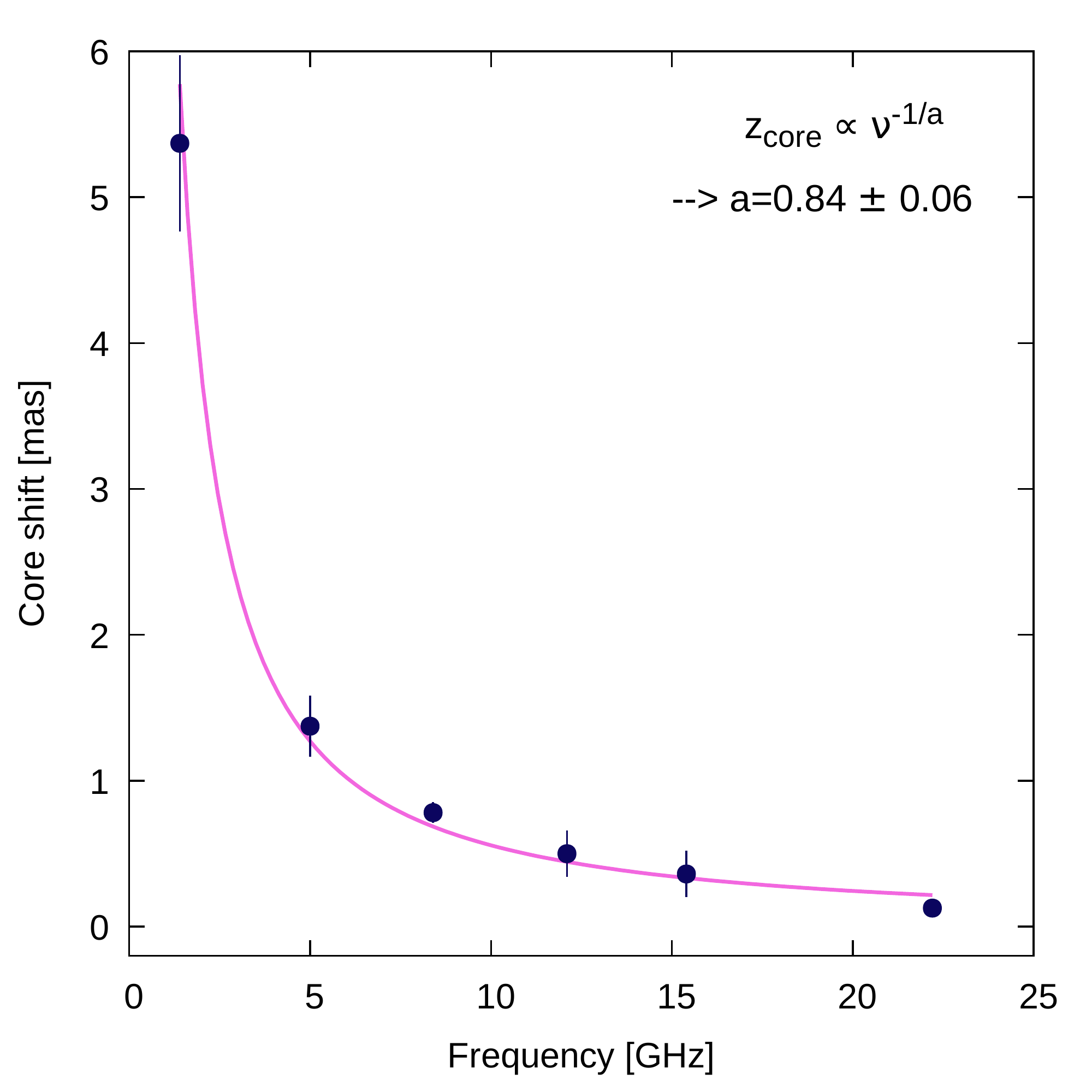}
 \caption{Core positions as a function of frequency relative to the mm-VLBI cores (43 GHz and 86 GHz).}
  \end{figure}
  
\subsection{Modeling}
Aiming at measuring the transverse width of the jet as a function of distance from its base, we have modeled the data through the MODELFIT sub-routine in DIFMAP by fitting circular Gaussian components to the visibilities. The derived MODELFIT parameters at each frequency and epoch are reported in Tables 1-10 of the Appendix.  
After the modeling, two corrections were applied to the position of each feature, reported in Columns 3 and 4 of these Tables. The positions were first shifted so that the brightest core component would be centered exactly at zero; then a second shift was applied to correct for synchrotron opacity, as described in Sect. 2.1 and Tables 3-4.  The shifted positions are shown in Columns 5 and 6 of Tables 1-10 in the Appendix. In addition to the integrated flux density, reported in Column 2, the component size $d$, assumed equal to the full width at half maximum (FWHM) of the Gaussian, is reported in Column 7 for each feature.  Since the errors on the size determined by MODELFIT are extremely small, as the dominant systematic errors associated to the calibration and imaging procedure are not taken into account, in the following we will assume a more conservative error equal to one fourth of the FWHM.

\section{Jet collimation in NGC\,315}

\begin{figure*}[!h]
 \centering
  \includegraphics[trim=0cm 0cm 0cm 0cm, clip=true, width=0.49\textwidth]{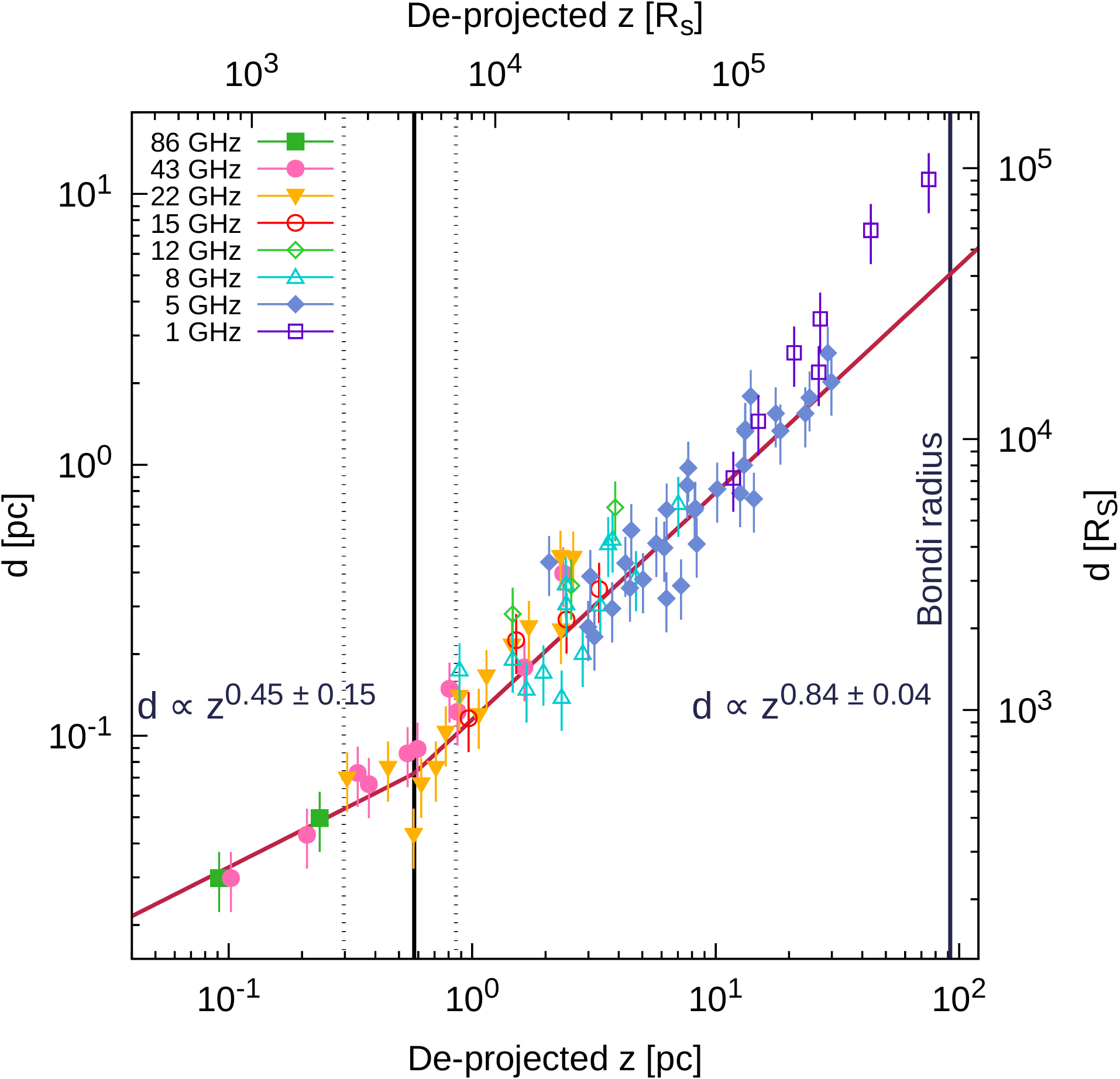}
    \includegraphics[trim=0cm 0cm 0cm 0cm, clip=true, width=0.49\textwidth]{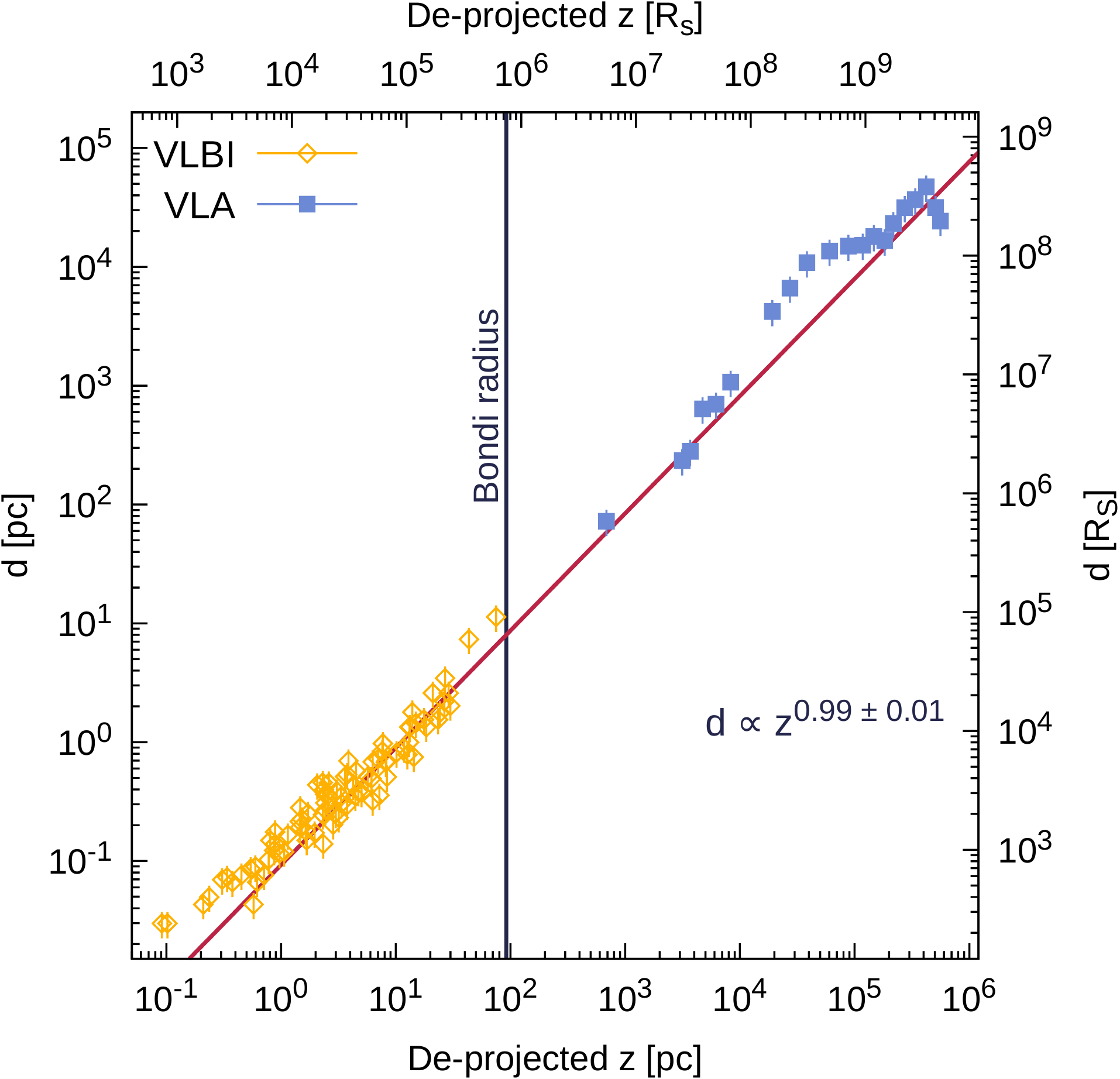}
  \caption{Jet collimation profile in NGC\,315. Left panel: VLBI data at different frequencies. A transition from a parabolical to conical jet shape is observed on sub-parsec scales. The fitted red line is a broken power-law. The dotted lines delimit the uncertainty interval of the transition distance. The latter is much smaller than the estimated Bondi radius. Right panel: VLBI and VLA data. After the transition, the global jet shape remains conical beyond the Bondi radius and up to Megaparsec distance from the black hole.}
 \end{figure*}
 
\subsection{Black hole mass}
The black hole mass $M_{\rm BH}$ in NGC\,315 has been estimated in several works. Based on the correlation between the mass and the stellar velocity dispersion (M-$\sigma$ relation), \cite{2002ApJ...579..530W} found $M_{\rm BH}{\sim}8\times10^8$\,$M_{\odot}$, \cite{2005ApJ...633...86S} reported $M_{\rm BH}{\sim}1.3\times10^9$\,$M_{\odot}$, \cite{2007ApJ...671L.105G} a lower value of $M_{\rm BH}{\sim}3.1\times10^8$\,$M_{\odot}$, \cite{2020ApJ...894..141I} derived $M_{\rm BH}{\sim}1.7\times10^9$\,$M_{\odot}$.  Larger values were obtained by \cite{2003A&A...399..869B}, who inferred $M_{\rm BH}{\sim}2.0\times10^9$\,$M_{\odot}$ based on the M-$\sigma$ relation and $M_{\rm BH}{\sim}3.4\times10^9$\,$M_{\odot}$ when considering the relation between mass and bulge luminosity. Estimates based on gas kinematic studies place the black hole mass of NGC\,315 in a similar range, since \cite{2007ApJ...663...71N} derived $M_{\rm BH}{\sim}1.5\times10^9$\,$M_{\odot}$, compatible with the upper limits obtained by \cite{2009ApJ...692..856B}. In the following, we will assume an intermediate mass value of $M_{\rm BH}{\sim}1.3\times10^9$\,$M_{\odot}$ \citep{2005ApJ...633...86S}. Then, for the adopted cosmology, $1$\,$\rm mas$ corresponds to $0.331$\,$\rm pc$ and $2662$\,$R_S$.

\subsection{Expansion profile}

In Figure 3, left panel, we examine the expansion profile of the approaching jet based on the analysis of the VLBI data set, which covers distances from the black hole ranging from the sub-parsec to the sub-kiloparsec. For the assumed black hole mass $M_{\rm BH}{\sim}1.3\times10^9$\,$M_{\odot}$, these distances translate to a range between ${\sim}10^2$ and ${\sim} 5\times10^5$\,$R_{S}$. For the de-projection we assume a viewing angle of $38^{\circ}$, derived by \cite{2005MNRAS.363.1223C} based on the large scale radio properties of the source. While a larger value of ${49.8}^{\circ}$ was obtained in a more recent study by \cite{2014MNRAS.437.3405L}, an angle of $38^{\circ}$ is in better agreement with VLBI constraints provided by \cite{2001ApJ...552..508G}, who derived an interval between $30^{\circ}-40^{\circ}$.

Not all the jet MODELFIT features (Tables 1-10 of the Appendix) were considered in our analysis. Those which were not fully resolved at a given frequency were filtered out. Specifically, we have excluded components with $\rm FWHM$ smaller than half of the beam minor axis. While the resolution limits may be smaller than this in some cases, this approach is justified by the large number of images and frequencies available, and ensures a reliable determination of the jet width on each scale. The reliability of these measurements is demonstrated by the fact that we obtain similar width values at different frequencies, thus at different resolutions. 
As explained in Sect. 2.1, the radial distances of each jet feature are relative to the positions of the 43 GHz and 86 GHz cores (origin of the axes in Fig. 3) while, ideally, the jet expansion profile should be described with respect to the black hole position. However, based on our core-shift study, the expected separation between the mm-VLBI core and the black hole is negligible with respect to the range of scales considered.

The presence of possibly two breaks in the jet expansion profile is suggested by the data in Figure 3, left panel. A flattening is observed in the inner jet, in the region described by the\,86 GHz and 43\,GHz data and, partially, by the 22\,GHz data, while a possible steepening is hinted by the two outermost data points at 1 GHz. As we are firstly interested in the properties of the innermost jet regions, we have tested the existence of a transition in the jet shape by excluding the two outermost points and by fitting a broken power law to describe the jet width $d$ as a function of de-projected distance $z$. This power law has the form:
\begin{equation}
d(z)=d_{\rm t}\,2^{(k_u-k_d)/h} \left(\frac{z}{z_{\rm t}}\right)^{k_u} \left[1+\left(\frac{z}{z_{\rm t}}\right)^{h}\right]^{(k_d-k_u)/h}  
\end{equation}
where $k_u$ and $k_d$ are respectively the upstream and downstream power-law coefficients, $d_{\rm t}$ is the width of the jet at the transition, $z_{\rm t}$ is the distance from the core at which the transition occurs, and $h$ is a parameter regulating the sharpness of the break. The fit was performed based on a nonlinear least-squares (NLLS) Marquardt-Levenberg algorithm, implemented in GNUPLOT.  The sharpness parameter was fixed, and several values were tested; the reduced $\chi^2$ ($\chi^2/\rm dof$) was found to be only weakly dependent on $h$, improving slightly for larger values. By fixing $h$ to 100 and letting all the other four parameters free, we achieved a reduced $\chi^2$ ($\chi^2/\rm dof$) of 1.35. A single power-law ($d\propto z^k$) fitted for comparison in the same region does not well describe the innermost data points, and the reduced $\chi^2$ is larger (1.44). 
The broken power-law fit yields a power-law index $k_u=0.45\pm0.15$ in the inner jet, which thus has a close to parabolic shape, while a close to conical shape is determined outwards, with $k_d=0.84\pm0.04$. The transition occurs at a distance $z_{\rm t}=0.58\pm0.28$ de-projected parsecs, corresponding to ${\sim}5\times10^3$\,$R_{\rm S}$.  

The two outermost VLBI data points at 1.4 GHz, which were so far excluded by our analysis, hint at a possible further steepening of the expansion profile. In order to examine how the profile evolves on larger scales, in the right panel of Fig. 3 we add the VLA data, which enable us to probe the jet up to a distance of almost one Megaparsec ($10^{9}-10^{10}$\,$R_{\rm S}$) from the central engine. The unresolved VLA MODELFIT components were filtered out following the same criteria described for the VLBI data. In the case of the 1 GHz data, which describe the entire large scale structure of the approaching jet, we have also excluded data points beyond the ninthy-degrees bend of the jet direction towards south (Figure 1). The analysis of the multi-scale data set confirms our result that the jet collimation is completed on small scales, and also shows that, except for some local oscillations of the jet width (e.g., between $10^4-10^5$\,$\rm pc$) the global jet shape remains conical for several orders of magnitude in distance after the initial collimation. A single power-law fit performed considering the entire jet after the transition (i.e., excluding the innermost $0.58\,\rm pc$) yields a power-law index $k=0.99\pm0.01$, only slightly steeper than the one derived based on the VLBI data set. The power-law index does not change significantly if the innermost parabolic region is included in the fit ($k=0.98\pm0.01$), but the reduced $\chi^2$ becomes larger (2.30 vs 2.07). 
In Figure 4, the residuals of a single power-law fit performed considering the entire data set as well as the VLBI data only are displayed. In both cases, a single power-law cannot well describe the expansion profile on sub-parsec scales, where the residuals are of positive sign. A broken power-law fit considering the entire VLA plus VLBI data set was also attempted. Even in this case, the results are consistent with a small scale shape transition. The fit becomes more sensitive to the assumed sharpness parameter (which we varied between 1 and 50, as the fit did not converge for larger values) even though the reduced $\chi^2$ does not change significantly ($\chi^2/\rm dof{\sim}1.76-1.79)$. The assumption of a soft break ($h$=1) yields a transition distance $z_{\rm t}=1.29\pm1.23$\,$\rm pc$ with $k_u=0.48\pm0.22$ and $k_d=1.01\pm0.01$, while the assumption of a sharp break ($h$=50) places the transition at $z_{\rm t}=6.32\pm2.67$\,$\rm pc$ with $k_u=0.74\pm0.05$ and $k_d=1.01\pm0.01$, thus with the inner jet shape deviating significantly from a parabola. By letting all the parameters, including the sharpness $h$, free to vary, we obtain $h=1.50\pm1.89$ and a transition distance $z_{\rm t}=2.75\pm2.88$\,$\rm pc$ with $k_u=0.62\pm0.19$ and $k_d=1.01\pm0.02$, in agreement with the VLBI results within the uncertainty interval. 
Overall, our findings match those obtained by \cite{2017MNRAS.468.4992P} and \cite{2020MNRAS.tmp.1273K}, who performed a detailed analysis of the 15 GHz VLBI stacked image, probing scales comprised between ${\sim}1-10$ de-projected parsecs. The power-law indices obtained by these authors by fitting a single power-law are $k=0.86\pm0.01$ and $k=1.07\pm0.05$, respectively. \cite{2020MNRAS.tmp.1273K} also tested the presence of a transition along the 15 GHz jet, with negative results, and indeed the transition we detect is suggested to occur on smaller scales, unresolved at 15 GHz.
In conclusion, the present data are compatible with the occurrence of a sub-parsec scale transition from parabolic to conical expansion, after which the global jet shape remains conical.

\begin{figure}[!h]
 \centering
  \includegraphics[trim=0cm 0cm 0cm 0cm, clip=true, width=0.4\textwidth]{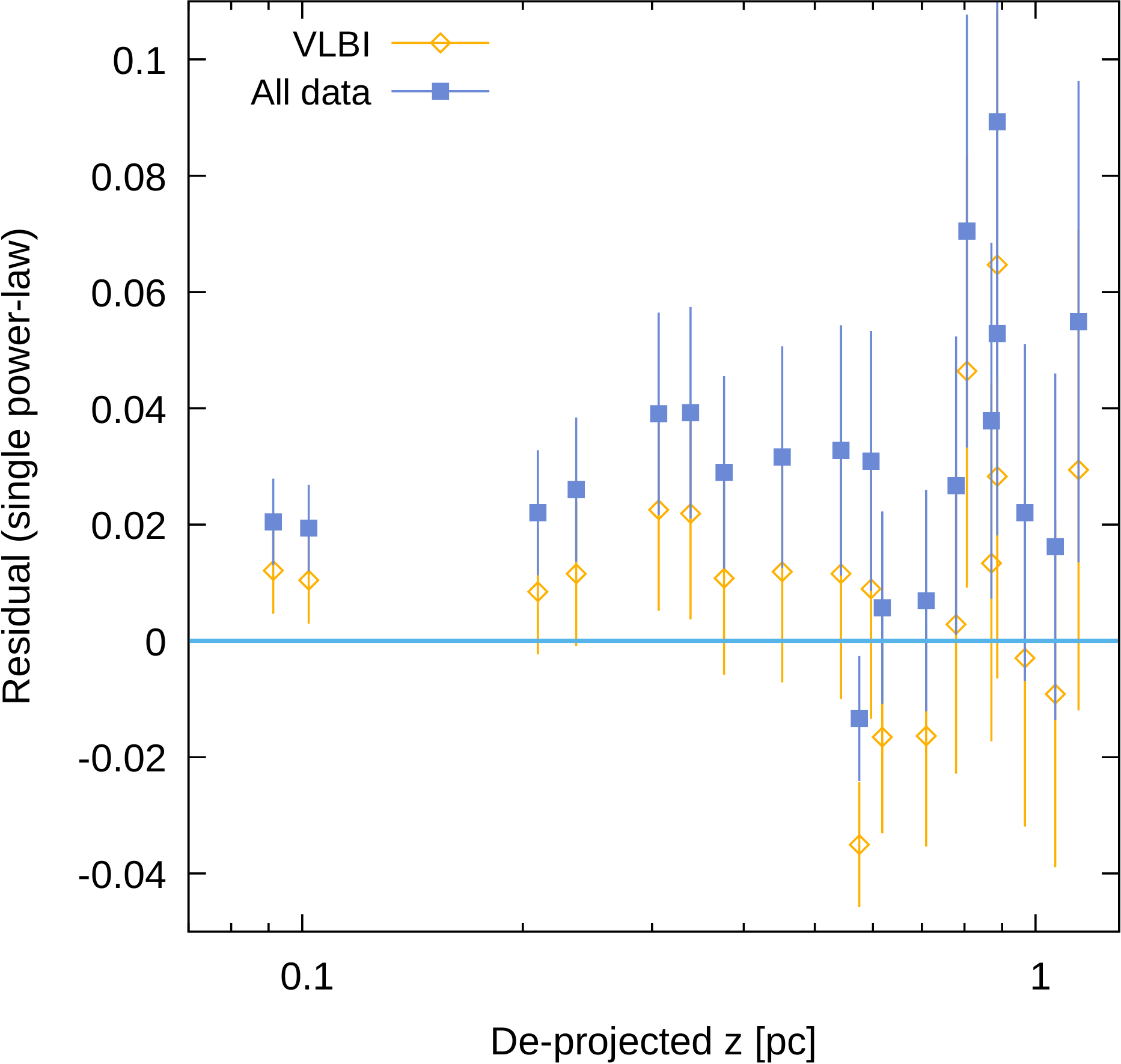}
  \caption{Residuals in the innermost jet region for a single power-law fit performed considering the VLBI data only and the entire data set (VLBI+VLA).}
 \end{figure}

 \subsection{Jet confinement from the environment}
 
As discussed in Sect. 1, the transition from parabolic to conical shape is observed in M87 and in several other sources on relatively large scales, at distances of the same order as the sphere of influence of the black hole. 
Like M\,87, NGC\,315 is a cluster member and, as shown by \cite{2007MNRAS.380....2W}, is powered by a radiatively inefficient nucleus. Under the assumption that the hot accretion flow can be well described by the Bondi theory, we can test the hypothesis that the jet shape transition occurs in the proximity of the Bondi radius. The Bondi radius $r_{\rm B}$ is the radius inside which the gravitational influence of the central black hole dominates over the thermal motion of the gas in the accretion flow.  Its expression is \citep[see e.g.,][]{frank_king_raine_2002}:
\begin{equation}
    r_{\rm B}=\frac{2GM_{\rm BH}}{c_{\rm s}^2},
\end{equation}
where $G$ is the gravitational constant and $c_{\rm s}$ is the sound speed.
Since the sound speed depends on the gas temperature $T$, assuming an adiabatic index $\gamma=5/3$ 
the Bondi radius can be also expressed, in convenient units, as \citep[see][]{2015MNRAS.451..588R}:
\begin{equation}
   \frac{r_{\rm B}}{\rm kpc}=0.031\left (\frac{kT}{\rm keV}\right)^{-1}\left (\frac{M_{\rm BH}}{10^9 M_{\odot}}\right).
\end{equation}
The expected pressure profile of the interstellar medium inside the Bondi sphere is $p\propto z^{-2}$ and, according to analytical and numerical models \citep{2008MNRAS.388..551T, 2009MNRAS.394.1182K, 2017MNRAS.472.3971B}, a jet propagating in a medium with such a pressure gradient develops the characteristic parabolic shape. Based on Eq. 3, estimating the Bondi radius requires the knowledge of the black hole mass and of the temperature of the accretion flow in the nuclear region. A detailed analysis of the X-ray emission in NGC\,315 was carried out by \cite{2003MNRAS.343L..73W, 2007MNRAS.380....2W} using sensitive Chandra data. The jet, which presents several X-ray-bright knots of synchrotron emission, is embedded in a hot gaseous atmosphere, as inferred from the presence of an X-ray thermal emission component in the spectrum.  Such a feature is also detected within a circle of 1 arcsecond radius centered around the nucleus, in addition to a dominant, mildly absorbed power-law component possibly associated to the jet. The nuclear hot atmosphere has a temperature $kT=0.44^{+0.08}_{-0.04}$\,$\rm keV$.  
Assuming a black hole mass of $1.3\times10^9$\,$M_{\odot}$, we then estimate $r_{\rm B}=92$\,$\rm pc$ (dark blue vertical line in Fig. 3). Even if the black hole mass was smaller by a factor of several, this radius would still be much larger (two orders of magnitude) than the distance at which we observe the jet shape transition from parabolic to conical, $z_{\rm t}=0.58\pm0.28$\,$\rm pc$. This result would also not be affected by varying the gas temperature within the given small uncertainty. By assuming slightly different parameters, a similar value for the Bondi radius in NGC\,315 was recently obtained by \cite{2020ApJ...894..141I}, who estimated an uncertainty of 50 percent\footnote{\cite{2020ApJ...894..141I} assumed slightly larger values for both the temperature and the black hole mass, but defined the Bondi radius as $r_{\rm B}=GM_{\rm BH}/c_{\rm s}^2$ (factor of 2 difference), thus obtaining $r_{\rm B}=(51\pm25)$\,$\rm pc$.}.
Then, if the transition from parabolic to conical shape in NGC\,315 is induced by a change in the external pressure gradient, this must occur not in the proximity of the black hole sphere of influence, but in the vicinity of the black hole, on sub-parsec scales. Based on the analysis of the VLA data in Fig. 3, right panel, we also note that no discontinuity is observed in the expansion profile after the jet crosses the Bondi radius: the jet shape is close to conical both inside and outside the Bondi sphere.

It is important to point out that the Bondi theory is likely over-simplified, not taking into account important aspects like viscosity, the presence of magnetic fields or a non-zero angular momentum of the accretion flow. In fact, simulations assuming more realistic physical conditions \citep[e.g.,][]{2013MNRAS.432.3401G} have suggested that, even in the case of radiatively inefficient AGN, the accretion is likely to be cold and chaotic, rather than hot and spherically symmetric. The presence of cold gas in NGC\,315 on scales of tens of parsecs was revealed by VLBI HI absorption studies \citep{2009A&A...505..559M}. However, there is no evidence for the existence of a dusty torus feeding the nucleus (X-ray data reveal only mild intrinsic absorption), nor of a cold thin disk. The nuclear emission in NGC\,315 may be adequately described by more complex models of hot accretion flows, such as advection-dominated accretion flows (ADAF). \cite{2007ApJ...669...96W} showed that the X-ray spectrum in NGC\,315 can be well fitted by an ADAF.  Interestingly, such thick disks are suggested to have an outer radius of the order of ${\sim}10^3$\,$R_{\rm S}$ \citep[e.g.,][]{1997ApJ...477..585M}, comparable to the distance at which the jet break is observed (${\sim}5\times10^3$\,$R_{\rm S}$). Thus it is possible that the jet in NGC\,315 is initially confined by the thick disk itself, and starts to freely expand beyond the outer disk radius. 
Alternatively, analytical and semi-analytical models \citep{2009ApJ...698.1570L, 2017MNRAS.472.3971B, 2020MNRAS.tmp.1273K} have shown that the transition to a conical jet shape can occur also in the presence of a single external pressure profile, and that the location of the transition may be strongly dependent on the initial jet magnetization. The external medium may not necessarily be the gas in the accretion flow. 
\cite{2016MNRAS.461.2605G} have recently extended the work of \cite{2009ApJ...698.1570L} to describe the case in which the Poynting-flux dominated jet is confined by a hydrodynamic wind layer. In this semi-analytic model, the wind layer was also shown to be characterized by a pressure profile $p\propto z^{-2}$, thus conferring the characteristic parabolic shape to the central jet filaments. The collimation was found to be effective for sufficiently high ratios ($>0.1$) of wind power to jet power, and the extension of the collimation region was shown to be dependent on the thickness and power distribution of the wind layer. The possibility that this and other jets are collimated by accretion disk winds is further discussed in the following.

\section{The parabolic jet in different sources}
 \begin{table*}
 \caption{Sample of sources considered for the analysis in Section 4, sorted by redshift. Column 1: Source name (B1950). Column 2: Other name. Column 3: Redshift. Column 4: Source classification. RG: Radio galaxy,  BL: BL Lac, NLSy1: Narrow-line Seyfert 1. Column 5: Log of the black hole mass $M_{\rm BH}$, expressed in units of solar masses. Column 6: Ratio between the X-ray luminosity in the 2-10\,$\rm KeV$ band (reference in the Table note) and the Eddington luminosity $L_{\rm Edd}=1.3\cdot10^{38}$\,$M_{\rm BH}/M_{\odot}$. Column 7: Classification as HEG: High-excitation galaxy or LEG: Low-excitation galaxy. A dividing limit $L_{\rm X}/L_{\rm Edd}=1.1\rm e{-3}$ was assumed. Column 8: Conversion factor, parsec per milliarcsecond. Column 9: Conversion factor, Schwarzschild radii per milliarcsecond. Column 10: Viewing angle. Column 11: Power-law index defining the inner jet shape. Column 12: Transition distance in units of projected milli-arcseconds. References as in Column 11. }
  \centering
   \small
 \begin{tabular}{c c c c c c c c c c c c}
 \hline
 \hline
B1950     &   Other   & z & Class & $\log{M_{\rm BH}}$   & $L_{\rm X}/L_{\rm Edd}$ & Accretion & $\rm pc$/$\rm mas$ &   $R_{\rm S}$/$\rm mas$ &$\theta$ & $\rm k$ & $z_{\rm t}$  \\
name &  name   &   &    & [$M_{\odot}$]  &   & mode  & & & $[\rm deg]$    &    & $[\rm mas]$    \\
\hline

1228+126  &   M\,87          &   0.0043   & RG   &  $9.81^{[1]}$    & 2.7e-8$^{[15]}$  & LEG   & 0.087  &    140     &  $16^{[22]}$    &   ${\sim}0.57^{[44]}$   &  131   \\
0238-084  &   NGC\,1052      &   0.0050   & RG   &  $8.19^{[2]}$    & 2.1e-5$^{[16]}$  & LEG   & 0.102  &    6880    &  $49^{[23]}$    &   $0.39\pm0.05^{[34]}$  &  3.73      \\
1216+060  &   NGC\,4261      &   0.0074   & RG   &  $8.69^{[3]}$    & 8.4e-7$^{[15]}$  & LEG   & 0.150  &    3201    &  $63^{[24]}$    &   $0.62\pm0.13^{[45]}$  &  ${\sim}4.5$      \\
0055+300  &   NGC\,315       &   0.0165   & RG   &  $9.12^{[4]}$    & 3.7e-6$^{[17]}$  & LEG   & 0.331  &    2662    &  $38^{[25]}$    &   $0.45\pm0.15$         &  1.07      \\
0316+413  &   3C\,84         &   0.0176   & RG   &  $8.49^{[2]}$    & 1.3e-4$^{[15]}$  & LEG   & 0.352  &    11910   &  $18^{[26]}$    &   $0.50\pm0.03^{[23]}$  &   -        \\
0313+411  &   IC\,310        &   0.0189   & BL   &  $8.48^{[5]}$    & -                & LEG   & 0.379  &    13208   &  $15^{[27]}$    &   $2.30\pm0.10^{[34]}$  &   -        \\
1142+198  &   3C\,264        &   0.0217   & RG   &  $8.67^{[6]}$    & 1.2e-5$^{[15]}$  & LEG   & 0.433  &    9673    &  $10^{[28]}$    &   $0.66\pm0.02^{[29]}$  &  ${\sim}4.5$\\
1637+826  &   NGC\,6251      &   0.0247   & RG   &  $8.78^{[7]}$    & 7.4e-5$^{[15]}$  & LEG   & 0.491  &    8513    &  $19^{[29]}$    &   $0.51\pm0.04^{[34]}$  &  2.13       \\
0305+039  &   3C\,78         &   0.0287   & RG&  $8.60^{[2]}$ & 4.5e-5$^{[15]}$  & LEG   & 0.567  &    14894   &  $ 20^{[30]}$    &   $0.99\pm0.12^{[34]}$  &   -           \\
1101+384  &   Mrk\,421       &   0.0300   & BL   &  $8.29^{[2]}$    & -                & LEG   & 0.593  &    31794   &  $4^{[31]}$     &   $1.49\pm0.11^{[34]}$  &   -           \\
0430+052  &   3C\,120        &   0.0330   & RG   &  $7.74^{[8]}$    & 1.4e-2$^{[18]}$  & HEG   & 0.650  &    122446  &  $19^{[23]}$    &   $0.56\pm0.07^{[34]}$  &  2.85         \\
1652+398  &   Mrk\,501       &   0.0337   & BL   &  $9.21^{[2]}$    & -                & LEG   & 0.662  &    4272    &  $4^{[32]}$     &   $0.57\pm0.02^{[34]}$  &  -             \\
2344+514  &   1ES\,2344+514  &   0.0440   & BL   &  $8.80^{[2]}$    & -                & LEG   & 0.855  &    14166   &  $2^{[33]}$            &   $1.33\pm0.15^{[34]}$  &   -           \\
1133+704  &   Mrk\,180       &   0.0453   & BL   &  $8.21^{[2]}$    & -                & LEG   & 0.879  &    56728   &  $5^{[34]}$     &   $0.53\pm0.04^{[34]}$  &  1.46         \\
1959+650  &   1ES\,1959+650  &   0.0470   & BL   &  $8.09^{[2]}$    & -                & LEG   & 0.910  &    77350   &  $2^{[35]}$              &   $0.81\pm0.26^{[34]}$  &   -           \\
1514$-$241&   AP\,Librae     &   0.0490   & BL   &  $8.10^{[2]}$    & -                & LEG   & 0.947  &    78579   &   $6^{[36]}$             &   $1.10\pm0.06^{[34]}$  &   -           \\
0415+379  &   3C\,111        &   0.0485   & RG   &  $8.26^{[9]}$    &5.1e-3$^{[18]}$   & HEG   & 0.937  &    54424   &  $17^{[23]}$    &   $0.47\pm0.03^{[34]}$  &  7.07         \\
1807+698  &   3C\,371        &   0.0510   & BL   &  $8.51^{[2]}$    & -                & LEG   & 0.983  &    31720   &  $7^{[23]}$     &   $0.39\pm0.09^{[34]}$  &  1.67         \\
1514+004  &   PKS\,1514+00   &   0.0525   & RG   &  $7.50^{[10]}$   &2.8e-3$^{[19]}$   & HEG   & 1.010  &    334163  &  $15^{[34]}$    &   $0.56\pm0.05^{[34]}$  &  3.39          \\
1727+502  &   I\,Zw\,187     &   0.0554   & BL   &  $7.86^{[2]}$    & -                & LEG   & 1.062  &    153359  &  $5^{[37]}$           &   $0.80\pm0.15^{[34]}$  &    -           \\
1957+405  &   Cygnus\,A      &   0.0561   & RG   &  $9.40^{[11]}$   &1.1e-3$^{[20]}$   & HEG   & 1.074  &    4491    &  $75^{[38]}$    &   $0.55\pm0.07^{[38]}$  &  ${\sim}5$   \\
1845+797  &   3C\,390.3      &   0.0561   & RG   &  $8.46^{[8]}$    &3.4e-3$^{[18]}$   & HEG   & 1.075  &    39025   &  $26^{[39]}$    &   $1.71\pm0.26^{[34]}$  &  -              \\
0321+340  &   1H\,0323+342   &   0.0610   & NLSy1&  $7.30^{[12]}$   &2.6e-2$^{[21]}$   & HEG   & 1.162  &    607436  &  $5^{[40]}$     &   $0.58\pm0.02^{[40]}$  &  ${\sim}7$   \\           
1219+285  &   W\,Comae       &   0.1020   & BL   &  $8.00^{[13]}$   &-                 & LEG   & 1.855  &    193940  &  $3^{[41]}$            &   $1.28\pm0.10^{[34]}$  & -                \\
1215+303  &   ON\,325        &   0.1300   & BL   &  $8.12^{[2]}$    &-                 & LEG   & 2.291  &    181458  &  $2^{[42]}$             &   $1.86\pm0.40^{[34]}$  &  -              \\
0806+524  &   1ES\,0806+524  &   0.1380   & BL   &  $8.90^{[2]}$    &-                 & LEG   & 2.411  &    31747   &  $4^{[37]}$              &   $1.32\pm0.42^{[34]}$  &  -               \\
1055+567  &   7C\,1055+5644  &   0.1433   & BL   &  $8.54^{[14]}$   &-                 & LEG   & 2.489  &    74993   &  $5^{[43]}$             &   $1.68\pm0.18^{[34]}$  &  -               \\
\hline
 \end{tabular}
\tablebib{
Column 5: 
 [1] \cite{2019ApJ...875L...1E}, [2]  \cite{2002ApJ...579..530W}, [3] \cite{1996ApJ...470..444F}, [4] \cite{2005ApJ...633...86S},  [5] \cite{2014A&A...563A..91A}, [6] \cite{2008A&A...486..119B}, [7] \cite{2003ApJ...589L..21M}, [8] \cite{2004ApJ...613..682P}, [9] \cite{2011ApJ...734...43C}, [10] \cite{2007ApJ...658..815S}, [11] \cite{2003MNRAS.342..861T}, [12] \cite{2017MNRAS.464.2565L}, [13] \cite{2004AJ....127...53X}, [14] \cite{2015MNRAS.451.4193C}.  References for the X-ray luminosity used in Column 6: [15] \cite{2018MNRAS.476.5535T}, [16] \cite{2019MNRAS.485..416B}, [17] \cite{2009A&A...506.1107G}, [18] \cite{2006ApJ...642..113G}, [19] \cite{2006MNRAS.365..688G}, [20]: \cite{2002ApJ...564..176Y}, [21]: \cite{2019A&A...632A.120B}. Columns 10, 11, and 12 (same as 11):  
 [22] \cite{2016A&A...595A..54M}, [23] \cite{2017MNRAS.468.4992P}, [24] \cite{2001AJ....122.2954P}, [25] \cite{2005MNRAS.363.1223C},  [26] \cite{2018NatAs...2..472G}, [27] \cite{2014A&A...563A..91A} infer an angle range of $10^{\circ}$- $20^{\circ}$, we assume the average value, [28] \cite{2019A&A...627A..89B}, [29] \cite{2016ApJ...833..288T}, [30] \cite{2015ApJ...798...74F},  [31] \cite{2012A&A...545A.117L},  [32] \cite{2004ApJ...600..127G}, [33] \citep{2020A&A...640A.132M}, [34] \cite{2020MNRAS.tmp.1273K}, [35] \cite{2013ApJS..206...11S}, [36] \cite{2006ApJ...646..801G},  [37] \cite{2002A&A...384...56C}, [38] \cite{2016A&A...585A..33B},
 [39] \cite{1994ApJS...90....1E}, [40] \cite{2018ApJ...860..141H}, [41] \cite{2014ApJ...794...54S}, [42] \cite{2020ApJ...891..170V}, [43] \cite{2010MNRAS.402..497G},  [44] \cite{2019MNRAS.489.1197N},  [45] \cite{2018ApJ...854..148N}.}
 \end{table*}
 
 \begin{figure*}[t]
\centering
\includegraphics[width=0.49\textwidth]{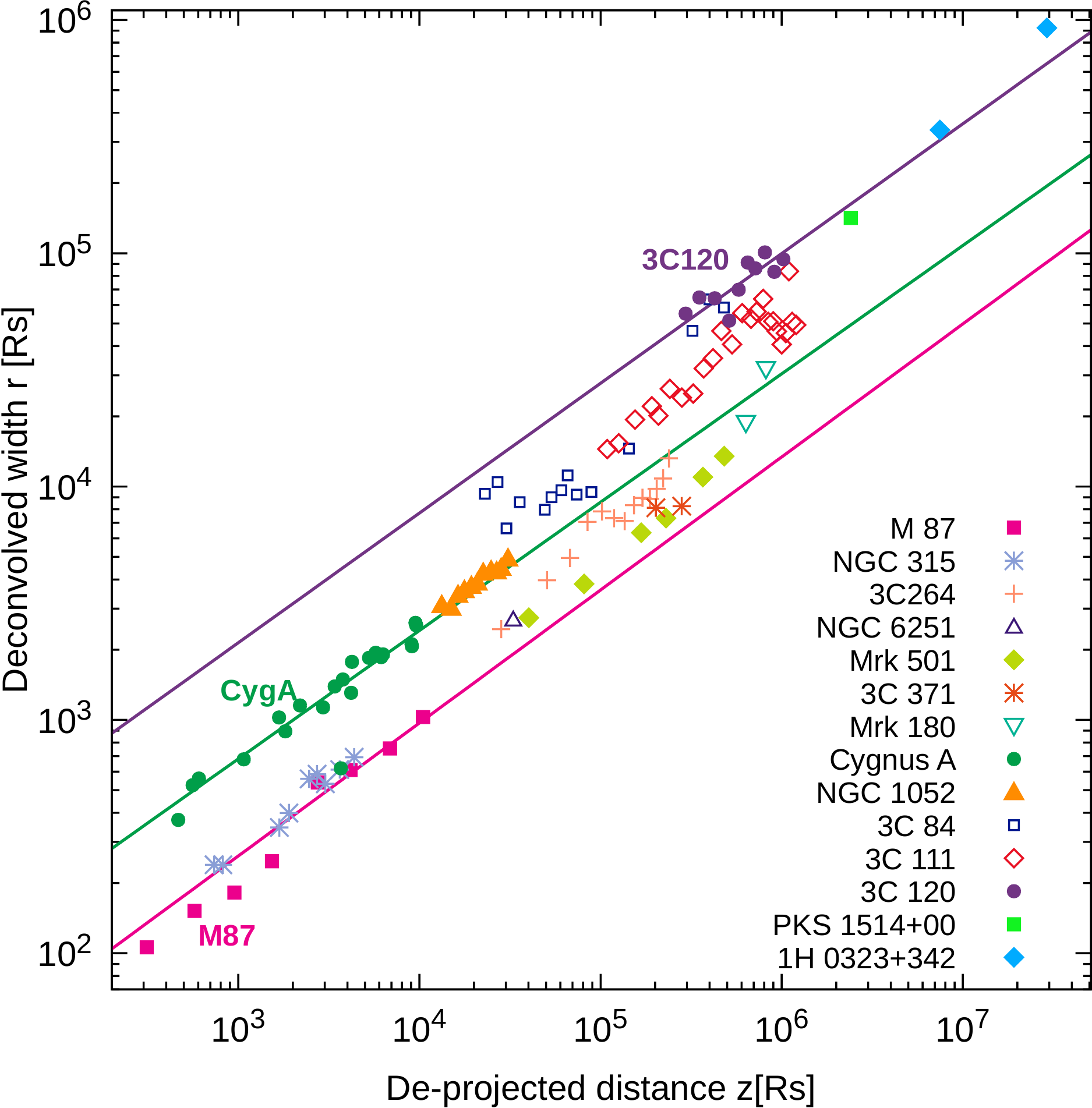}
\includegraphics[width=0.49\textwidth]{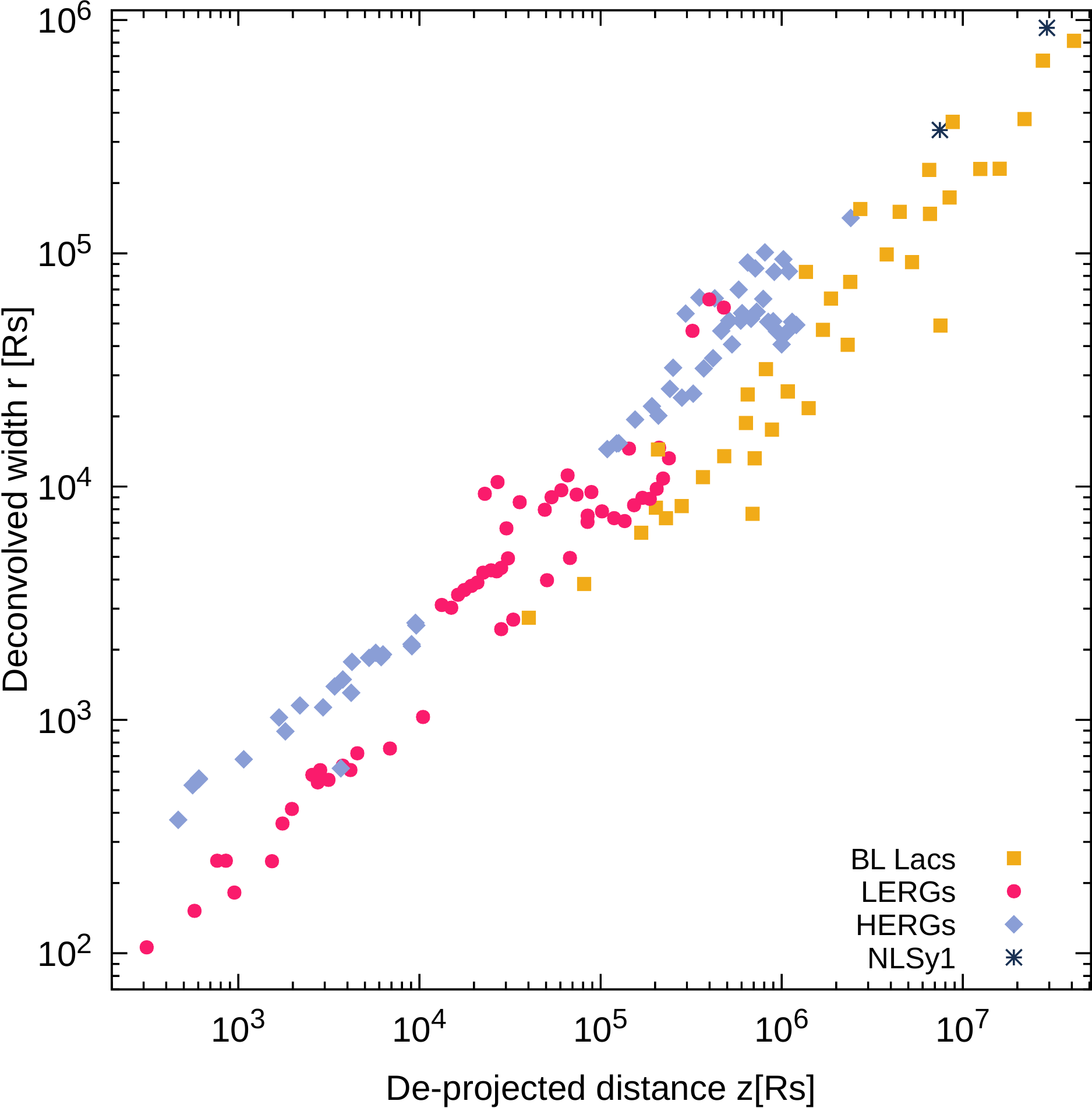}
\caption{Jet collimation profiles of the sources in Table 5. For jets showing a parabolic expansion at their base, we only report the parabolic region. For NGC\,315, we show results from this paper; for Cygnus\,A, we report 43 GHz data from \cite{2016A&A...585A..33B}; for 3C\,264, 15 GHz data from \cite{2019A&A...627A..89B}; for the rest of the sources we use 15 GHz data based on published MOJAVE results \citep{2019ApJ...874...43L}. In the case of M\,87, and possibly of 3C\,84 and Mrk\,501, the 15 GHz data describe only part of the parabolic region. Left panel: profile of the parabolically expanding jets in the sample. Each source is labeled using a different color and point type. The solid lines represent fits to the data of M\,87, Cygnus\,A, and 3C\,120 (magenta, green, and violet line, respectively).  Right panel: profile of all the sources in the sample, divided per classes. Light blue points are assigned to the HERGs, magenta points to the LERG, orange points to the BL\,Lacs, and the dark blue points to the NLSy1.}
\end{figure*}

Substantial observational evidence has accumulated so far concerning the existence of parabolically expanding jet bases in AGN. The study of NGC\,315 presented in this paper shows another example of this, but also indicates that the jet collimation can be completed already on sub-parsec scales. In this section we investigate how the properties of the jet collimation region vary in different sources, combining our findings with results from the literature. In particular, one open question which we would like to address concerns how the presence of disk-winds affects jet collimation, and how are the properties of these winds related to the nature of the accretion flow. 
Millimiter-VLBI observations of M\,87 indicate that the jet base is anchored in the vicinity of the ergosphere, having a transverse size of the order of few $R_{\rm S}$ \citep[e.g.,][]{2012Sci...338..355D, 2018A&A...616A.188K}. On the other hand, imaging of the two-sided jet in Cygnus\,A has revealed that its base is quite broad, with a transverse size of ${\sim}230$\,$R_{\rm S}$ \citep{2016A&A...588L...9B}, thus part of the jet may be launched from the outer regions of the accretion disk. M\,87 and Cygnus\,A clearly differ both in the jet power, of ${\sim}10^{43}\rm{erg}\,\rm{s^{-1}}$ in the former and ${\sim}10^{46}\rm{erg}\,\rm{s^{-1}}$ in the latter, and in the type of the accretion disk they host, the former being powered by an ADAF \citep[e.g.,][]{2003ApJ...582..133D} and the latter by a heavily absorbed cold disk \citep[e.g.,][]{2002ApJ...564..176Y, 2015ApJ...808..154R}. The two sources also differ in the jet large scale morphology, FR\,I and FR\,II respectively. In the following we compare the mass-scaled properties of the jet collimation zone in NGC\,315 and in several other sources of both high- and low-power. We take into account results from studies of individual nearby objects, namely Cygnus\,A \citep{2016A&A...585A..33B}, 3C\,264 \citep{2019A&A...627A..89B}, and NGC\,4261 \citep{2018ApJ...854..148N}, as well as results obtained for the MOJAVE sub-sample analyzed by \cite{2017MNRAS.468.4992P, 2020MNRAS.tmp.1273K}, limiting our analysis to sources with redshift smaller than $0.15$ and with known black hole mass. We exclude BL Lacertae, whose peculiar properties will be discussed in a dedicated paper (C. Casadio et al. in preparation). While observational constraints exist also for high-redshift sources based on the analysis of the MOJAVE sample, in this case we prefer to focus on nearby objects, for which several estimates of the black hole mass are given in literature and a higher spatial resolution can be achieved in VLBI observations. 

A summary of the properties of the twenty-seven sources in our sample, thirteen radio galaxies, thirteen BL Lac objects, and one Narrow-line Seyfert\,1, is presented in Table 5. In Column 7 we report the classification as low-excitation (LEG) or high-excitation (HEG) galaxy. As mentioned in Sect. 1, this division is indicative of the AGN accretion regime and is based on the optical spectroscopic properties of the nucleus. Since this optical classification is not available in the literature for all the sources in our sample, in order to adopt a uniform criterion we consider instead the X-ray luminosity $L_{\rm X}$ as a proxy of the accretion activity. Following the work of \cite{2020MNRAS.493.4355M}, who have investigated the X-ray properties of a large sample of high-excitation (HERG) and low-excitation (LERG) radio galaxies, we assume a ratio $L_{\rm X}/L_{\rm Edd}=1.1\rm e{-3}$\footnote{This is the average value found by \cite{2020MNRAS.493.4355M} in high-excitation radio galaxies. The median is $L_{\rm X}/L_{\rm Edd}=6.7\rm e{-4}$.}  as the limit above (below) which the source is classified as HEG (LEG). 
This method is applied to all the radio galaxies and to the NLSy1; the BL\,Lacs are instead assumed to be LEG since they form the beamed parent population of LERG. In the following we use the abbreviations HEG and LEG when we generically refer to high- and low-excitation galaxies respectively, and HERG and LERG when we limit our analysis to radio galaxies. 

The jet viewing angles, reported in Column 10, are assumed based on results from the literature. In the case of several BL\,Lacs for which more stringent constraints are unavailable, the adopted angles were used in the cited works to reproduce the broadband spectral energy distribution.
The jet shape is defined by the power-law index $k$, reported in Column 11. For jets showing a parabolic shape in the innermost regions, $k$ is the power-law index in the parabolic region. About half of the sources in the sample, mostly radio galaxies, show parabolic expansion ($0.3 \leq k \leq 0.7$), while the shape of the other objects, mainly BL\,Lacs, is either conical or hyperbolic ($k \geq 0.7$). For most of these parabolic jets, a possible location for the transition distance has been suggested in the literature through the observation of a jet break in the expansion profile and/or based on the jet kinematic properties. The transition distance expressed in projected milli-arcseconds is reported in Column 12. Note that some sources in the MOJAVE sub-sample (e.g., 3C\,84, NGC\,1052, NGC\,6251) have also been studied by other authors \citep[e.g.,][]{2016ApJ...833..288T, 2018NatAs...2..472G, 2020AJ....159...14N} but we refer to the MOJAVE results for uniformity. The dedicated studies are discussed in the text. 

\subsection{Literature data}

For the visualization of the 15 GHz data from the MOJAVE survey we follow a different but complementary approach with respect to the one used by \cite{2017MNRAS.468.4992P, 2020MNRAS.tmp.1273K}. These authors have analyzed the jet expansion by examining transverse intensity profiles in stacked images, and have determined the jet shape. For the purposes of our comparison, we consider instead the results from Gaussian fitting in the visibility domain, presented by \cite{2019ApJ...874...43L}. The availability of many observing epochs for each source implies that several measurements of the jet width at similar distance from the core are provided. Since one of our main goals is to compare the full transverse width of different jets, and the jet cross-section may be not fully visible in a single epoch, we adopt the following empirical method, which was tested against the stacking\footnote{ In general, the two methods yield comparable results. Our tests indicate that stacking is to be preferred when the jet ridge line varies strongly over time (e.g., in the diffuse regions of an edge-brightened BL\,Lac like Mrk\,501), but in other cases an even larger part of the cross-section may be recovered through the described method.}: for every five measurements of the jet width we select the maximum; then, to reduce the scatter, we compute a 5-point average of the maxima. In the single case of 1ES\,1959+650, where the number of points is insufficient, we select the maximum every 3 measurements and we compute a 3-point average. When the jet is two-sided, we consider width measurements in both sides. To ensure that we are taking into account resolved jet regions, we exclude data points in the inner $0.5\,\rm mas$. For sources showing a parabolic shape on the MOJAVE scales, we also exclude data points beyond the transition distance, whenever this was determined (Column 12 in Table 5). In the case of M\,87, and possibly of 3C\,84 and Mrk\,501, the MOJAVE data describe only part of the parabolic region. The distances of the jet features reported by \cite{2019ApJ...874...43L} are computed with respect to the VLBI core position, while we ideally want to compare the width of different jets at the same distance from the black hole. Even though this offset is usually negligible on the considered scales, we apply the correction when the offset is known \citep[based on the analysis from][Column 7 in Table 3]{2020MNRAS.tmp.1273K}. No shift was applied in the case of Mrk\,501 and I\,Zw\,187, for which a negative offset was determined. 

The radio galaxy 3C\,264 is also part of the MOJAVE sample, but was not analyzed by \cite{2019ApJ...874...43L}. For this source we include data from \cite{2019A&A...627A..89B}, Fig. 9, where the jet expansion has been investigated in the image domain by considering the stacked 15 GHz image. We take into account data for the innermost jet region until the recollimation at ${\sim}11$\,$\rm pc$ and we apply the same method described above, except that, since in the stacked image we already recover most of the jet cross-section, we do not average maxima but the data points directly. 

For Cygnus\,A, we report 43 GHz data from \cite{2016A&A...585A..33B}, while for NGC\,4261 the data points are not publicly available and we refer to Fig. 8 of \cite{2018ApJ...854..148N}.

\subsection{Mass-scaled expansion profiles}

The mass-scaled jet expansion profiles are shown in Fig. 5, left and right panels. In the left panel we compare the profiles of fourteen sources showing a parabolic expansion. The data points are coded with a different color and symbol for each source. A power law of the form $r\propto z^k$ is also fitted to the data of M\,87, Cygnus\,A, and 3C\,120 (magenta, green, and violet line, respectively in Fig. 5, left panel), with the coefficient $k$ fixed to the value reported in Table 5 for each source. Indeed, our aim is not to determine the jet shape, an information that already exists in the literature, but to examine where these profiles lie with respect to each other.  The comparison shows that the sources are not all aligned along the same profile. NGC315 and M\,87 present the "thinnest" jet, and sources like NGC\,6251, 3C\,264, 3C\,371, Mrk\,501, and Mrk\,180 lie on a similar or slightly upshifted profile. On the contrary, the rest of the radio galaxies, as well as the narrow-line Seyfert\,1 1H\,0323+342, present "thicker" jets, with the maximum width being observed in 3C\,120 and in 1H\,0323+342. None of the radio galaxies classified as HERG lies on the M\,87 profile, but two of the LERG (3C\,84, NGC\,1052) are aligned with the profiles of Cygnus\,A or 3C\,111.  While, in Fig. 5, we do not report data for the LERG NGC\,4261, \cite{2018ApJ...854..148N} showed that this object presents a jet width which is intermediate between the one of Cygnus\,A and M\,87. 

The relation between the source classification and the collimation profile is further explored in the right panel, where all the sources in Table 5, not only those with parabolic shape, are compared. As evident from the figure, the BL\,Lacs are on average probed on larger scales than radio galaxies, which is the most likely reason for the observation of mainly conical shapes (i.e., the jet collimation occurred on unresolved scales). By expanding faster, the BL\,Lac jets reach, at the largest distances (${\sim}10^7$\,$R_{\rm S}$), widths which are comparable with those of the HERG (green points) and of the NLSy1 1H\,0323+342 (dark blue points). However, the BL\,Lacs conical profiles smoothly connect to the parabolic profiles of M\,87 and other LERG (magenta points) observed on smaller scales. If we assume that the jet expansion in the collimation zone is described by a single parabolic profile from the launching up to the transition region, as observed for M\,87 \citep[e.g.,][]{2019MNRAS.489.1197N}, than these results suggest that the HERG and the NLSy1 in our sample have jets launched at larger disk radii than jets in BL\,Lacs and in most of the LERG. 

Concerning the two ``outliers'', NGC\,1052 and 3C\,84, we note that these are both peculiar LERG. NGC\,1052 presents features which are unusual for a low-luminosity AGN, since the nucleus is obscured by a high column density torus  \citep{2004A&A...426..481K}, and a broad iron line of unclear origin has been detected based on X-ray observations \citep{2009ApJ...698..528B}. \cite{2018MNRAS.478L.122R} proposed that that NGC\,1052 hosts a hybrid accretion disk in a transition regime, i.e. formed by a central ADAF and an outer thin disk \citep[e.g.,][]{1997ApJ...489..865E}. 
A disk rather than a spherical accretion flow has been proposed also for 3C\,84 \citep{2014ApJ...797...66P} based on a study of Faraday rotation in the nucleus, and space-VLBI observations with the RadioAstron telescope have shown that the jet base is indeed wide, and the jet may be anchored to the outer accretion disk \citep{2018NatAs...2..472G}. As discussed further in Sect. 4.5, both these sources show signs of young and/or restarted activity. 

\subsubsection{Caveats}

Before discussing further the possible implications of the results in Fig. 5, we comment on some caveats to be taken into account, and that are also relevant for the comparison in the next Sect. 4.3. Firstly, there exists the possibility that some of the profiles in Fig. 5 are misplaced. The main source of uncertainty is in this case the black hole mass. The assumed values may be incorrect, and a different mass will cause a given profile to shift along the $x$ and $y$ axis by the same amount. Since the Schwarzschild radius is directly proportional to the mass, the shift will also be directly proportional to the change in mass, and the profiles will shift along parallel lines. Another source of uncertainty is related to the jet viewing angle. While the mass-scaled transverse widths ($y$ axis) only depend on the mass, the de-projected distances on the $x$ axis depend, in addition, on the assumed $\theta$, being inversely proportional to $\sin\theta$. Our choice to limit the sample to nearby and well studied objects is aimed at mitigating the impact of the uncertainty on the mass and jet orientation, but incorrect assumptions are possible. In the case of 1H\,0323+342, in particular, \cite{2014ApJ...795...58L} have proposed a mass larger by one order of magnitude; the radio properties of some of the BL\,Lacs in the sample are also compatible with a more misaligned jet orientation \citep{2004ApJ...613..752G, 2006ApJ...646..801G}

Concerning the conclusions we can draw on the jet origin, based on a back-extrapolation of the jet profiles, it should be noted that the assumption that the jet expands following a single parabolic profile in the collimation region may not be valid. While this is observed in the best studied case of M\,87, we cannot exclude that jets in other sources present more complex profiles at their base. In the case of NGC\,1052, one of the two LERG placed on the HERG region, high resolution VLBI observations with the GMVA \citep{2016A&A...593A..47B} do not clearly resolve the jet base, thus it is plausible that this jet was initially narrow and has experienced a fast expansion at larger distances. Moreover, \cite{2020AJ....159...14N} did not confirm a parabolic expansion for this source, suggesting instead a transition from cylindrical to conical shape at a distance of ${\sim}10^4$\,$R_{\rm S}$.  3C\,84, i.e., the other LERG in the HERG region, shows evidence for a more complex profile as well. In this case the jet base was well resolved by space-VLBI observations \citep{2018NatAs...2..472G}, and the inner-jet shape was found to also approach a cylinder.

Finally, another element to consider is the impact of relativistic effects. If a wide and mildly relativistic, disk-launched component was present in some of the BL\,Lacs, its detection could be prevented by the strong relativistic Doppler boosting, which amplifies the emission from the narrow and fast spine. However the good alignment between the BL\,Lacs profiles and those of their misaligned parent population, i.e. the low-power radio galaxies, suggests that BL\,Lacs intrinsically miss this component. Similarly, the jet orientation could influence the observed difference between HERGs and LERG, as a mildly relativistic jet sheath could appear more or less prominent in radio galaxies seen at different angles. This also seems to be not a concern, given that different widths are observed in radio galaxies seen at similar angles. For instance, the jets in M\,87 and 3C\,120 have very similar orientation ($\theta\sim16^{\circ}-19^{\circ}$), but the mass-scaled jet widths differ by one order of magnitude. In fact, unlike M\,87, 3C\,120 hosts a powerful classic disk with high Eddington ratio \citep{2006ApJ...642..113G,2009MNRAS.392.1124V, 2012ApJ...752L..21C}, and a disk origin for this jet was suggested based on a combined X-ray and radio monitoring by \cite{2002Natur.417..625M}.
To test further the impact of relativistic effects, it would be interesting to examine where FSRQs, not included in our sample due to the redshift cut, would be placed in Fig. 5.  The only high-excitation source with a blazar-like jet orientation is 1H\,0323+342, whose jet profile appears to be well aligned with that of 3C\,120. This would indicate that a jet orientation towards the line of sight does not prevent the observer from detecting an extended jet sheath, if this is present.

Ultimately, a more solid investigation of all these aspects will be possible in the future by considering larger samples at the highest possible resolution, in order to reduce the impact of the uncertainties related to the mass and viewing angle and to the back-extrapolation of the profiles, and to test how the appearance of the jet structure depends on relativistic effects.

\subsection{Transition distance}
\begin{figure*}
\centering
\includegraphics[width=0.49\textwidth]{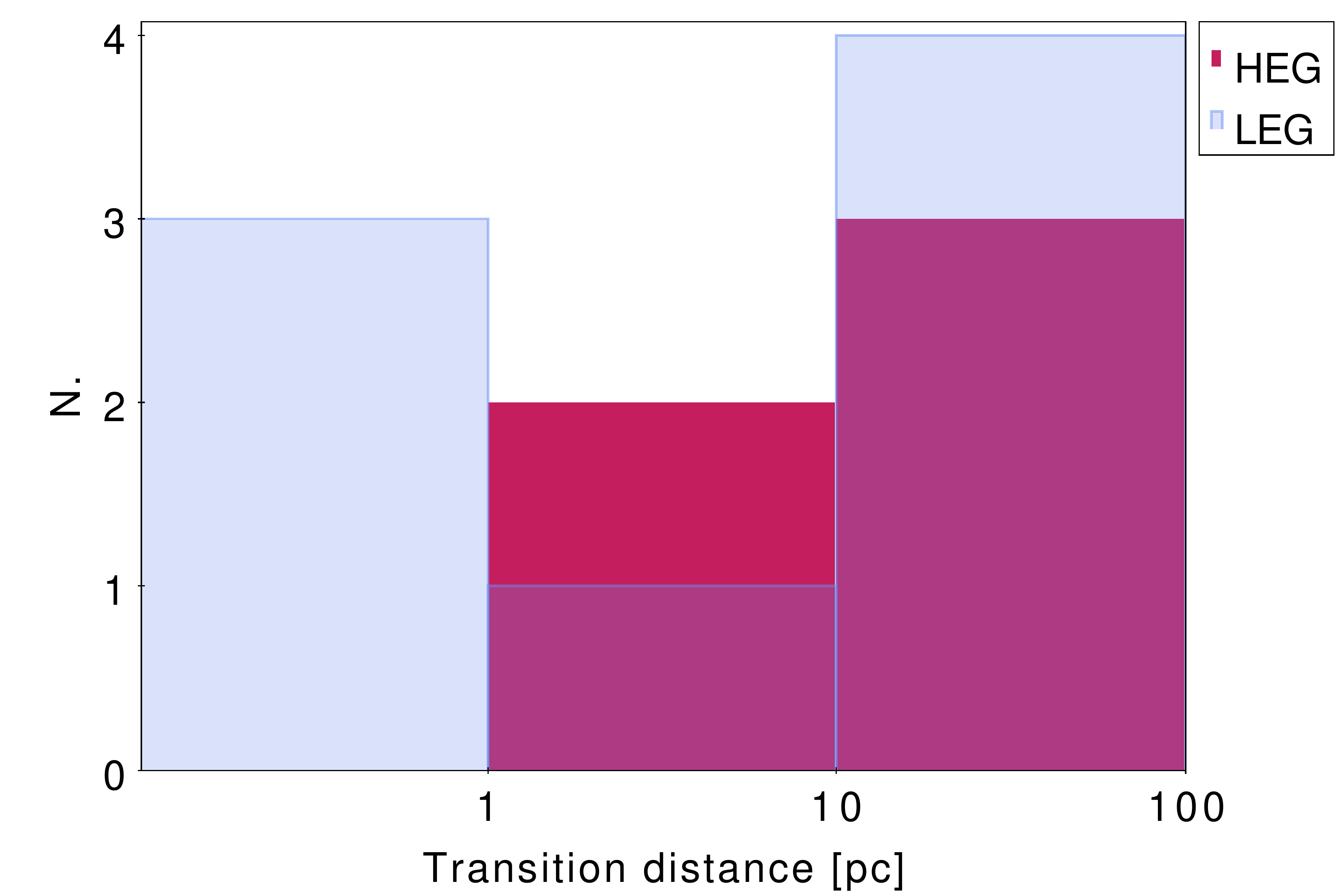}
\includegraphics[width=0.49\textwidth]{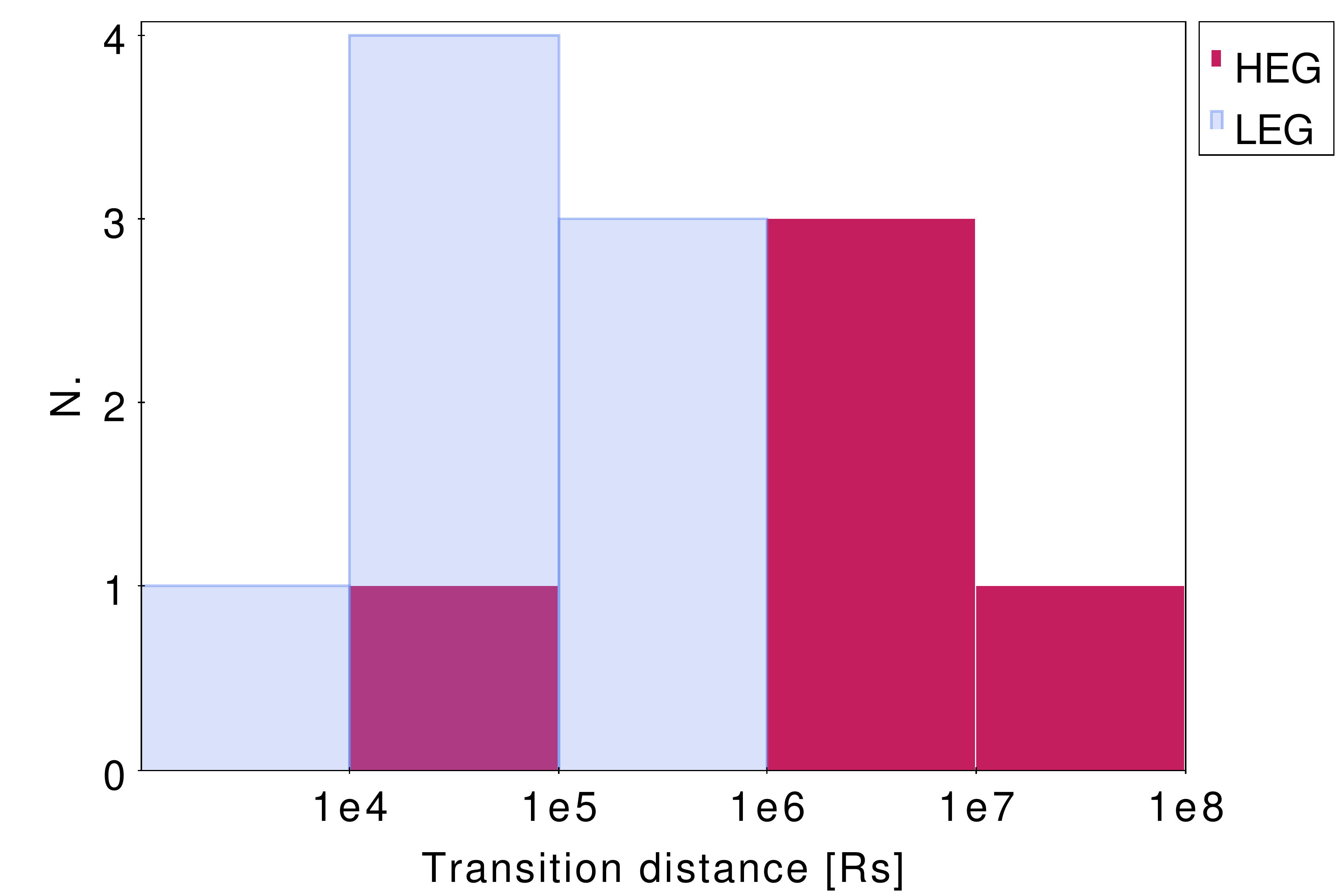}
\caption{Distribution of the de-projected transition distance in low-excitation (LEG) and high-excitation (HEG) galaxies. Left panel: the distance is expressed in units of parsecs. Right panel: the distance is expressed in units of Schwarzschild radii.}
\end{figure*}
Limiting our analysis to sources which show a transition in the jet shape and/or in the kinematic properties of the jet (Col. 11 in Table 5), we examine how the transition distances are distributed based on the AGN classification as HEG or LEG. With the exception of Mrk\,180 and 3C\,371, this excludes most of the BL\,Lacs, which are either already freely expanding or show a purely parabolic profile on the examined scales (this is the case for Mrk\,501 only). For the latter reason, 3C\,84 is also excluded.

Results are presented in the histograms in Fig. 6. In the left panel, the transition distance is expressed in de-projected parsecs. When considering the low-excitation sources, it is evident that NGC\,315 is not the only object showing a sub-parsec scale transition, which is observed also in other two low-luminosity jets, NGC\,1052 and NGC\,4261. At least in the case of NGC\,4261, this distance is also much smaller than the Bondi radius, estimated by \cite{2008A&A...486..119B} to be ${\sim}32$\,$\rm pc$, but it may match the actual extent of the hot gas phase as inferred from a spectral fitting performed by \cite{2010PASJ...62.1177K}.
None of the jets in HEGs presents a sub-parsec scale transition, and the transition distances seem to be shifted to larger radii than in LEGs. However, a two-sample Kolmogorov-Smirnov (KS) test 
does not allow us to conclude that the two classes are drawn from a different population, as the p-value\footnote{Here the p-value is the largest probability that, given two random samples drawn from identical populations (null hypothesis), the difference between the two cumulative distributions is at least as extreme as the observed one. The null hypothesis is assumed to be rejected for $p<0.05$. See, however, the discussion on p-values by \cite{pvalue}.}
is $p_{\rm ks}=0.32$. In the right panel of Fig. 6, the transition distance is expressed in de-projected Schwarzschild radii. With the exception of Cygnus\,A, all the jets in HEGs show a transition on scales $>10^6$\,$R_{\rm S}$, while all the LEGs are below this limit. In this case, the separation between the two classes becomes more evident, and the KS-test indicates that they are drawn from a different population ($p_{\rm ks}=0.02$). 

Additional results from the literature for sources not included in our sample are consistent with this picture. In the LERG Centaurus\,A \citep{2011A&A...530L..11M}, the jet was observed to be freely expanding already on scales of ${\sim}0.1\,\rm{pc}$, or ${\sim}10^4$\,$R_{\rm S}$ for $M_{\rm BH}=5\times10^7$\,$M_{\odot}$ \citep{2010PASA...27..449N}, in agreement with our findings for NGC\,315 and other LERG. On the other hand, studies of the jet shape in some high redshift FSRQs \citep{2019ApJ...886...85A, 2020A&A...634A.112T} locate the shape transition at distances $>10^6$\,$R_{\rm S}$, as found for the HEGs in our sample. 

The different extent of the collimation zone in low- and high-luminosity sources is well matched by the findings of \cite{2015MNRAS.453.4070P}, who have investigated the extent of the acceleration region in a sample of blazars. FSRQs were shown to reach their terminal Lorentz factor on scales larger than $10^5$\,$R_{\rm S}$, while shorter distances were inferred for BL\,Lacs. This also supports the idea that, both in high-excitation and in low-excitation sources, the jet acceleration and collimation processes are co-spatial, as expected based on theoretical predictions for magnetically-driven cold outflows \citep{2007MNRAS.380...51K,2008MNRAS.388..551T,2009ApJ...698.1570L}.

\subsection{Implications for the disk-jet connection}

Keeping in mind the caveats discussed in Sect. 4.2, a relation between the properties of the jet collimation region and the properties of the accretion disk is suggested by Fig. 5 and 6. According to theoretical models and simulations, both thin disks \citep[e.g.,][]{1982MNRAS.199..883B, 2006ApJ...651..272F, 2019MNRAS.487..550L} and geometrically-thick hot disks \citep[e.g.,][]{1999MNRAS.303L...1B, 2006MNRAS.368.1561M, 2012MNRAS.420.2912B, 2013A&A...559L...3M, 2016A&A...596A..13M} can launch collimated outflows. Due to the higher mass loading and lower speed, the disk-driven jet is expected to dominate the emission in radio galaxies with respect to the de-boosted black hole-launched jet. This is confirmed in observations by the direct imaging of limb-brightened jet structures \citep[e.g.,][]{2016A&A...585A..33B, 2016A&A...595A..54M,2018NatAs...2..472G} as well by kinematic studies of radio galaxies, which generally show much lower intrinsic speeds than measured in blazars \citep{2019ApJ...874...43L}. As these properties are observed in high-luminosity and low-luminosity radio galaxies alike, a jet sheath must be produced from disks spanning different accretion regimes. Our results, however, indicate that the disk-driven jet in LERG originates at small disk radii (few $R_S$, as measured in M\,87), and indeed the expansion profiles of most of the LERG are well aligned with those of BL\,Lacs, which are expected to be dominated by the black hole-launched spine \citep[see e.g.,][]{2014Natur.515..376G}. This result is in broad agreement with models of jet launching from ADAFs \citep[e.g.,][and references therein]{2011ApJ...737...94C, 2014ARA&A..52..529Y}, which predict the formation of a thin and mildly-relativistic outer layer. ADAF models also predict the launch of a non-relativistic disk-wind component carrying the bulk of the disk mass outflow and spanning a large solid angle. There is no evidence for such a component based on the analyzed VLBI images, at least in the considered frequency regime. The jet profiles in HEGs, on the other hand, are all shifted upwards, and a back-extrapolation down to the jet base suggests that the jet sheath is launched at larger disk radii. Taking as a reference Cygnus\,A, which shows the thinnest jet among HEGs and for which an initial jet width of ${\sim}200$\,$R_{\rm S}$ was measured based on GMVA observations \citep{2016A&A...588L...9B}, the present data suggest that thin disks could launch collimated winds with an initial outer radius $\gtrsim100$\,$R_{\rm S}$. This possible difference in the outer radius of the jet sheath is accompanied by a different extent of the collimation region in HEGs and LEGs (Fig. 6). Modeling of jet collimation by disk winds, presented by \cite{2016MNRAS.461.2605G}, revealed a direct link between the wind outer radius and the collimation radius: for a given wind power, larger wind radii correspond to more extended collimation zones. A sufficiently high ratio ($>0.1$) of wind power to jet power is required for this process to be efficient. When this condition is verified in reality, is a matter of debate. In recent simulations presented by \cite{2017A&A...606A.103H}, the diverse kinematic behavior of VLBI knots in blazars of different powers could be well explained by varying this ratio. Except for the least powerful class among BL\,Lacs (that of the High-frequency peaked BL Lacs, HBLs), whose properties could be reproduced by assuming an absent or very weak wind, ratios larger than 0.3 were suggested for blazars. A question remains concerning the portion of these winds which is actually detected in VLBI observations. When attempting to model the M\,87 jet collimation profile, \cite{2016MNRAS.461.2605G} have suggested that the radio emission is produced in the shocked interface between the relativistic jet and the outer wind, which is undetected. Observational constraints on extended disk winds may be provided through other methods. For instance, we note that for all the HERGs in our sample (except PKS\,1514+00) the detection of ultra-fast outflows was reported based on X-ray observations \citep{2010ApJ...719..700T, 2014MNRAS.443.2154T, 2015ApJ...808..154R}. These outflows, whose launching mechanism is unclear, are suggested to be characterized by mildly relativistic speeds, to originate at disk radii of $10^2-10^4$\,$R_{\rm S}$ (in agreement with our findings), and to carry a significant fraction of the jet kinetic power. Thus collimation via the action of disk winds, where by disk winds we mean a mildly relativistic jet sheath plus possible broader outflows, appears to be a viable mechanism, especially for high-luminosity sources.

\subsection{Implications for the FRI-FRII dichotomy}

In this final section we discuss the possible impact that a more or less prominent disk-launched jet may have on the long-term jet stability and evolution. A surrounding jet sheath \footnote{Please note that we refer to sheaths or winds surrounding fast jet spines at sub-pc and pc scales, in contrast to well developed mixing layers in already decelerated jets at kpc-scales.}, characterized by a smaller sound speed with respect to the central hot spine, provides further stability to relativistic outflows due to the increase in the jet inertia and the drop in instability growth rates
\citep{2002ApJ...576..204H, 2003ApJ...583..116H, 2005A&A...443..863P, 2007ApJ...664...26H, 2007A&A...469L..23P, 2007ApJ...662..835M, 2019A&A...627A..79V, 2019MNRAS.482.3718P}. Furthermore, its presence mitigates the impact of 
surface perturbations that can be induced by the penetration of stars \citep[which has been recently appointed as a possible triggering mechanism for jet deceleration of FRI jets, see][]{2020MNRAS.494L..22P}, since its smaller sound speed implies a slower propagation of the turbulent layer towards the jet axis. This provides the jet channel with time to reach the intergalactic medium without being decelerated \citep[see the discussions in, e.g.,][]{2012IJMPS...8..241P, 2016AN....337...18P}. In summary, a surrounding sheath or wind component shields the inner spine against entrainment from the ambient interstellar medium. 

According to this, we could expect a relation between the jet widths plotted in Fig. 5 and the large-scale morphologies of those jets. At zero order, this is the case, with the lower line populated by FRI jets alone, in addition to BL\,Lacs. Among the upper lines corresponding to thicker jets we observe the FRII radio galaxies Cygnus\,A, 3C\,111, PKS\,1514+00, and 3C390.3, plus other sources that do not develop clear FRI morphologies. NGC\,1052 is a young source, and the large scale morphology may be consistent with that of a young FRII \citep[see][]{1984ApJ...284..531W}; 3C\,84 is a recently reactivated source showing a bright hotspot-like feature \citep[e.g.,][]{2017ApJ...849...52N} which would approach it to a (temporary, at least) FRII morphology; 3C\,120, while is more often classified as an FRI, presents a peculiar morphology, with a strongly bent jet and edge-brightened lobe \citep{1987ApJ...316..546W}.

The origin of the FRI/FRII dichotomy was widely discussed in the literature \citep[e.g.][]{2007A&A...470..531W, 2013MNRAS.430.3086G, 2016MNRAS.461L..46T,  2019MNRAS.488.2701M}, being attributed to a diversity in the nuclear properties or in the environmental conditions of the host. The different characteristics of disk winds ejected from HEGs and LEGs, suggested by our work, may provide a crucial link between these two scenarios, since the nature of the accretion disk has a direct impact on the properties of the environment, i.e. the wind, by which the jet is collimated and stabilized. We note that, while FRI and FRII morphologies usually develop in LERG and HERGs respectively, cross-populations formed by FRI-HERGs and FRII-LERG also exist, but differ significantly in size. FRI jets powered by powerful nuclei are highly infrequent (3C\,120 is one of the few examples), while FRII morphologies are often found in low-luminosity galaxies  \citep[][and references therein]{2014ARA&A..52..589H, 2020MNRAS.493.4355M}. This may suggest that when the jet is confined by a strong wind, the FRI morphology rarely develops. On the contrary, sources hosting inefficiently accreting disks that can still launch a relatively powerful wind may manage to develop an FRII morphology as long as the jet reaches the intergalactic medium with a high degree of collimation.

\section{Summary}

In this paper we have presented a detailed study of the jet collimation in the low-luminosity radio galaxy NGC\,315, based on a multi-frequency VLBI and VLA data set. At 86 GHz we have imaged the innermost jet base on scales of only ${\sim}160$\,$R_{\rm S}$. We have then compared our findings to those obtained for other nearby sources, considering a sample of 27 objects classified as low-excitation (LEG) or high-excitation (HEG) galaxies.
This classification reflects a different nature of the accretion, hot and radiatively inefficient in the first case, cold and radiatively efficient in the second case. The results are summarized in the following.
\begin{itemize}
    \item The jet collimation in NGC\,315 is completed on sub-parsec scales. A transition from a parabolic to conical jet shape is detected at a de-projected distance $z_{\rm t}=0.58\pm0.28$\,$\rm pc$ (or ${\sim}5\times10^3$\,$R_{\rm S}$), which is much smaller than the Bondi radius estimated based on X-ray data, $r_{\rm B}{\sim}92$\,$\rm pc$. While most of the jets analyzed in the literature collimate on larger scales, a similar behavior is observed in other low-luminosity galaxies (NGC\,1052, NGC\,4261, Cen\,A), whose jets are freely expanding at a distance of less than one parsec from the black hole. If the transition to a conical jet shape is induced by a change in the external pressure profile, this must occur in the nuclear regions. An initial confinement from a thick disk extending out to ${\sim}10^3$-$10^4$\,$R_{\rm S}$ is possible for such objects. 
    \item We have compared the mass-scaled expansion profiles of the jets in our sample. Most jets in radio galaxies show a parabolic shape while most jets in BL\,Lacs, which are probed on large scales on average, show a conical shape at their base. At the same de-projected distance from the black hole, HEGs present "thicker" jets, while most of the jets in LEGs (including NGC\,315 and M87) expand following profiles which are well aligned with those observed in BL\,Lacs at larger distances. We suggest that, while both hot and cold disks can launch collimated winds, jets in HEGs are surrounded by more prominent outer sheaths, with an outer launch radius $>100$\,$R_{\rm S}$. On the contrary, jet sheaths launched by hot disks, e.g., by ADAFs, are anchored in the innermost disk regions, as measured in M\,87. 
   \item Jet collimation in HEGs tends to proceed over larger scales (${>}10^6$\,$R_{\rm S}$) than in LEGs. This result matches the findings of \cite{2015MNRAS.453.4070P} obtained for blazars, since jets in FSRQs have been shown to accelerate over a more extended region (${>}10^5$\,$R_{\rm S}$) with respect to jets in BL\,Lacs. This supports the idea that, both in high-luminosity and in low-luminosity sources, the jet acceleration and collimation processes are co-spatial, as expected based on theoretical models for magnetically-driven cold outflows \citep{2007MNRAS.380...51K,2008MNRAS.388..551T,2009ApJ...698.1570L}.
  \item The possibility that relativistic jets are collimated by disk winds is discussed. The observation of more extended collimation zones in jets surrounded by thicker sheaths is in agreement with theoretical modeling describing the case of a Poynting-flux dominated jet confined by a wind layer   \citep{2016MNRAS.461.2605G}. This mechanism requires the wind to carry a significant fraction (${>}10\%$) of the total jet power. According to recent simulations aimed at explaining the diverse kinematic behavior of VLBI knots in blazars of different powers \citep{2017A&A...606A.103H} this condition is verified in all cases, except for the weakest BL\,Lacs (HBLs). The X-ray detection of ultra-fast outflows in most of the HEGs in our sample \citep{2010ApJ...719..700T, 2014MNRAS.443.2154T, 2015ApJ...808..154R} supports the existence of disk winds originating at large radii and carrying significant kinetic power. Thus, particularly for high-luminosity sources, jet collimation by disk winds may be viable mechanism.
 \item Motivated by the observation of mostly FRII morphologies among the sources presenting thicker jets, we have discussed the possible role of disk winds in the origin of the FRI/FRII dichotomy. A powerful sheath stabilizes the inner spine by shielding it against entrainment from the interstellar medium \citep{2012IJMPS...8..241P, 2016AN....337...18P}, thus enabling the jet to reach the intergalactic medium with a high degree of collimation. This may explain the observed formation of FRII morphologies in HEGs but also in some LEGs -- those producing sufficiently powerful winds -- as well the rare occurrence of FRI morphologies in HEGs.

\end{itemize}

 \begin{acknowledgements} 
The authors would like to thank the anonymous referee for her/his useful comments. BB acknowledges the financial support of a Otto Hahn research group from the Max Planck Society.
MP acknowledges financial support from the Spanish Ministry of Science through Grants PID2019-105510GB-C31, PID2019-107427GB-C33 and AYA2016-77237-C3-3-P, and from the Generalitat Valenciana through grant PROMETEU/2019/071. C. Casadio acknowledges support from the European Research Council (ERC) under the European Union Horizon 2020 research and innovation program under the grant agreement No 771282. This research has made use of data obtained with the Global Millimeter VLBI Array (GMVA), coordinated by the VLBI group at the Max-Planck-Institut f{\"u}r Radioastronomie (MPIfR). The GMVA consists of telescopes operated by MPIfR, IRAM, Onsala, Metsahovi, Yebes, the Korean VLBI Network, the Green Bank Observatory and the Very Long Baseline Array (VLBA). The VLBA is a facility of the National Science Foundation under cooperative agreement by Associated Universities, Inc.
The GMVA data were correlated at the MPIfR in Bonn, Germany. This research has made use of data from the MOJAVE database that is maintained by the MOJAVE team \citep{2018ApJS..234...12L}.
The European VLBI Network is a joint facility of European, Chinese, South African and other radio astronomy institutes funded by their national research councils.   

 \end{acknowledgements}

\bibliographystyle{aa}
\bibliography{reference.bib}
\clearpage
\appendix

\begin{figure*}[!h]
    \centering
    \includegraphics[height=6.4cm]{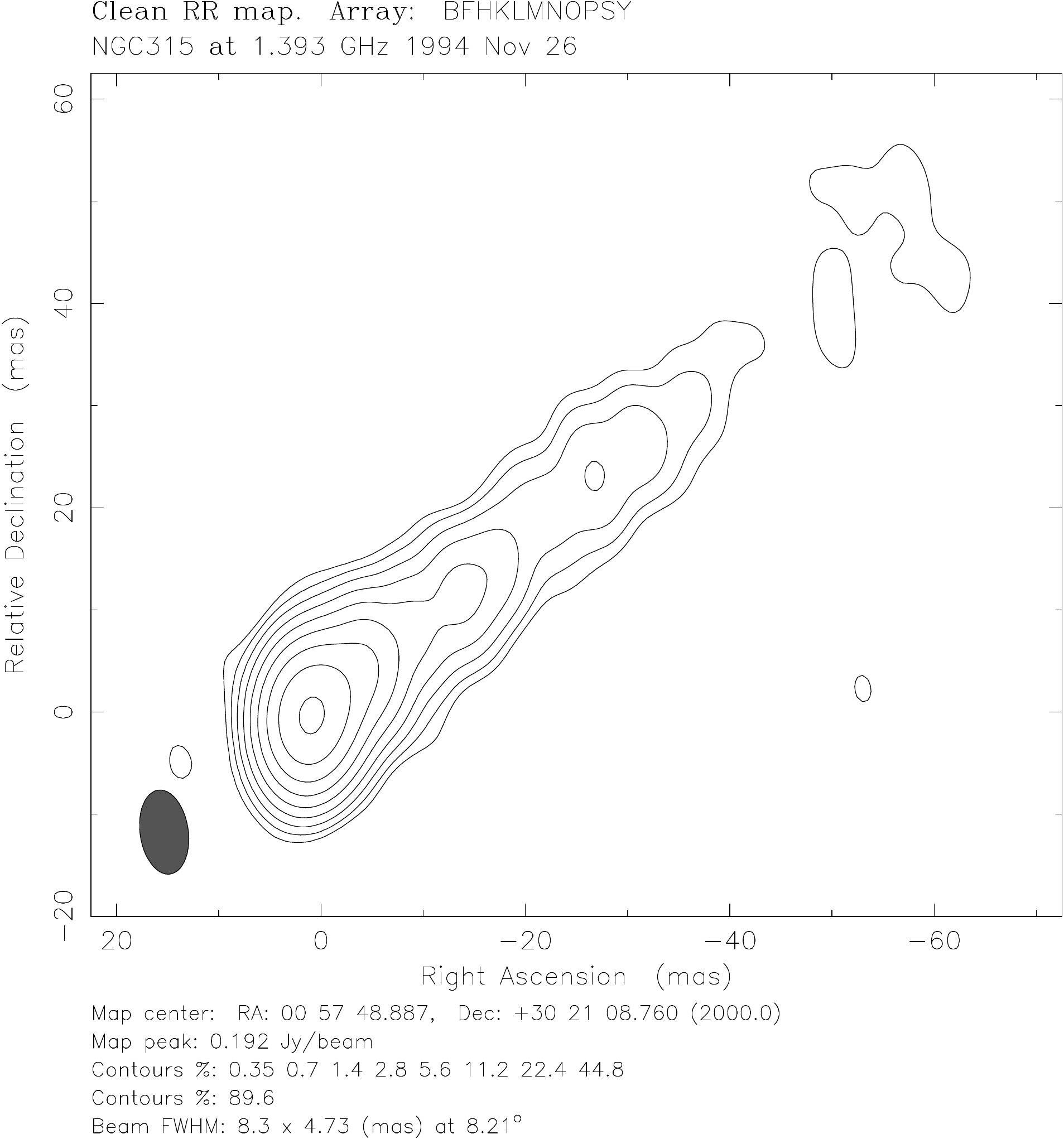}\quad\quad
    \includegraphics[height=6.4cm]{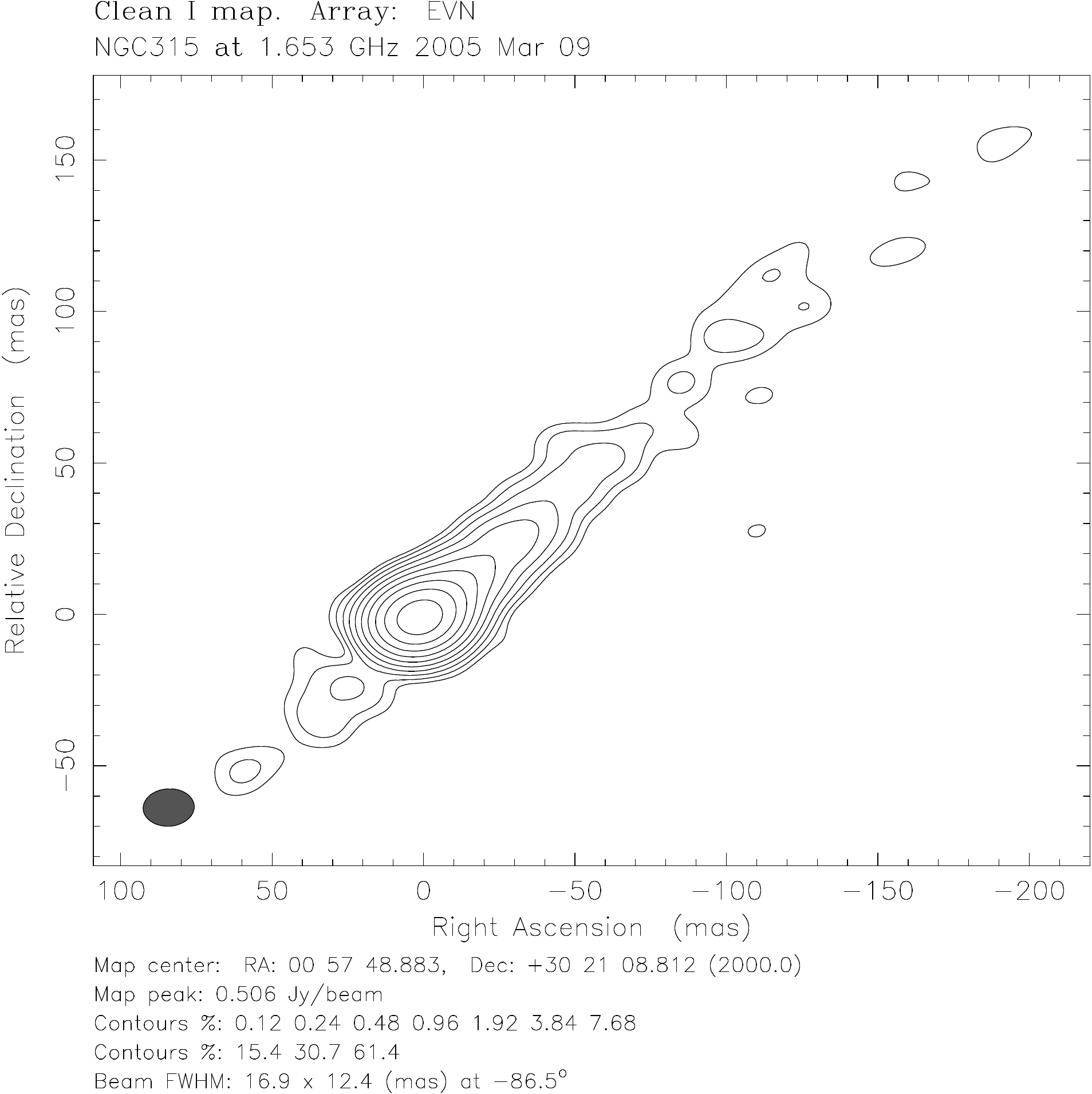}\\
    \caption{VLBI images of NGC\,315 at 1 GHz.}
\vspace{1cm}
\end{figure*}

\begin{figure*}[!h]
    \centering
    \includegraphics[height=6.4cm]{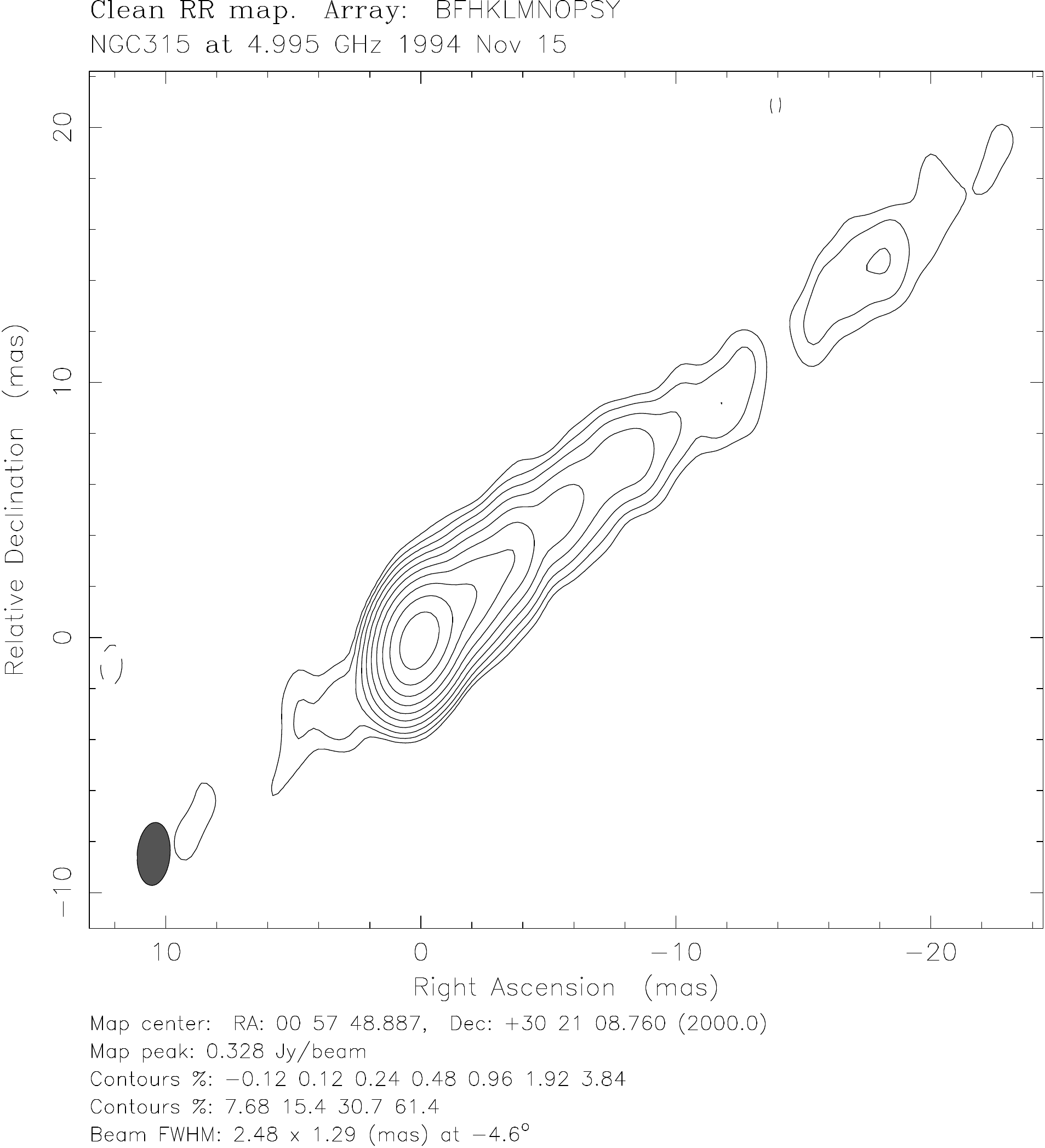}\quad
        \includegraphics[height=6.4cm]{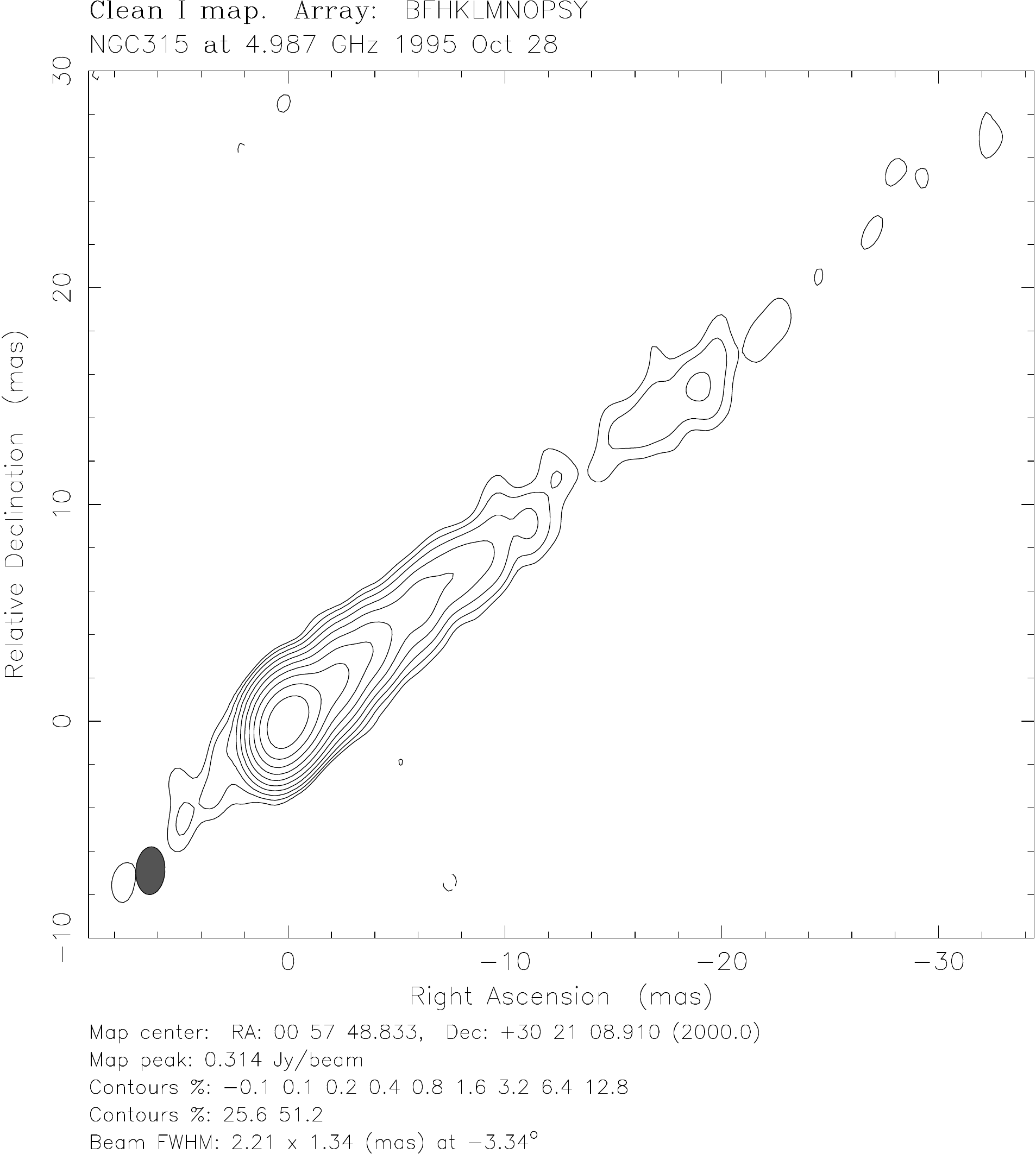}\quad
    \includegraphics[height=6.4cm]{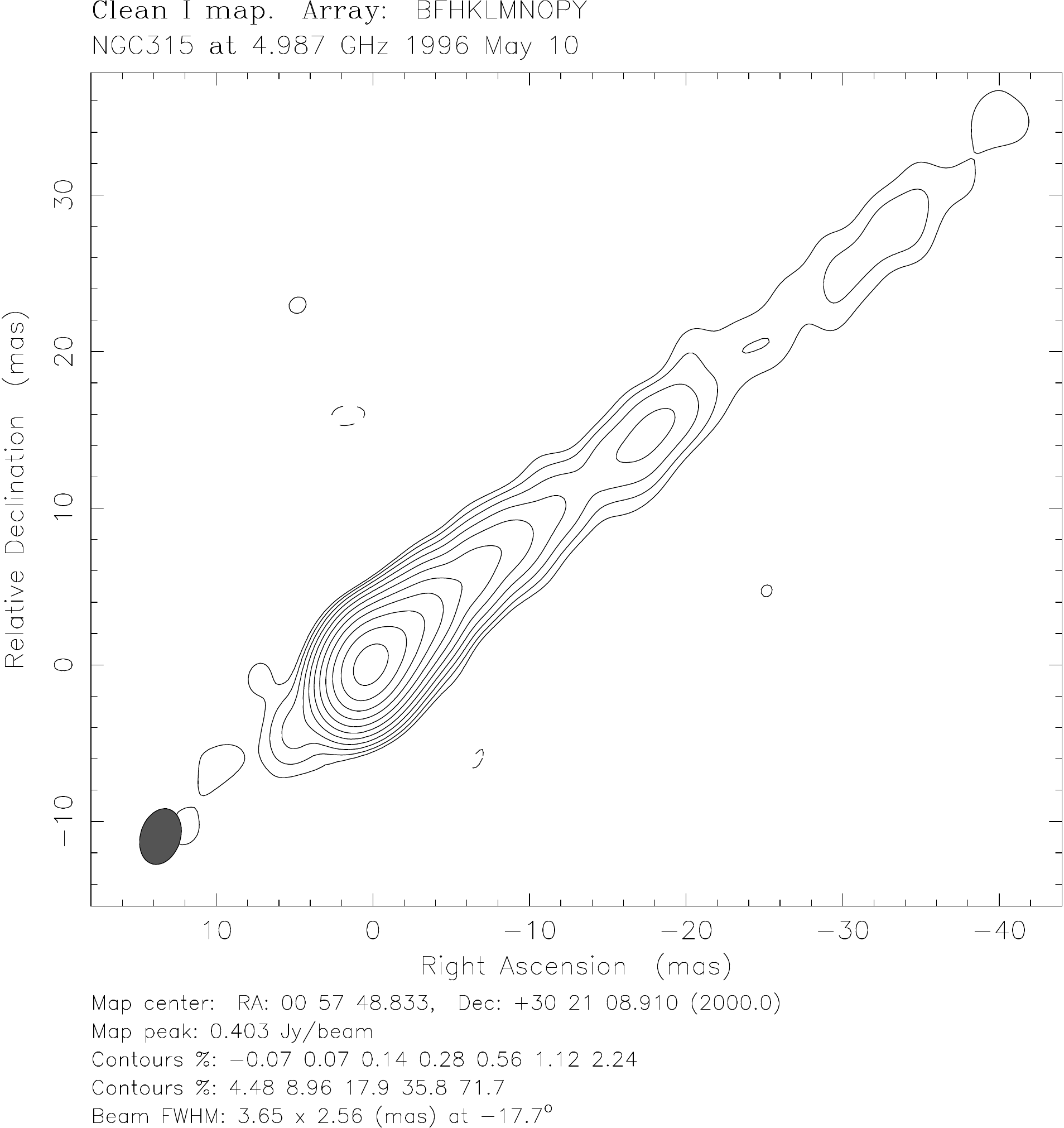}\\
    \vspace{1cm}
        \includegraphics[height=6.4cm]{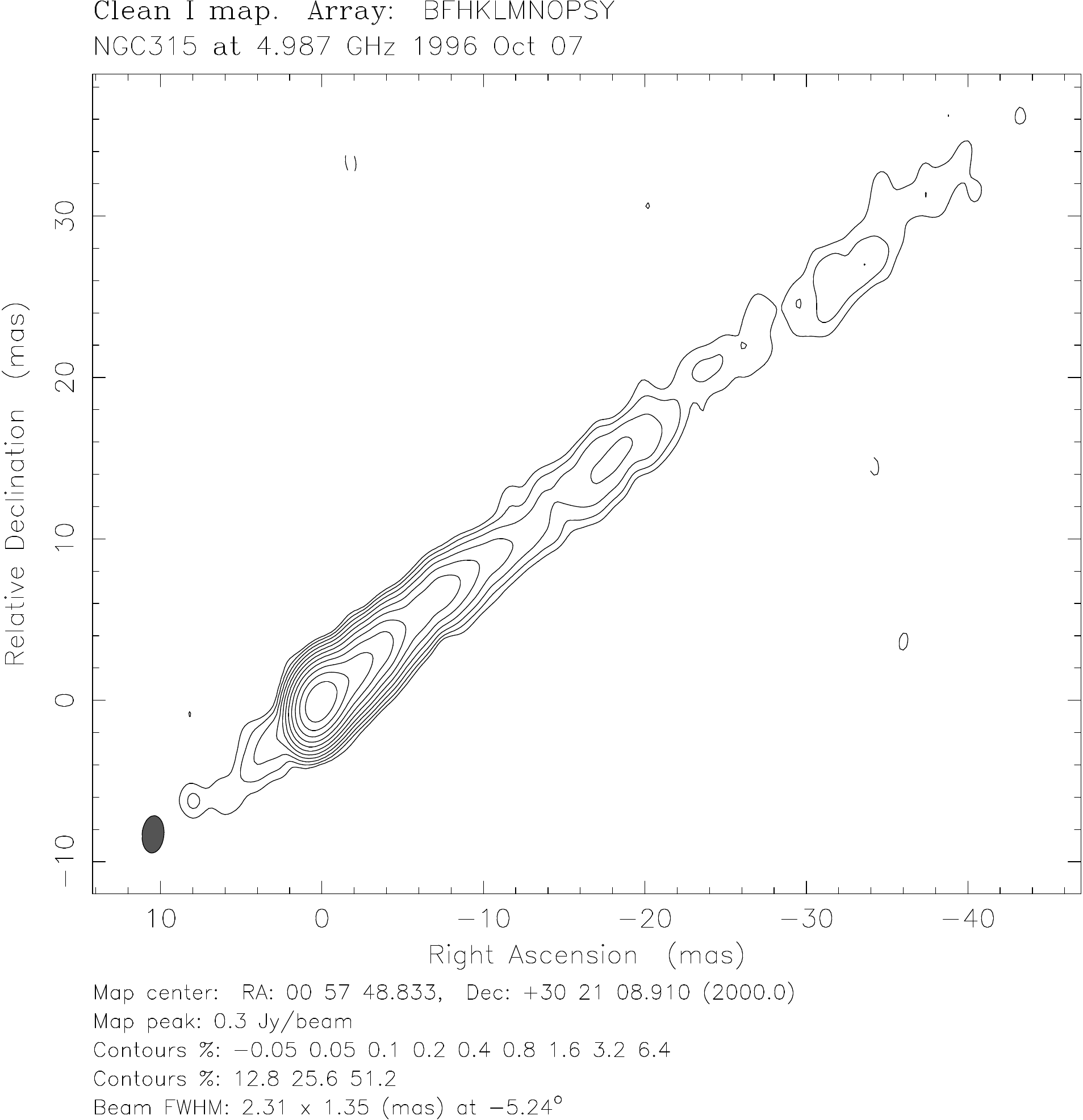}\quad\quad
    \includegraphics[height=6.4cm]{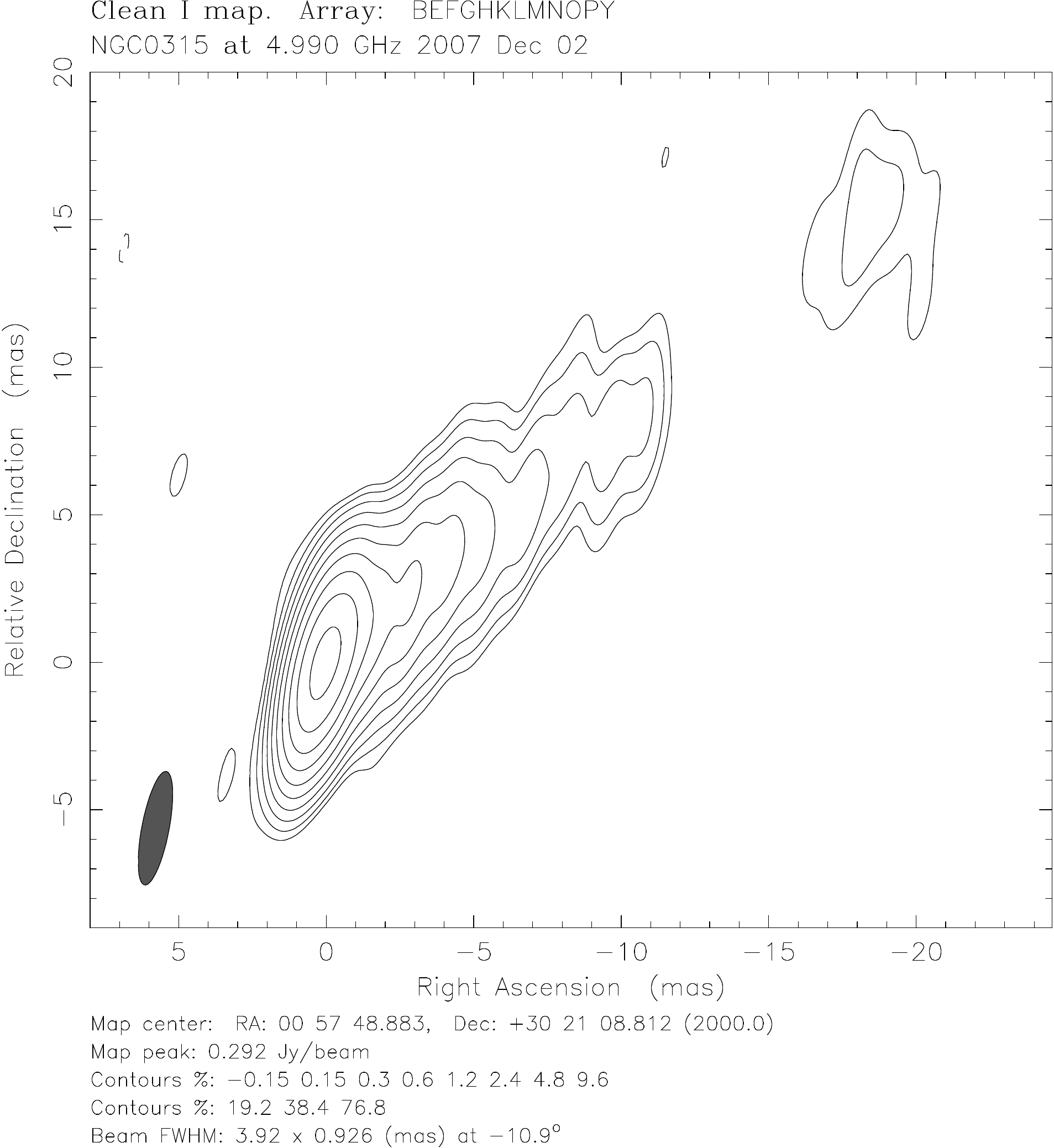}
    \caption{VLBI images of NGC\,315 at 5 GHz.}
\vspace{1cm}
\end{figure*}

\begin{figure*}[!h]
    \centering
    \includegraphics[height=6.4cm]{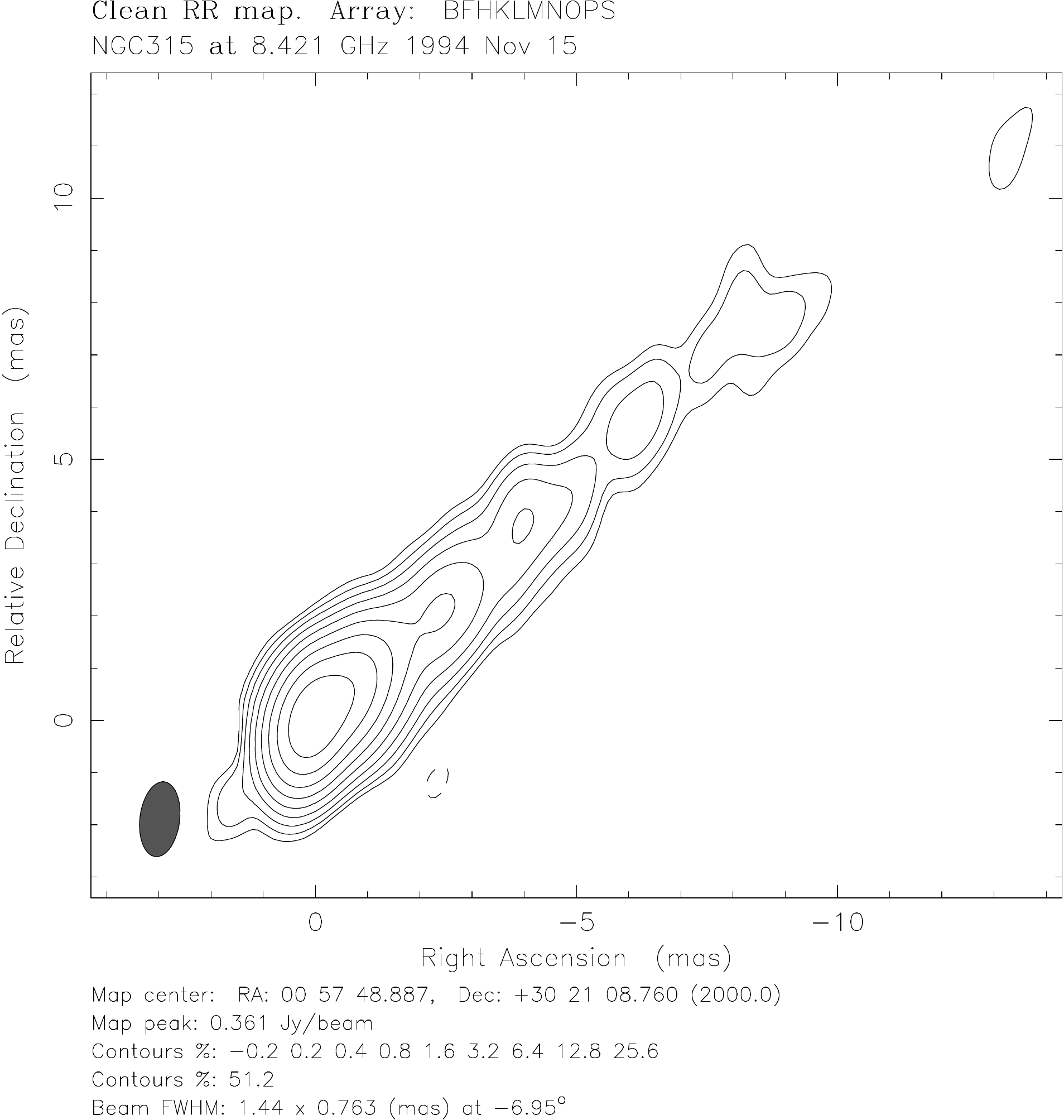}\quad
        \includegraphics[height=6.4cm]{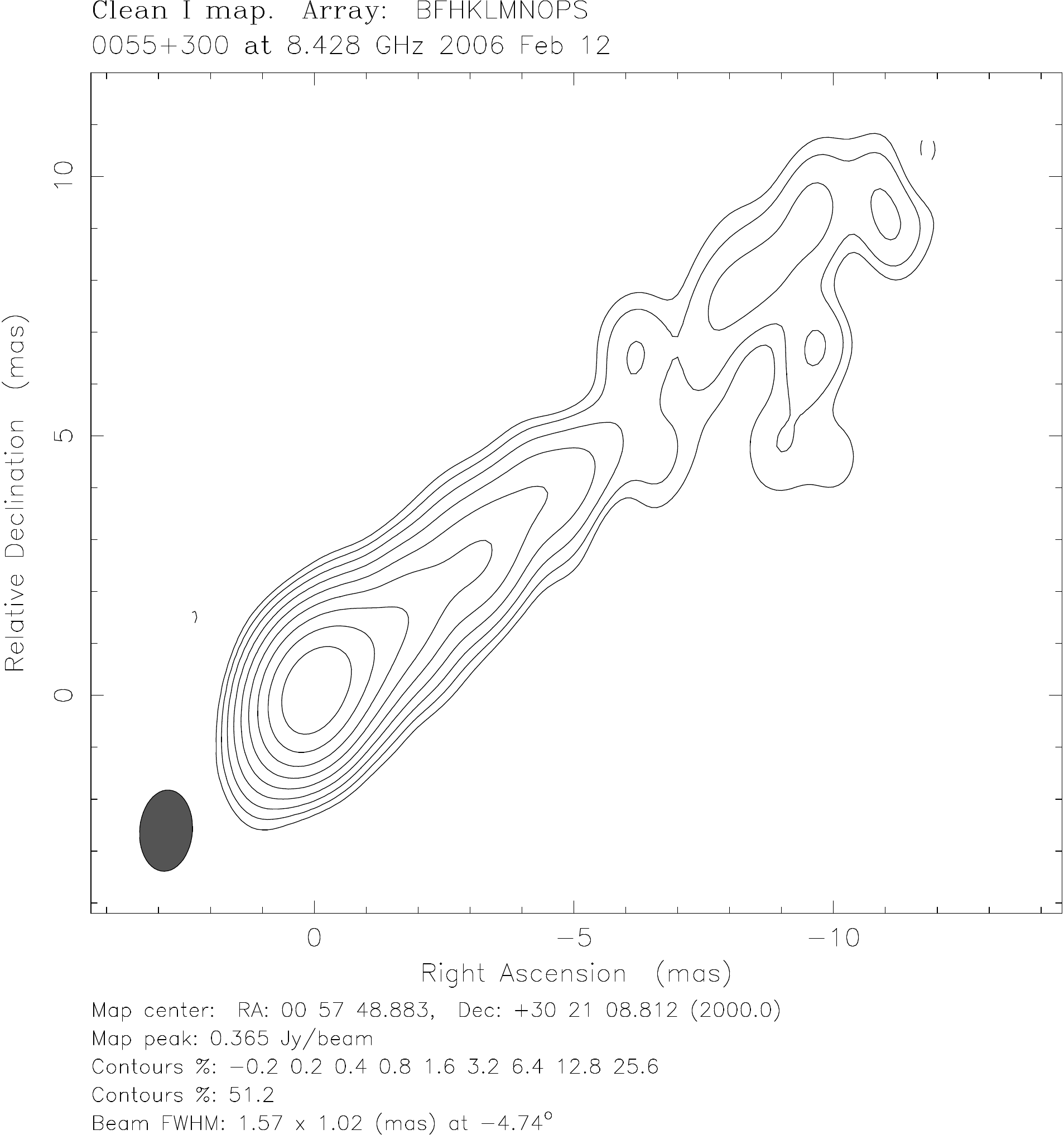}\quad
    \includegraphics[height=6.4cm]{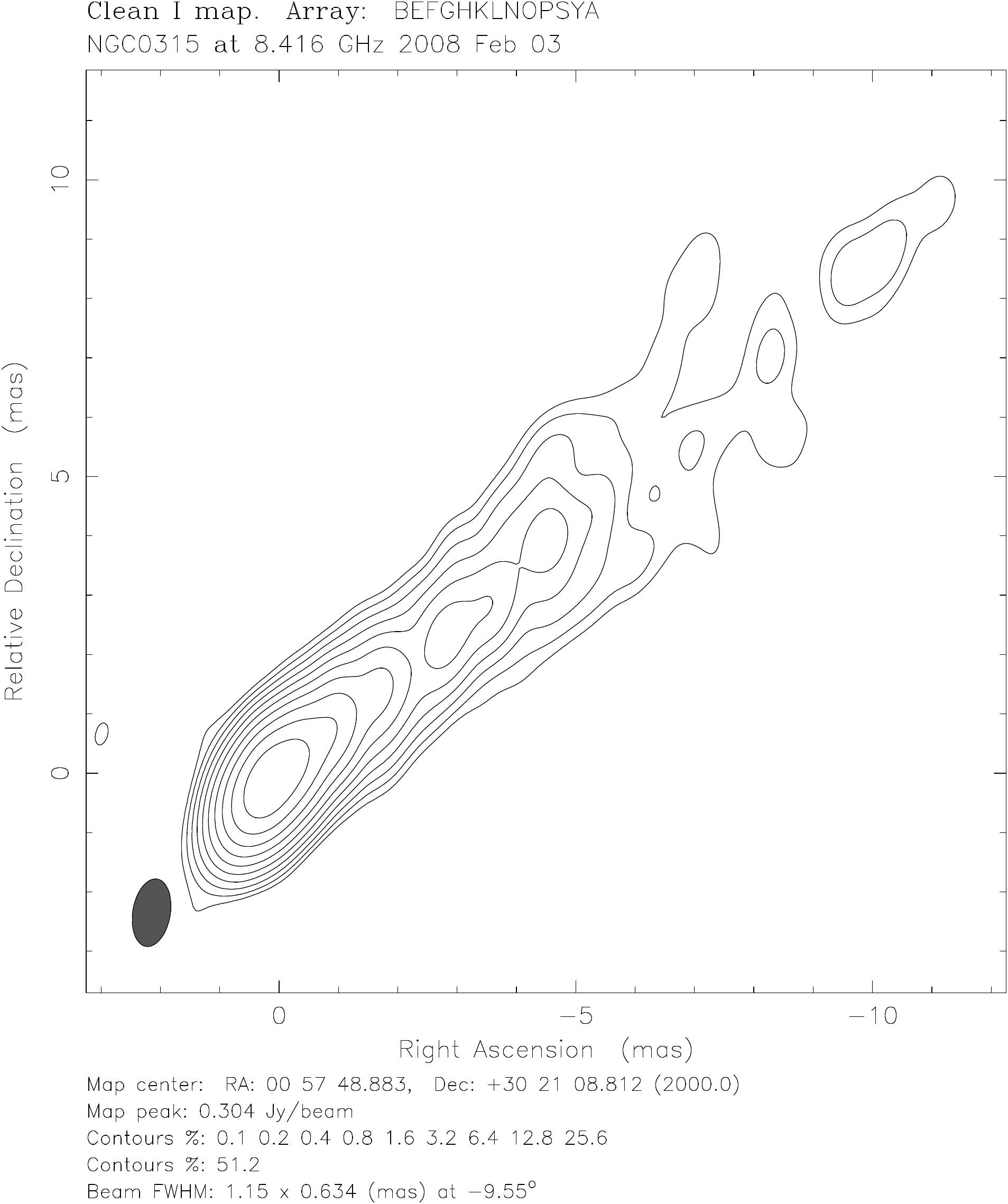}
    \caption{VLBI images of NGC\,315 at 8 GHz.}
    \vspace{0.8cm}
\end{figure*} 

\begin{figure*}[!h]
    \centering
    \includegraphics[height=6.4cm]{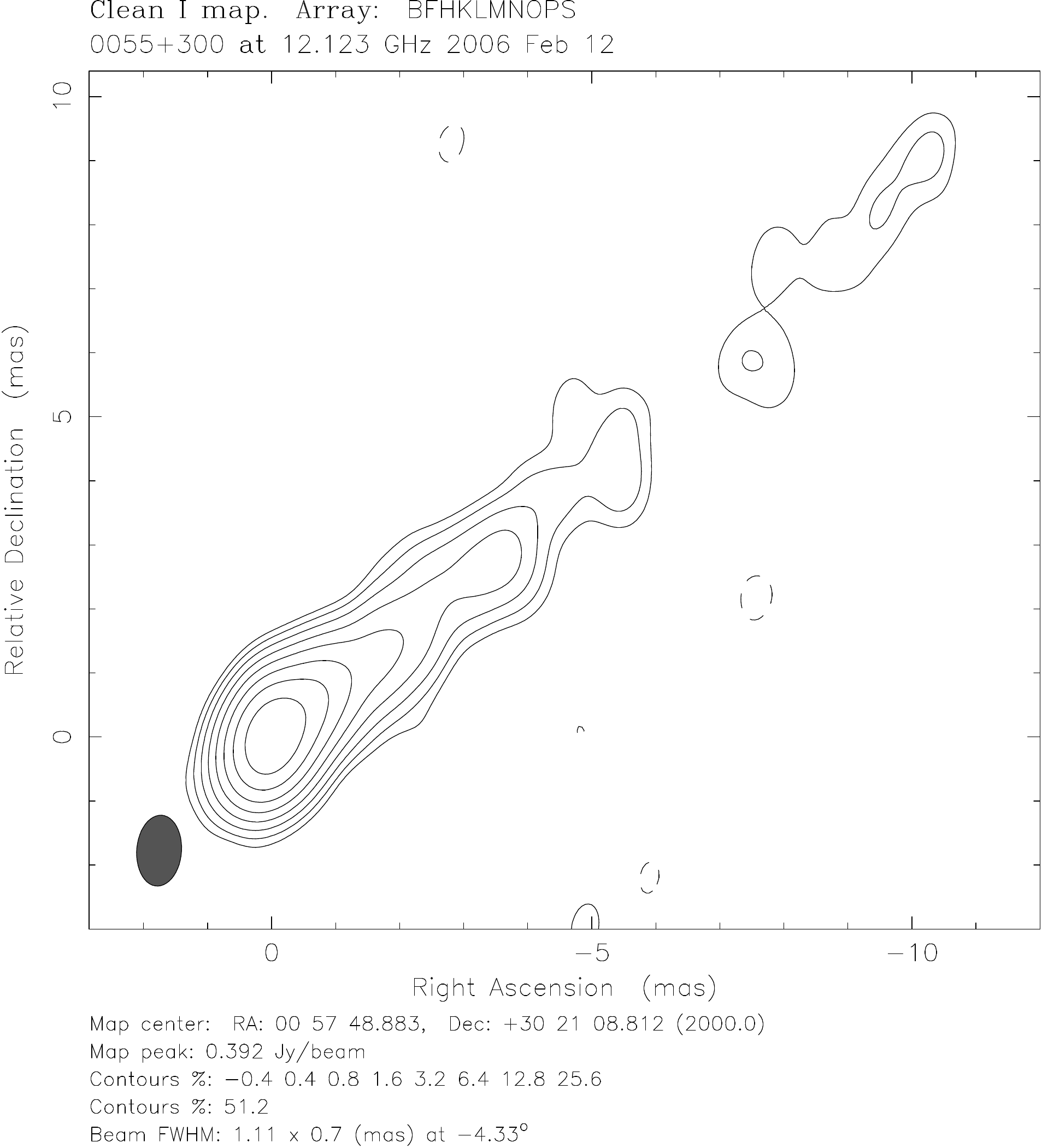}\quad\quad
    \includegraphics[height=6.4cm]{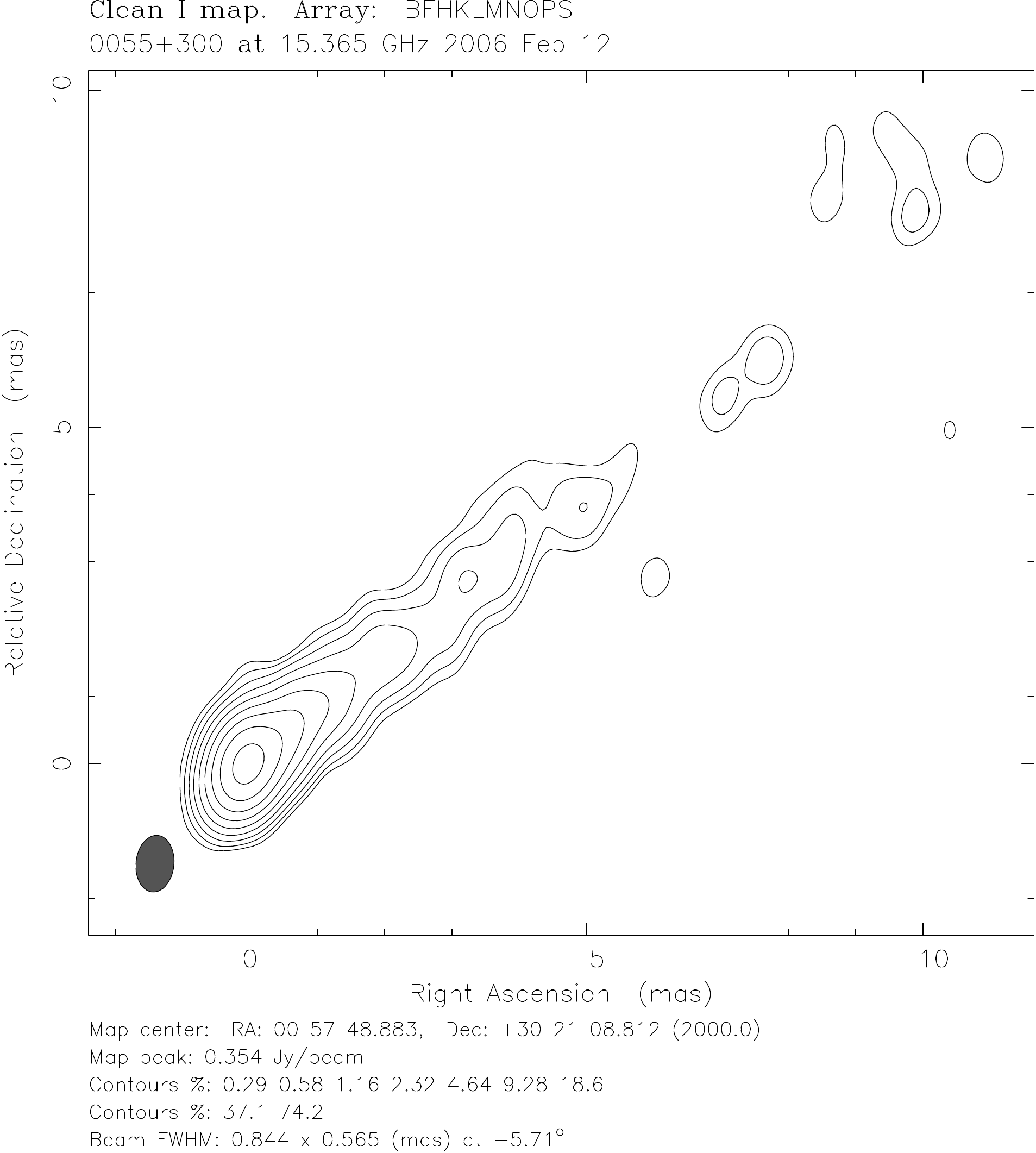}\\
    \caption{VLBI images of NGC\,315 at 12 GHz (left) and 15 GHz (right).}
 \vspace{1cm}
\end{figure*} 

\begin{figure*}[!h]
    \centering
    \includegraphics[height=6.4cm]{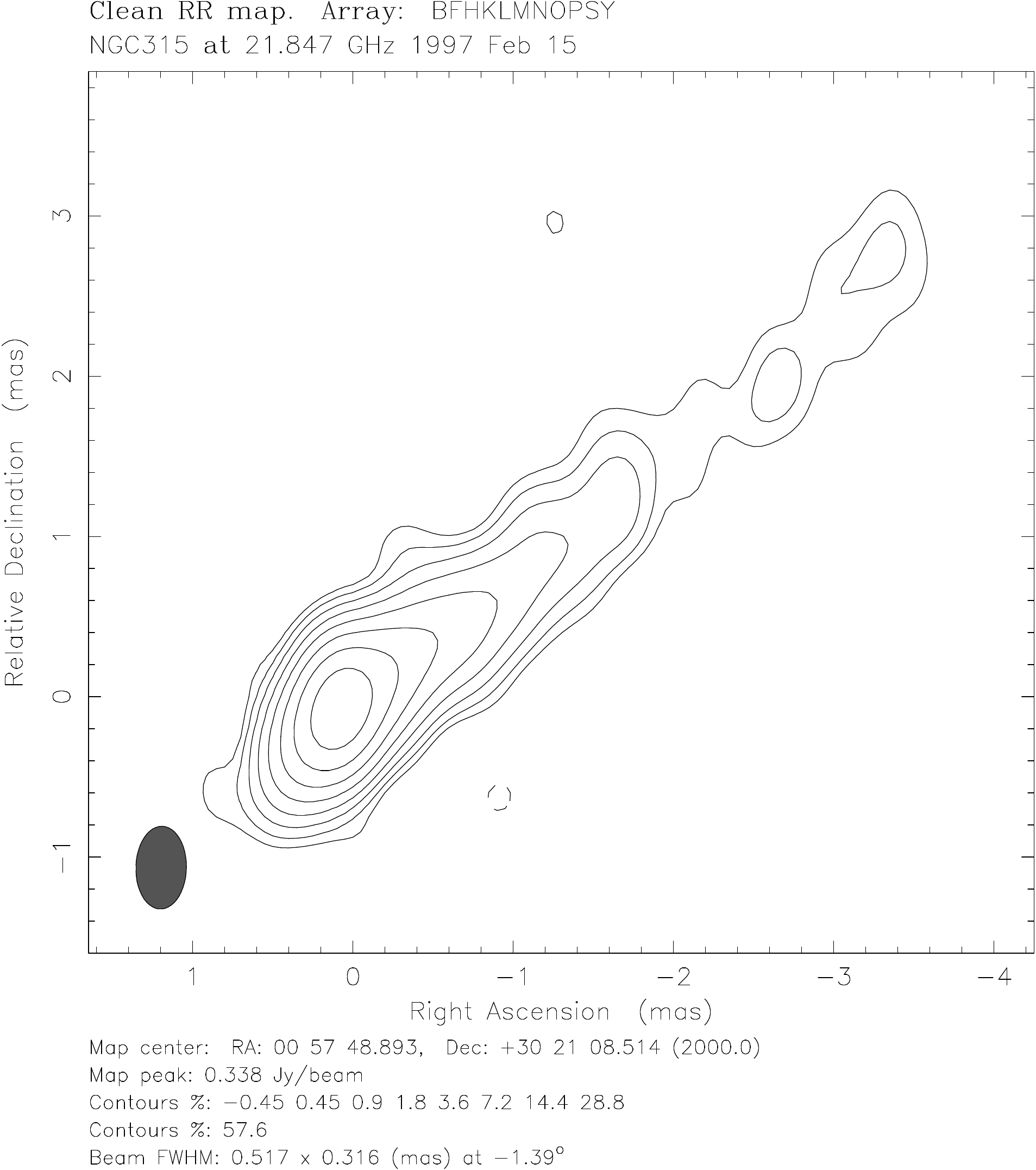}\quad
    \includegraphics[height=6.4cm]{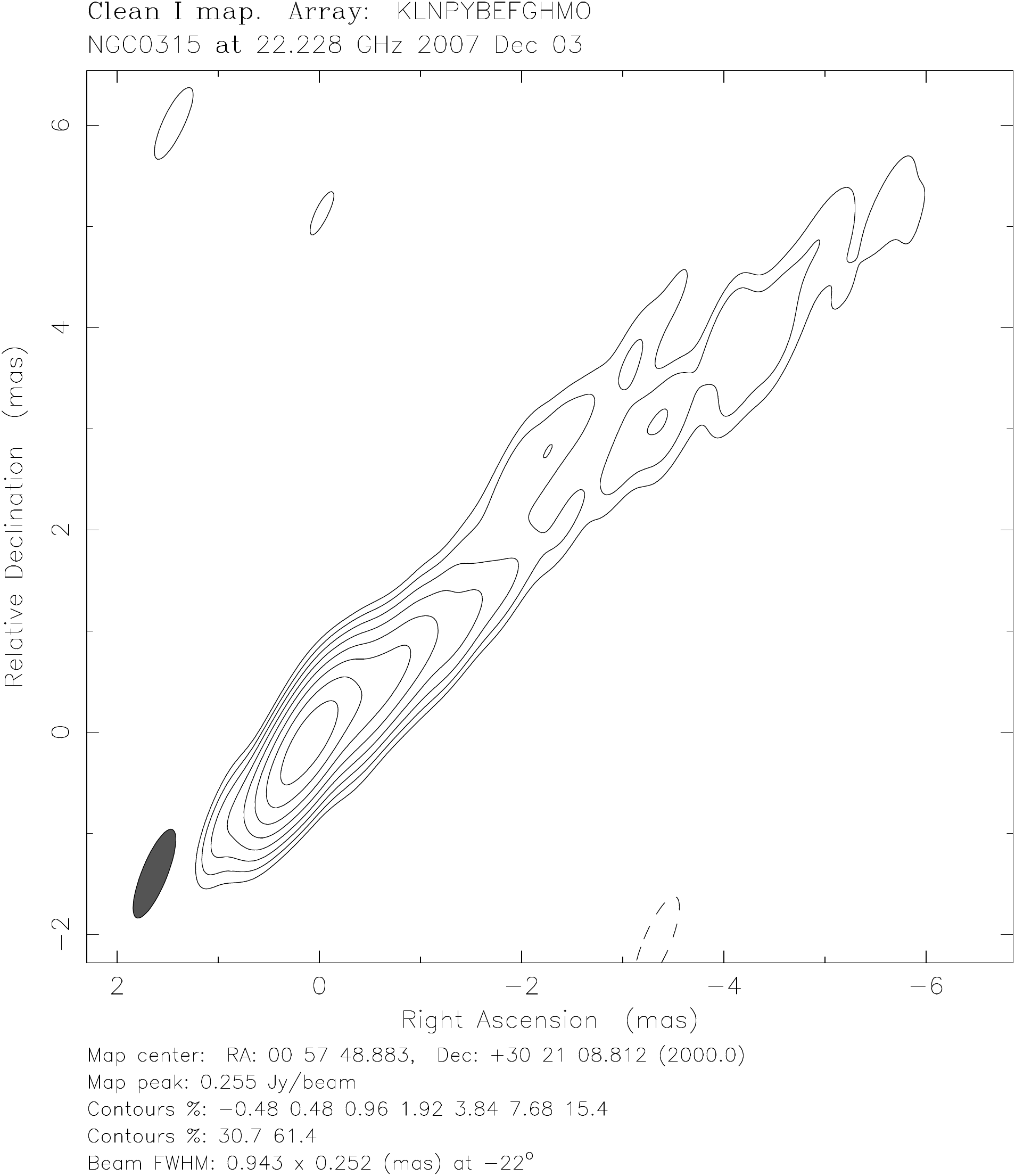}\quad
     \includegraphics[height=6.4cm]{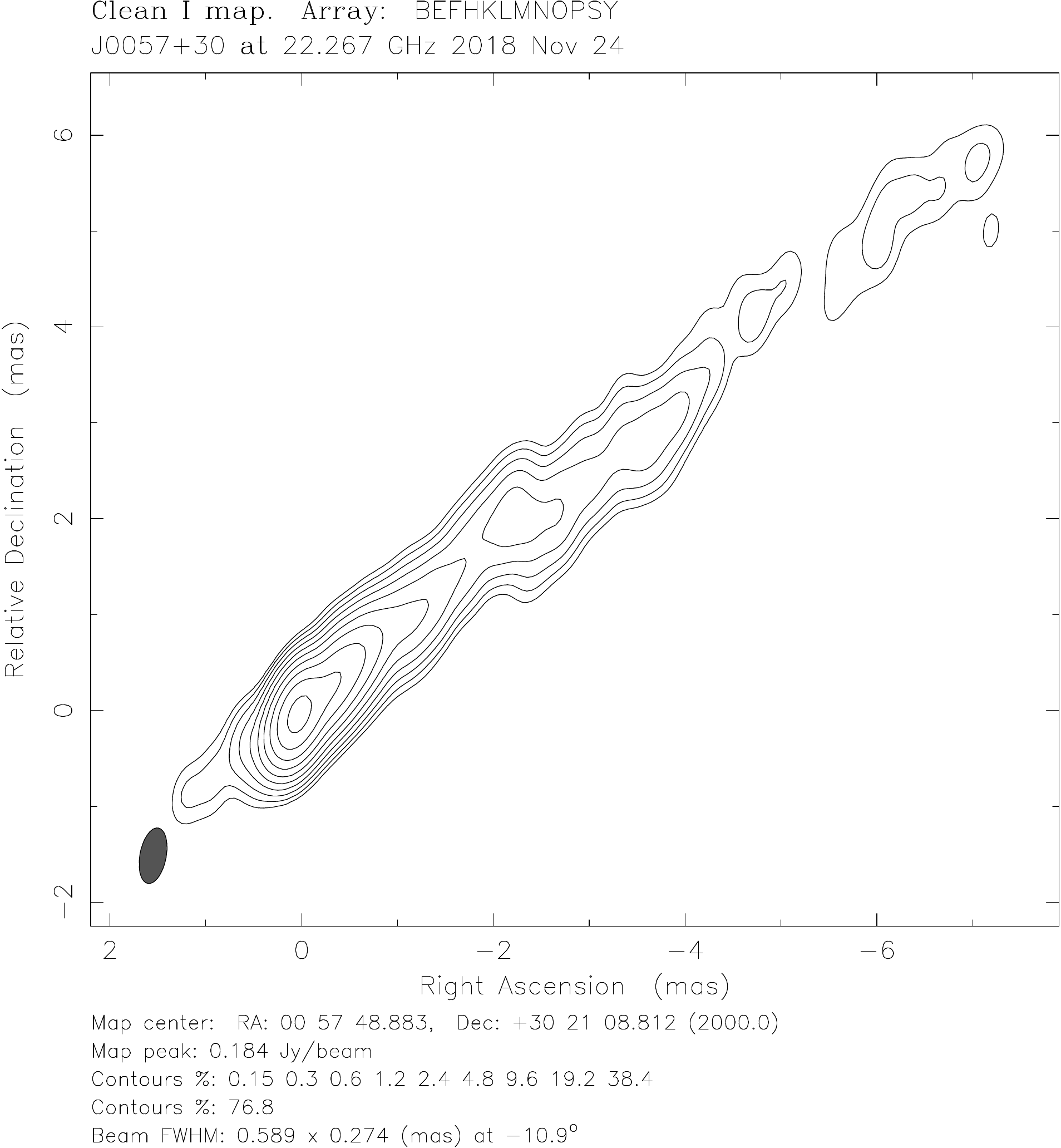}\\
    \caption{VLBI images of NGC\,315 at 22 GHz.}
   \vspace{1cm}
\end{figure*} 

\begin{figure*}[!h]
    \centering
    \includegraphics[height=6.4cm]{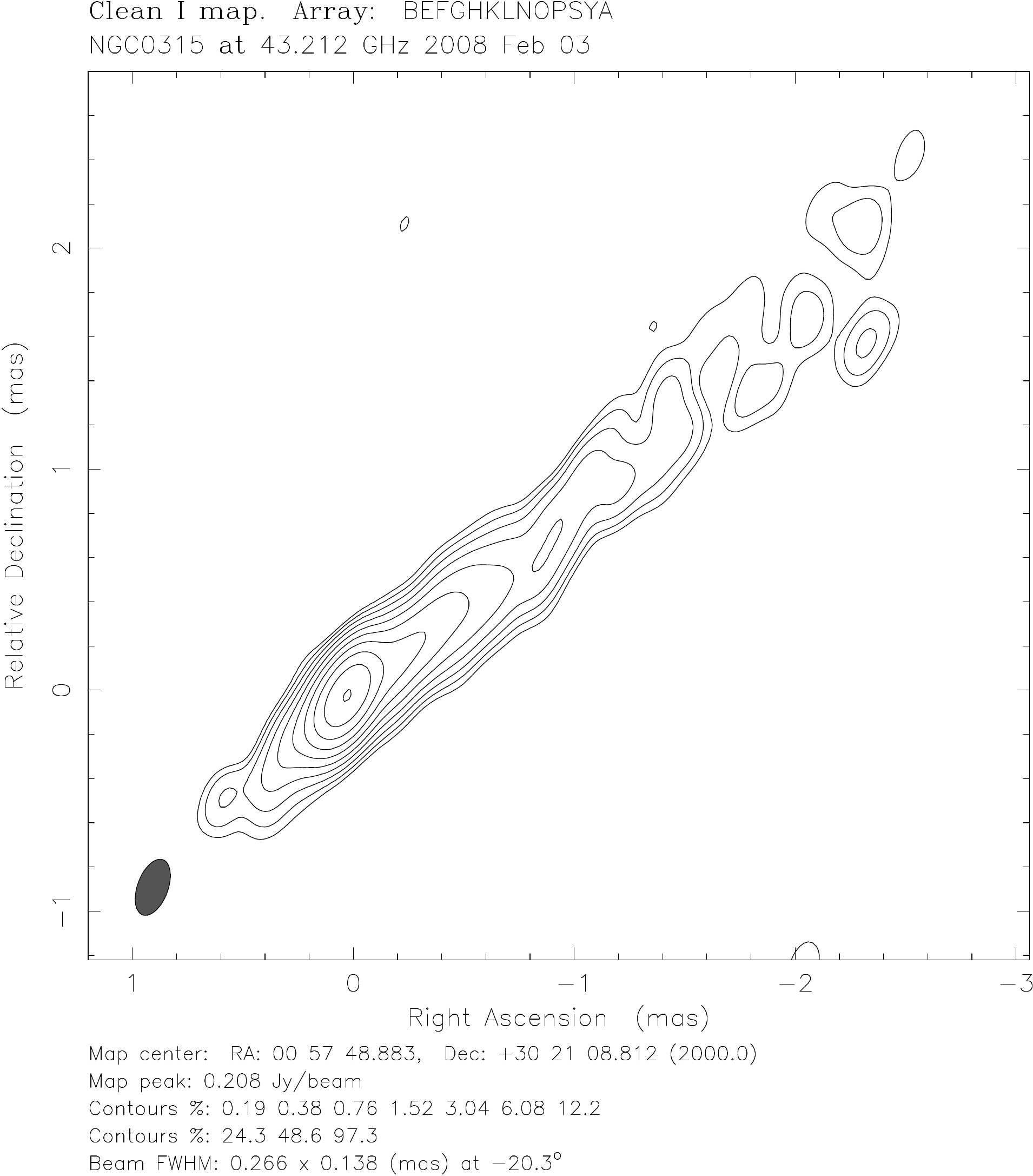}\quad\quad
    \includegraphics[height=6.4cm]{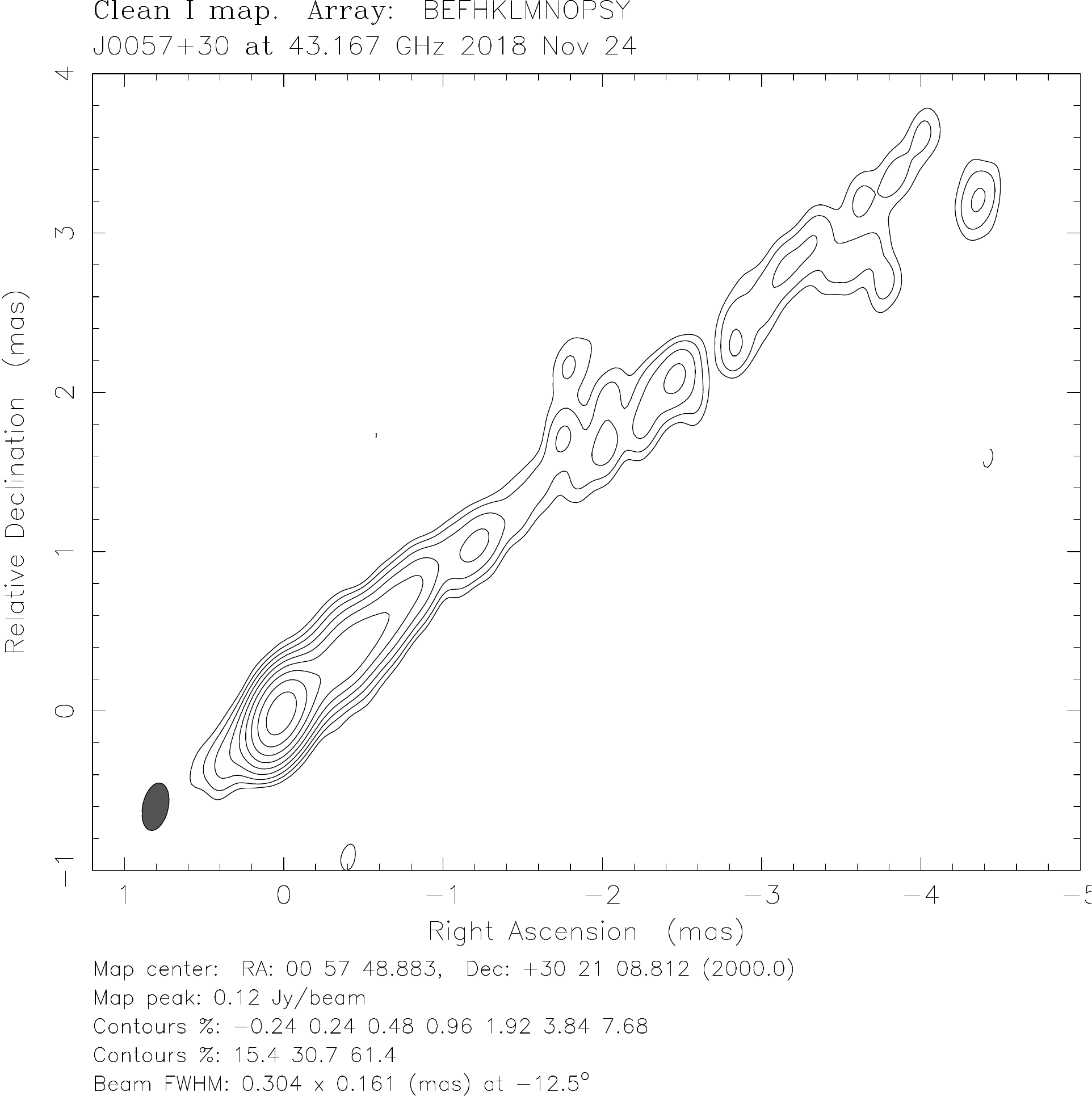}\\
    \caption{VLBI images of NGC\,315 at 43 GHz.}
    \label{fig:my_label}
\end{figure*} 

\begin{figure*}[!h]
    \centering
    \includegraphics[height=6.4cm]{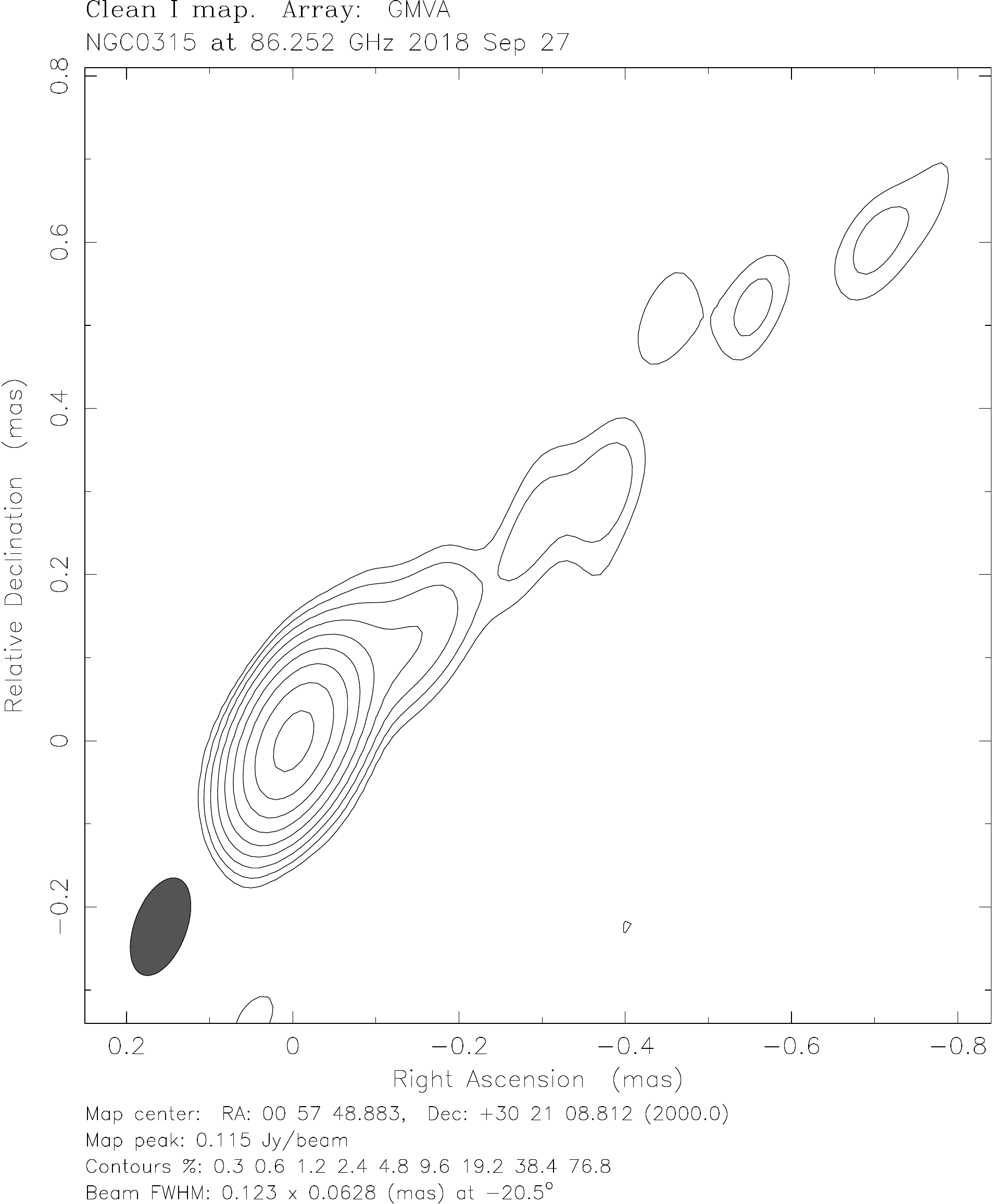}
    \caption{VLBI image of NGC\,315 at 86 GHz.}
   \vspace{1cm}
\end{figure*} 

\begin{figure*}
    \centering
    \includegraphics[height=8.2cm]{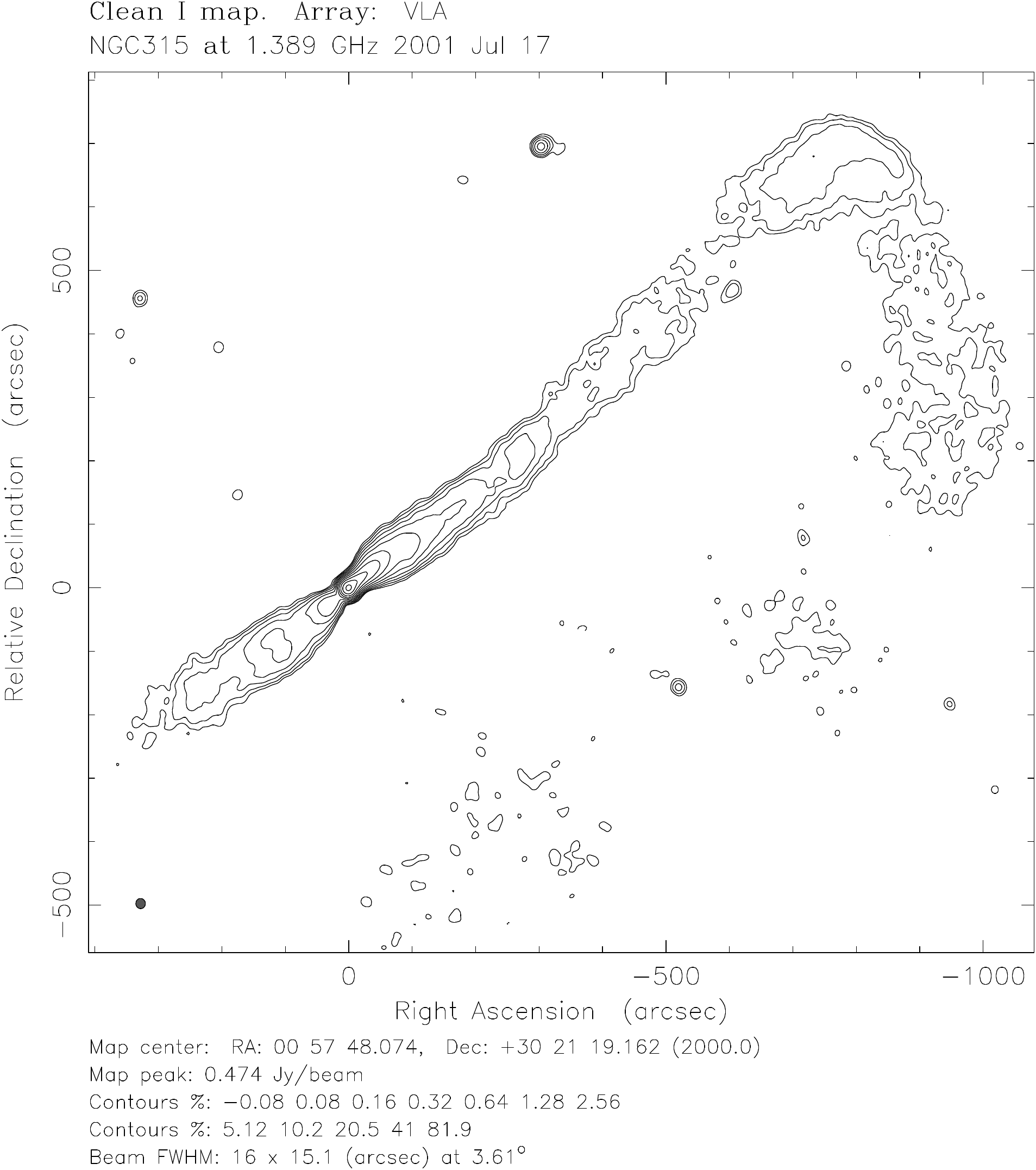}\quad\quad\quad
        \includegraphics[height=8.2cm]{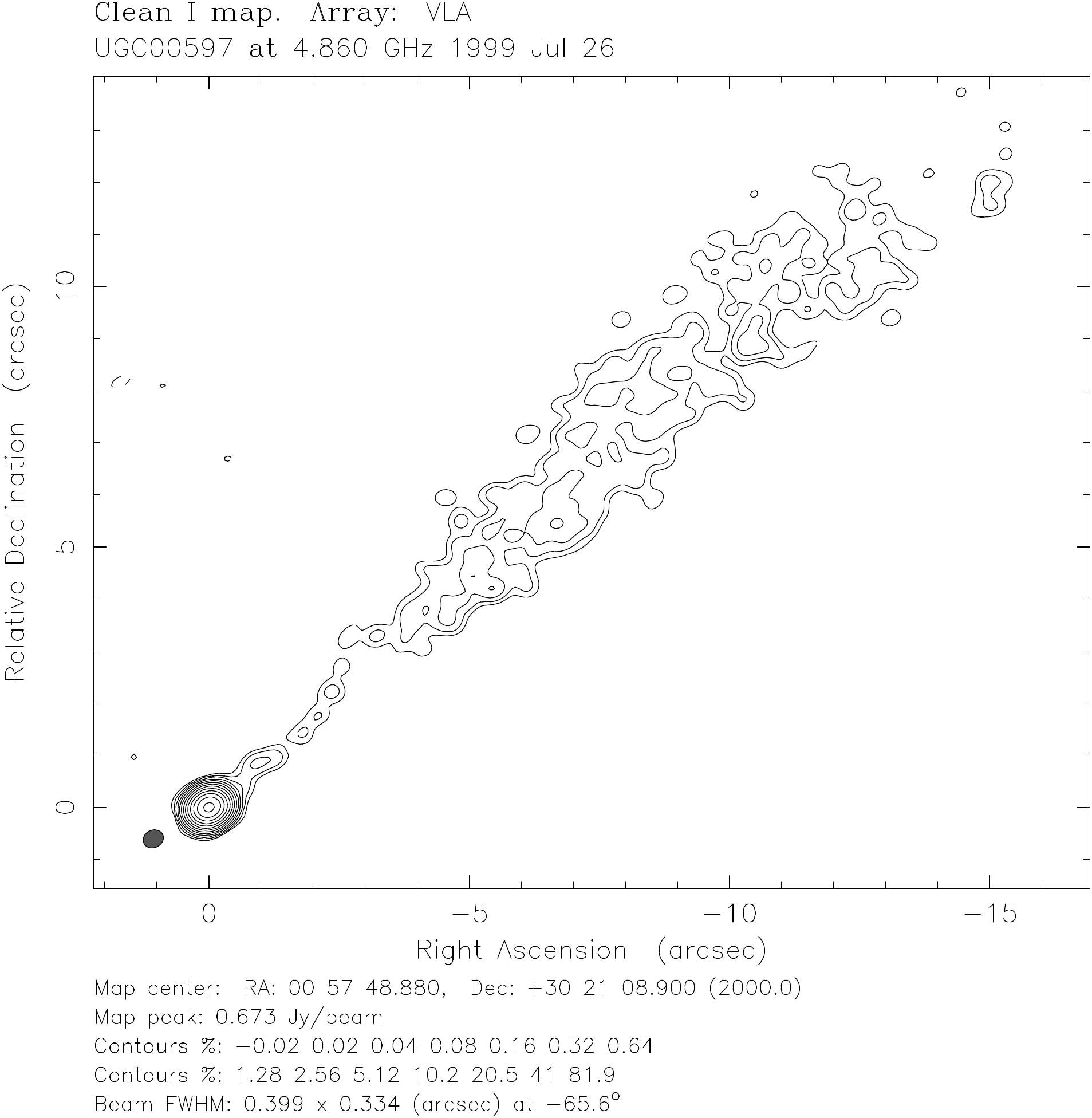}
    \caption{VLA images of NGC\,315 at 1 GHz (left) and 5 GHz (right).}
\end{figure*} 

\clearpage

\begin{table}[]
\scriptsize
\centering
\caption{Modelfit parameters of the circular Gaussian features describing the 1 GHz VLBI structure of NGC\,315 at each epoch. Column 1: Date of observation. Column 2: Integrated flux density. Column 3: Radial distance. Column 4: Position angle. Column 5: Radial distance with respect to the mm-core (shift applied as described in Sect. 2.2). Column 6: Position angle after the shift. Column 7: Full Width at Half Maximum. }
\begin{tabular}{ccccccc}
\hline
\hline
Epoch      & $\rm S$   & $\rm z$  &$\rm pa$  & $\rm z_{\rm sh}$  &  $\rm pa_{\rm sh}$  & $\rm d$  \\
           & $\mathrm {[mJy]}$& $\mathrm {[mas]}$&  $\rm{[deg]}$& $\mathrm {[mas]}$&  $\rm{[deg]}$ &  $\mathrm {[mas]}$  \\
           \hline
1994-11-26  &117.34     &-2.76      & 125.76    &-2.22         & -58.88     &  0.95    \\
            &124.05     &0.57       & -8.26     &5.37          & -51.81     &  -       \\
            &60.29      &5.28       & -43.90    &10.19         & -49.92     &  1.67    \\
            &16.87      &10.37      & -47.45    &15.31         & -50.32     &  0.94    \\
            &16.56      &17.02      & -50.30    &21.98         & -51.66     &  2.70    \\
            &9.51       &22.92      & -49.21    &27.86         & -50.47     &  4.37    \\
            &15.81      &34.12      & -49.77    &39.07         & -50.60     &  7.82    \\
            &8.31       &45.08      & -49.80    &50.02         & -50.44     &  10.44   \\
            \hline
2005-03-09  &5.81       &-42.91     & 133.06    &-38.05        & 133.98     &  6.15    \\
            &227.18     &-5.23      & 133.92    &-0.78         & 194.67     &  1.88    \\
            &338.58     &0.51       & -29.39    &5.37          & -51.81     &  2.06    \\
            &49.98      &13.12      & -49.04    &18.00         & -50.42     &  2.17    \\
            &21.42      &28.01      & -50.21    &32.90         & -50.79     &  5.29    \\
            &15.51      &44.44      & -50.39    &49.33         & -50.75     &  6.63    \\
            &12.13      &75.84      & -47.72    &80.71         & -48.11     &  22.16   \\
            &7.20       &134.71     & -47.59    &139.58        & -47.82     &  34.16   \\
            \hline
            \end{tabular}
\end{table}

\begin{table}[]
\scriptsize
\centering
\caption{Modelfit parameters of the circular Gaussian features describing the 5 GHz VLBI structure of NGC\,315 at each epoch. See the description in Table .1.}
\begin{tabular}{ccccccc}
\hline
\hline
Epoch      & $\rm S$   & $\rm z$  &$\rm pa$  & $\rm z_{\rm sh}$  &  $\rm pa_{\rm sh}$  & $\rm d$   \\
           & $\mathrm {[mJy]}$& $\mathrm {[mas]}$&  $\rm{[deg]}$& $\mathrm {[mas]}$&  $\rm{[deg]}$ &  $\mathrm {[mas]}$  \\
          \hline 
1994-11-15 &2.49     &  -4.25    &   134.07   &   -2.97   & 133.44   &   1.89  \\
           &170.46   &  -0.76    &   130.74   &    0.52   & -37.44   &   0.24  \\
           &224.80   &   0.12    &   -87.15   &    1.37   & -47.94   &   0.20  \\
           &79.24    &   1.18    &   -52.74   &    2.45   & -48.42   &   0.50  \\
           &49.92    &   2.79    &   -49.61   &    4.07   & -47.99   &   0.52  \\
           &30.57    &   4.63    &   -49.11   &    5.91   & -48.10   &   0.70  \\
           &18.00    &   7.10    &   -48.39   &    8.38   & -47.79   &   1.73  \\
           &12.38    &   10.43   &   -48.54   &    11.70  & -48.09   &   2.06  \\
           &3.95     &   14.08   &   -51.98   &    15.35  & -51.35   &   2.09  \\
           &4.95     &   23.03   &   -50.64   &    24.31  & -50.31   &   3.01  \\
           \hline 
1995-10-28 &1.61     &  -4.50    &   135.05   &   -3.17   & 136.04   &   0.90  \\
           &149.75   &  -0.74    &   127.77   &    0.60   & -41.23   &   0.31  \\
           &207.96   &   0.05    &   -65.83   &    1.37   & -47.93   &   0.31  \\
           &109.82   &   1.06    &   -52.20   &    2.39   & -49.49   &   0.44  \\
           &55.35    &   2.77    &   -49.79   &    4.10   & -48.99   &   0.64  \\
           &24.59    &   4.24    &   -48.97   &    5.57   & -48.57   &   0.76  \\
           &14.83    &   5.66    &   -48.17   &    6.99   & -48.01   &   0.89  \\
           &11.47    &   8.03    &   -47.86   &    9.36   & -47.78   &   1.14  \\
           &7.16     &   10.11   &   -48.95   &    11.44  & -48.76   &   1.49  \\
           &5.36     &   12.90   &   -49.55   &    14.23  & -49.34   &   2.54  \\
           &5.26     &   23.29   &   -51.10   &    24.61  & -50.90   &   4.09  \\
           \hline 
1996-05-10 &2.94     &  -5.27    &   128.84   &   -3.13   & 129.80   &   1.20  \\
           &254.18   &  -0.78    &   119.40   &    1.37   & -47.97   &   0.30  \\
           &210.28   &   0.67    &   -44.16   &    2.81   & -50.57   &   0.43  \\
           &82.35    &   3.41    &   -47.24   &    5.55   & -49.29   &   0.66  \\
           &20.88    &   6.46    &   -47.12   &    8.60   & -48.48   &   0.52  \\
           &11.57    &   9.72    &   -47.84   &    11.86  & -48.69   &   0.69  \\
           &7.07     &   13.10   &   -49.29   &    15.25  & -49.75   &   2.05  \\
           &7.56     &   22.46   &   -50.28   &    24.60  & -50.48   &   4.01  \\
           &1.99     &   32.18   &   -49.67   &    34.32  & -49.85   &   4.03  \\
           &3.13     &   43.07   &   -50.46   &    45.22  & -50.56   &   5.34  \\
           &2.06     &   53.49   &   -49.73   &    55.63  & -49.84   &   6.11  \\
           \hline 
1996-10-07 &2.09     &  -5.31    &   131.55   &   -3.71   & 131.40   &   2.72  \\
           &1.13     &  -3.27    &   129.41   &   -1.68   & 127.04   &   0.70  \\
           &88.80    &  -1.00    &   129.23   &    0.60   & -43.68   &   0.21  \\
           &214.78   &  -0.22    &   130.99   &    1.37   & -47.95   &   0.23  \\
           &123.97   &   0.80    &   -53.53   &    2.40   & -49.92   &   0.34  \\
           &51.16    &   2.26    &   -52.22   &    3.86   & -50.51   &   0.45  \\
           &39.27    &   3.61    &   -49.09   &    5.20   & -48.78   &   0.62  \\
           &24.87    &   5.13    &   -48.76   &    6.73   & -48.60   &   0.55  \\
           &10.17    &   6.68    &   -48.49   &    8.27   & -48.42   &   1.06  \\
           &6.59     &   8.37    &   -47.99   &    9.96   & -48.01   &   0.58  \\
           &9.06     &   10.09   &   -48.19   &    11.68  & -48.18   &   0.97  \\
           &4.34     &   11.82   &   -49.34   &    13.41  & -49.20   &   1.08  \\
           &2.97     &   13.96   &   -50.53   &    15.56  & -50.28   &   1.54  \\
           &2.68     &   17.29   &   -50.17   &    18.89  & -50.00   &   2.46  \\
           &3.88     &   21.87   &   -50.45   &    23.47  & -50.29   &   2.37  \\
           &2.91     &   25.11   &   -50.19   &    26.71  & -50.07   &   2.26  \\
           &2.22     &   31.23   &   -50.61   &    32.83  & -50.49   &   4.67  \\
           &3.48     &   41.84   &   -50.66   &    43.44  & -50.57   &   4.67  \\
           &2.63     &   52.13   &   -49.38   &    53.72  & -49.34   &   7.81  \\
           \hline  
2007-12-02 &226.05   &  -0.50    &   132.22   &    1.37   & -47.95   &   0.52  \\
           &170.46   &   0.48    &   -40.28   &    2.35   & -46.35   &   0.35  \\
           &36.87    &   1.42    &   -50.87   &    3.29   & -49.18   &   -     \\
           &40.59    &   1.98    &   -53.69   &    3.85   & -50.88   &   1.32  \\
           &40.67    &   3.82    &   -49.19   &    5.69   & -48.77   &   1.17  \\
           &28.04    &   6.05    &   -48.63   &    7.92   & -48.46   &   1.31  \\
           &9.19     &   8.74    &   -49.62   &    10.61  & -49.32   &   1.55  \\
           &12.65    &   12.49   &   -48.78   &    14.36  & -48.67   &   2.94  \\
           &10.35    &   24.07   &   -51.97   &    25.94  & -51.68   &   5.41  \\
           &12.65    &   12.49   &   -48.78   &    14.36  & -48.67   &   2.94  \\
           &10.35    &   24.07   &   -51.97   &    25.94  & -51.68   &   5.41  \\
           \hline                                                           
            \end{tabular}
\end{table}   

\begin{table}[]
\scriptsize
\centering
\caption{Modelfit parameters of the circular Gaussian features describing the 8 GHz VLBI structure of NGC\,315 at each epoch. See the description in Table .1.}
\begin{tabular}{ccccccc}
\hline
\hline
Epoch      & $\rm S$   & $\rm z$  &$\rm pa$  & $\rm z_{\rm sh}$  &  $\rm pa_{\rm sh}$  & $\rm d$  \\
           & $\mathrm {[mJy]}$& $\mathrm {[mas]}$&  $\rm{[deg]}$& $\mathrm {[mas]}$&  $\rm{[deg]}$ &  $\mathrm {[mas]}$  \\
           \hline
1994-11-15 &151.91   &  -0.39    &   117.89   &     0.32  & -40.98   &   0.26   \\ 
           &245.20   &   0.09    &   -29.95   &     0.78  & -50.20   &   0.07   \\ 
           &49.17    &   0.63    &   -47.69   &     1.33  & -50.29   &   -      \\ 
           &157.67   &   0.95    &   -50.93   &     1.65  & -51.66   &   0.53   \\ 
           &35.90    &   1.78    &   -48.22   &     2.48  & -49.47   &   0.31   \\ 
           &27.63    &   2.95    &   -49.87   &     3.65  & -50.40   &   0.52   \\ 
           &21.68    &   3.83    &   -48.72   &     4.53  & -49.33   &   0.93   \\ 
           &23.13    &   5.55    &   -47.41   &     6.25  & -48.00   &   0.92   \\ 
           &9.34     &   8.07    &   -47.25   &     8.76  & -47.68   &   1.16   \\ 
           \hline                                                                  
2006-02-12 &140.58   &  -0.43    &   123.29   &     0.21  & -55.69   &   -      \\ 
           &259.63   &   0.16    &   -24.60   &     0.78  & -50.22   &   0.19   \\ 
           &52.75    &   2.09    &   -49.42   &     2.73  & -51.05   &   0.58   \\ 
           &107.13   &   0.94    &   -47.49   &     1.58  & -51.09   &   0.29   \\ 
           &47.19    &   3.87    &   -49.48   &     4.51  & -50.46   &   1.10   \\ 
           &26.03    &   6.09    &   -49.33   &     6.73  & -50.00   &   1.55   \\ 
           &16.53    &   12.43   &   -46.29   &     13.06 & -46.78   &   2.18   \\ 
           \hline                                                                  
2008-02-03 &69.58    &  -0.83    &   131.43   &     0.23  & -51.87   &   -      \\ 
           &232.05   &  -0.28    &   133.23   &     0.78  & -50.19   &   0.06   \\ 
           &167.42   &   0.28    &   -56.61   &     1.34  & -50.83   &   0.11   \\ 
           &88.16    &   1.06    &   -50.48   &     2.12  & -49.88   &   0.22   \\ 
           &38.38    &   2.05    &   -50.59   &     3.11  & -50.15   &   0.45   \\ 
           &28.02    &   5.97    &   -49.44   &     7.02  & -49.42   &   1.61   \\ 
           &15.26    &   3.29    &   -52.63   &     4.34  & -51.81   &   0.42   \\ 
           &17.11    &   4.23    &   -51.00   &     5.29  & -50.66   &   0.61   \\ 
           \hline
            \end{tabular}
\end{table}  

\begin{table}[]
\scriptsize
\centering
\caption{Modelfit parameters of the circular Gaussian features describing the 12 GHz VLBI structure of NGC\,315 at each epoch. See the description in Table .1.}
\begin{tabular}{ccccccc}
\hline
\hline
Epoch      & $\rm S$   & $\rm z$  &$\rm pa$  & $\rm z_{\rm sh}$  &  $\rm pa_{\rm sh}$  & $\rm d$  \\
           & $\mathrm {[mJy]}$& $\mathrm {[mas]}$&  $\rm{[deg]}$& $\mathrm {[mas]}$&  $\rm{[deg]}$ &  $\mathrm {[mas]}$  \\
           \hline
2006-02-12 & 103.73  &  -0.40    &   131.11   &     0.07  & -70.98    &   0.10  \\  
           & 327.38  &   0.04    &   -66.58   &     0.50  & -53.14    &   0.16  \\  
           & 118.71  &   0.62    &   -52.95   &     1.09  & -52.59    &   0.16  \\  
           & 54.62   &   1.24    &   -51.37   &     1.70  & -51.57    &   0.23  \\  
           & 58.43   &   2.26    &   -52.32   &     2.73  & -52.28    &   0.85  \\  
           & 42.66   &   4.29    &   -51.40   &     4.75  & -51.47    &   1.08  \\  
           & 22.11   &   6.73    &   -50.46   &     7.20  & -50.57    &   2.10  \\  
           \hline
\end{tabular}
\end{table}  

\begin{table}[]
\scriptsize
\centering
\caption{Modelfit parameters of the circular Gaussian features describing the 15 GHz VLBI structure of NGC\,315 at each epoch. See description in Table .1.}
\begin{tabular}{ccccccc}
\hline
\hline
Epoch      & $\rm S$   & $\rm z$  &$\rm pa$  & $\rm z_{\rm sh}$  &  $\rm pa_{\rm sh}$  & $\rm d$  \\
           & $\mathrm {[mJy]}$& $\mathrm {[mas]}$&  $\rm{[deg]}$& $\mathrm {[mas]}$&  $\rm{[deg]}$ &  $\mathrm {[mas]}$  \\
           \hline
2006-02-12 & 87.09   & -0.38     &    129.61  &     -0.06 & 229.60    &   0.11  \\ 
           & 274.28  & -0.04     &     92.04  &      0.36 & -56.38    &   -     \\ 
           & 94.65   &  0.35     &    -51.74  &      0.75 & -55.84    &   0.05  \\ 
           & 63.46   &  0.83     &    -48.83  &      1.22 & -52.27    &   0.17  \\ 
           & 35.66   &  1.41     &    -49.76  &      1.80 & -51.89    &   0.35  \\ 
           & 22.57   &  4.15     &    -51.38  &      4.54 & -52.08    &   0.81  \\ 
           & 33.26   &  2.43     &    -51.58  &      2.82 & -52.69    &   0.68  \\ 
           & 14.46   &  5.79     &    -48.94  &      6.18 & -49.62    &   1.05  \\ 
           \hline
\end{tabular}
\end{table}  

\begin{table}[]
\scriptsize
\centering
\caption{Modelfit parameters of the circular Gaussian features describing the 22 GHz VLBI structure of NGC\,315 at each epoch. See the description in Table .1.}
\begin{tabular}{ccccccc}
\hline
\hline
Epoch      & $\rm S$   & $\rm z$  &$\rm pa$  & $\rm z_{\rm sh}$  &  $\rm pa_{\rm sh}$  & $\rm d$  \\
           & $\mathrm {[mJy]}$& $\mathrm {[mas]}$&  $\rm{[deg]}$& $\mathrm {[mas]}$&  $\rm{[deg]}$ &  $\mathrm {[mas]}$  \\
           \hline
1997-02-15 & 125.18   &-0.34     &    132.29  &   -0.12   &    125.64    &     0.04   \\
           & 276.06   &-0.10     &    136.58  &    0.13   &    -44.93    &     -      \\
           & 111.00   & 0.21     &    -56.10  &    0.43   &    -49.92    &     0.11   \\
           & 69.39    & 0.62     &    -54.72  &    0.84   &    -51.95    &     0.23   \\
           & 36.33    & 1.10     &    -56.06  &    1.32   &    -54.07    &     0.23   \\
           & 31.23    & 1.90     &    -53.32  &    2.13   &    -52.36    &     0.50   \\
           & 26.57    & 4.07     &    -50.32  &    4.29   &    -50.01    &     1.38   \\
           \hline               
2007-12-02 & 20.62    &-0.77     &    132.50  &   -0.63   &    129.70    &     0.08   \\
           & 187.74   &-0.34     &    131.58  &   -0.21   &    122.28    &     0.08   \\
           & 195.40   &-0.03     &    193.93  &    0.13   &    -45.10    &     0.10   \\
           & 84.46    & 0.43     &    -56.83  &    0.57   &    -51.45    &     0.21   \\
           & 49.33    & 1.01     &    -50.31  &    1.15   &    -48.43    &     0.20   \\
           & 28.54    & 1.52     &    -52.85  &    1.65   &    -51.33    &     0.42   \\
           & 20.10    & 2.57     &    -48.83  &    2.71   &    -48.10    &     0.65   \\
           & 17.78    & 4.17     &    -48.39  &    4.31   &    -47.95    &     0.74   \\
           \hline               
2018-11-24 & 39.45    &-0.29     &    133.83  &   -0.11   &    126.83    &     0.09   \\
           & 161.55   &-0.06     &    143.84  &    0.13   &    -45.19    &     0.07   \\
           & 67.69    & 0.24     &    -52.75  &    0.42   &    -48.08    &     0.10   \\
           & 56.21    & 0.56     &    -48.56  &    0.75   &    -46.96    &     0.07   \\
           & 29.17    & 0.88     &    -48.14  &    1.07   &    -47.09    &     0.13   \\
           & 17.18    & 1.26     &    -48.42  &    1.45   &    -47.61    &     0.31   \\
           & 15.32    & 1.80     &    -49.58  &    1.98   &    -48.88    &     0.36   \\
           & 28.00    & 2.99     &    -48.21  &    3.18   &    -47.85    &     0.76   \\
           & 23.48    & 4.65     &    -50.82  &    4.84   &    -50.48    &     1.37   \\
           \hline
\end{tabular}
\end{table}  

\begin{table}[]
\scriptsize
\centering
\caption{Modelfit parameters of the circular Gaussian features describing the 43 GHz VLBI structure of NGC\,315 at each epoch. See the description in Table .1.}
\begin{tabular}{ccccccc}
\hline
\hline
Epoch      & $\rm S$   & $\rm z$  &$\rm pa$  & $\rm z_{\rm sh}$  &  $\rm pa_{\rm sh}$  & $\rm d$  \\
           & $\mathrm {[mJy]}$& $\mathrm {[mas]}$&  $\rm{[deg]}$& $\mathrm {[mas]}$&  $\rm{[deg]}$ &  $\mathrm {[mas]}$  \\
           \hline
2008-02-03 & 9.91     &-0.41      &   133.14  &  -0.40     &    133.20    &    0.03   \\
           & 85.16    &-0.13      &   131.40  &  -0.12     &    131.46    &    0.05   \\
           & 166.24   &-0.01      &   129.78  &   0.00     &    89.01     &    -      \\
           & 43.11    & 0.18      &   -57.10  &   0.19     &    -56.74    &    0.09   \\
           & 23.17    & 0.39      &   -53.54  &   0.40     &    -53.44    &    0.06   \\
           & 30.98    & 0.62      &   -53.87  &   0.63     &    -53.80    &    0.22   \\
           & 17.29    & 1.00      &   -56.10  &   1.01     &    -56.04    &    0.26   \\
           & 20.32    & 1.49      &   -50.14  &   1.50     &    -50.13    &    0.45   \\
           \hline                                                                   
2018-11-24 & 79.18    & 0.09      &   131.66  &   0.00     &    200.94    &    0.05   \\
           & 74.82    & 0.04      &   -53.10  &   0.14     &    -50.58    &    0.03   \\
           & 23.56    & 0.30      &   -50.24  &   0.39     &    -50.04    &    0.13   \\
           & 31.10    & 0.61      &   -48.36  &   0.70     &    -48.49    &    0.20   \\
           & 15.91    & 1.02      &   -46.99  &   1.11     &    -47.19    &    0.27   \\
           & 8.74     & 1.53      &   -48.86  &   1.62     &    -48.89    &    0.37   \\
           & 8.08     & 2.96      &   -47.28  &   3.05     &    -47.35    &    0.54   \\
           & 9.16     & 4.31      &   -47.23  &   4.40     &    -47.27    &    1.20   \\
            \hline                                                                         
\end{tabular}
\end{table}  

\begin{table}[]
\scriptsize
\centering
\caption{Modelfit parameters of the circular Gaussian features describing the 86 GHz VLBI structure of NGC\,315 at each epoch. See the description in Table .1. In addition to the unresolved components (see Sect. 3.2) we exclude from Fig. 3 the outermost jet feature at ${\sim}0.8$\,$\rm mas$, which clearly occupies only a small part of the jet cross-section. }
\begin{tabular}{ccccccc}
\hline
\hline
Epoch      & $\rm S$   & $\rm z$  &$\rm pa$  & $\rm z_{\rm sh}$  &  $\rm pa_{\rm sh}$  & $\rm d$  \\
           & $\mathrm {[mJy]}$& $\mathrm {[mas]}$&  $\rm{[deg]}$& $\mathrm {[mas]}$&  $\rm{[deg]}$ &  $\mathrm {[mas]}$  \\
          \hline
2018-09-27 &96.88     & 0.01      & 137.26     &  0.00       &    55.30    &   0.01   \\
           &29.87     & 0.03      & -55.72     &  0.04       &   -52.74    &   0.02   \\
           &6.30      & 0.16      & -48.52     &  0.17       &   -48.15    &   0.09   \\
           &3.55      & 0.43      & -50.74     &  0.44       &   -50.56    &   0.15   \\
           &2.16      & 0.81      & -47.26     &  0.82       &   -47.19    &   0.07   \\
           \hline
\end{tabular}
\end{table} 

\begin{table}[]
\scriptsize
\centering
\caption{Modelfit parameters of the circular Gaussian features describing the 1.4 GHz VLA structure of NGC\,315. See the description in Table .1. }
\begin{tabular}{ccccccc}
\hline
\hline
Epoch      & $\rm S$   & $\rm z$  &$\rm pa$  & $\rm z_{\rm sh}$  &  $\rm pa_{\rm sh}$  & $\rm d$  \\
           & $\mathrm {[mJy]}$& $\mathrm {['']}$&  $\rm{[deg]}$& $\mathrm {['']}$&  $\rm{[deg]}$ &  $\mathrm {['']}$  \\
          \hline
2001-07-17  &50.60  &    -285.82   &      125.39 &    284.76   &     125.37 &   73.57    \\
            &52.25  &    -197.23   &      127.44 &    196.16   &     127.42 &   64.45    \\
            &112.99 &    -140.75   &      129.17 &    139.68   &     129.16 &   56.84    \\
            &54.74  &    -76.48    &      131.43 &    75.40    &     131.44 &   42.40    \\
            &54.28  &    -45.78    &      130.83 &    44.70    &     130.83 &   21.31    \\
            &17.05  &    -28.61    &      130.77 &    27.54    &     130.77 &   99.97    \\
            &22.90  &    -7.67     &      130.06 &    6.59     &     129.93 &   25.22    \\
            &429.12 &    -1.07     &      130.88 &    0.00     &     -51.81 &   1.82     \\
            &157.33 &     11.06    &      -48.00 &    12.13    &     -48.10 &   2.15     \\
            &102.15 &     21.90    &      -50.09 &    22.97    &     -50.04 &   5.63     \\
            &113.37 &     34.55    &      -50.78 &    35.62    &     -50.73 &   12.73    \\
            &129.54 &     49.79    &      -49.74 &    50.86    &     -49.73 &   20.01    \\
            &146.81 &     70.30    &      -49.23 &    71.38    &     -49.23 &   32.72    \\
            &163.73 &     111.21   &      -53.42 &    112.28   &     -53.38 &   40.98    \\
            &80.39  &     163.09   &      -53.33 &    164.17   &     -53.30 &   45.10    \\
            &66.45  &     218.39   &      -52.92 &    219.46   &     -52.90 &   45.95    \\
            &44.67  &     272.91   &      -53.77 &    273.98   &     -53.75 &   54.26    \\
            &56.97  &     339.59   &      -52.45 &    340.67   &     -52.44 &   50.14    \\
            &35.60  &     403.46   &      -51.16 &    404.53   &     -51.16 &   69.94    \\
            &49.84  &     508.44   &      -49.76 &    509.52   &     -49.76 &   95.12    \\
            &64.86  &     627.26   &      -49.16 &    628.34   &     -49.16 &   111.00   \\
            &43.51  &     783.21   &      -47.91 &    784.29   &     -47.91 &   142.20   \\
            &89.16  &     943.58   &      -47.85 &    944.65   &     -47.85 &   95.32    \\
            &49.81  &     1040.40  &      -47.85 &    1041.47  &     -47.85 &   73.28    \\
            &55.22  &     1068.35  &      -52.00 &    1069.43  &     -51.99 &   94.70    \\
            &68.66  &     1016.04  &      -62.31 &    1017.09  &     -62.30 &   157.20   \\
            &110.51 &     963.84   &      -75.23 &    964.80   &     -75.20 &   190.00   \\
            &204.51 &     709.79   &      -95.66 &    710.53   &     -95.66 &   356.60   \\
            &168.65 &     460.18   &      -142.87 &   460.11   &     -142.87&   373.50   \\
           \hline
\end{tabular}
\end{table} 

\begin{table}[]
\scriptsize
\centering
\caption{Modelfit parameters of the circular Gaussian features describing the 5 GHz VLA structure of NGC\,315. See the description in Table .1. }
\begin{tabular}{ccccccc}
\hline
\hline
Epoch      & $\rm S$   & $\rm z$  &$\rm pa$  & $\rm z_{\rm sh}$  &  $\rm pa_{\rm sh}$  & $\rm d$  \\
           & $\mathrm {[mJy]}$& $\mathrm {['']}$&  $\rm{[deg]}$& $\mathrm {['']}$&  $\rm{[deg]}$ &  $\mathrm {['']}$  \\
          \hline
1999-07-26  &674.11  &   0.00   &    -88.49  &  0.00   &     -47.79   &   0.02   \\ 
            &0.88    &   1.28   &    -48.24  &  1.28   &     -48.22   &   0.22   \\
            &3.24    &   5.85   &    -49.70  &  5.85   &     -49.70   &   0.71   \\
            &5.58    &   6.86   &    -49.00  &  6.86   &     -48.99   &   0.85   \\
            &18.85   &   8.79   &    -48.15  &  8.79   &     -48.15   &   1.93   \\
            &21.31   &   11.51  &    -46.46  &  11.51  &     -46.45   &   2.11   \\
            &28.40   &   15.44  &    -48.74  &  15.45  &     -48.74   &   3.23   \\
           \hline
\end{tabular}
\end{table} 

\end{document}